\documentclass[floats,floatfix,showpacs,amssymb,prd,twocolumn,superscriptaddress,nofootinbib,nolongbibliography,reprint]{revtex4-2}

\usepackage{amssymb,amsmath,verbatim,mathtools,needspace,enumitem,etoolbox,graphicx,physics,microtype,afterpage,xspace,tabularx,lmodern,multirow,mathrsfs}
\usepackage{gensymb}
\usepackage[normalem]{ulem}
\usepackage[dvipsnames, usenames]{xcolor}
\definecolor{linkcolor}{rgb}{0.0,0.3,0.5}
\usepackage[unicode, colorlinks=true, linkcolor=linkcolor, citecolor=linkcolor, filecolor=linkcolor, urlcolor=linkcolor, linktocpage, breaklinks]{hyperref}
\usepackage[all]{hypcap}
\usepackage{subfigure}
\usepackage[T1]{fontenc}
\usepackage[utf8]{inputenc}
\usepackage[usenames,dvipsnames]{xcolor}

\usepackage{braket}

\hypersetup{colorlinks=true,citecolor=teal,linkcolor=teal,urlcolor=teal}

\definecolor{teal}{RGB}{0,128,128}
\newcommand{\nn}{\nonumber}

\newcommand{\be}{\begin{equation}}
\newcommand{\ee}{\end{equation}}

\usepackage{aas_macros}
\usepackage{makecell}
\usepackage{soul}
\usepackage{amssymb}
\usepackage{longtable}
\usepackage{lipsum}

\newcommand{\bea}{\begin{eqnarray}}
\newcommand{\eea}{\end{eqnarray}}
\newcommand{\ba}{\begin{align}}
\newcommand{\ea}{\end{align}}

\def\p{\partial}
\renewcommand{\i}{{\rm i}}
\renewcommand{\d}{\,\textnormal{d}} 

\def\s{\bar s}

\newcommand{\rh}{r_{\rm h}}

\newcolumntype{Y}{>{\centering\arraybackslash}X}

\def\tp{\check\tau}
\def\sigp{\check\sigma}
\def\phip{\check\varphi}
\def\xp{\check x}

\def\scri{\mathscr{I}}

\begin{document}

\title{\textbf{Bilinear products and the orthogonality of quasinormal modes on hyperboloidal foliations}}

\author{Marica Minucci}
\email{marica.minucci@nbi.ku.dk}
\author{Rodrigo Panosso Macedo}
\email{rodrigo.macedo@nbi.ku.dk}
\affiliation{Center of Gravity, Niels Bohr Institute, Blegdamsvej 17, 2100 Copenhagen, Denmark}

\author{Christiana Pantelidou}
\email{christiana.pantelidou@ucd.ie}
\affiliation{School of Mathematics and Statistics, University College Dublin, Dublin, Ireland}

\author{Laura Sberna}
\email{laura.sberna@nottingham.ac.uk}
\affiliation{School of Mathematical Sciences, University of Nottingham, University Park, Nottingham NG7 2RD, United Kingdom}

\begin{abstract}
We explore the properties of bilinear products for black-hole quasinormal modes (QNMs) formulated on hyperboloidal foliations. We find that, although QNM solutions are smooth and finite on future-directed hyperboloids, the integrand of the bilinear form with respect to which the modes are orthogonal is still divergent. This is a result of the reflection (equivalently, CPT) transformation required in the definition of the products, which modifies the behaviour of the integrand at the boundaries.
We present several regularisation procedures that yield a finite and well-defined bilinear form. In addition, we examine an alternative definition of the bilinear products that incorporates flux contributions, discussing its advantages and limitations.
Finally, we define the QNM excitation factors and coefficients within the hyperboloidal framework in terms of the bilinear products, and compute them explicitly for a choice of mode numbers and constant initial data. 
For concreteness, we work with the QNMs associated to scalar perturbations of the Schwarzschild family of spacetimes.

\end{abstract}

\maketitle
\tableofcontents

\newpage
\section{Introduction}
When two black holes merge, the remnant is expected to be a spinning, distorted black hole, with the corresponding spacetime emitting gravitational waves in a characteristic ``ringdown.'' The timescales associated with this exponentially damped and oscillatory signal are encoded in the so-called quasinormal modes (QNMs). These modes form a set of complex frequencies $\omega_{\ell m n}$, labelled by discrete indices $(\ell,m,n)$, where $(\ell,m)$ capture the angular distribution pattern of the radiation, while $n$ accounts for the tower of allowed vibration and decay timescales (overtones). Quasinormal modes are fingerprints of the final black hole spacetime, carrying information not only on astrophysical observables but also on the fundamental nature of gravity. Measuring these frequencies is the main goal of the black hole spectroscopy program, as this information enables precision tests of general relativity, particularly when multiple modes are resolved---see \cite{Berti:2025hly} and references therein for a recent and comprehensive review. With current observations already investigating the presence of overtones~\cite{LIGO2025,LIGOScientific:2026wpt}, next-generation detectors promise the sensitivity required for precision black hole spectroscopy \cite{Berti:2016lat,Maggiore:2019uih,Cabero:2019zyt,Toubiana:2023cwr}.

In addition to the challenges of detecting QNMs in gravitational-wave signals, fundamental conceptual issues within the underlying theory remain unresolved. Formally, the QNM problem arises in black hole perturbation theory once the underlying wave equations are solved with the physically relevant boundary conditions at the horizon and at future null infinity. However, within this linear response, the QNM frequencies dominate the signal only during an intermediate phase. The early-time dynamics are characterised by a prompt response that describes freely propagating radiation, while the late-time behaviour is dominated by power-law tails~\cite{Leaver1986,Price:1971fb,Gundlach:1993tp}. Strategies to extract QNMs from data rely on the separation of these different regimes, and therefore there has recently been a surge in analytical methods to better model them \cite{DeAmicis:2024not,Arnaudo:2025uos,Arnaudo:2025kit,DeAmicis:2025xuh,DeAmicis:2026tus,Su:2026fvj,Ma:2026qbq}.

Another theoretical issue arises when analysing the system in the frequency domain. Upon performing a Fourier transform of the original field equation, the QNM frequencies are identified as eigenvalues $\omega_{\ell m n}$---with associated eigenfunctions---of the corresponding evolution operator. In conservative (e.g. Hermitian) systems, such eigenfunctions form a complete basis in an appropriate functional space. Thus, the system is naturally endowed with a scalar product, allowing the projection of an arbitrary function onto that basis. However, this property is absent from the QNM problem, due to the leakage of energy through the spacetime boundaries.
It has long been known that QNM eigenfunctions blow up at the event horizon and in the infinitely far region, a feature that has historically prevented the definition of an appropriate functional space with $L^2$-type inner products. As a consequence, projecting solutions onto QNMs, an essential step for analysing nonlinear mode interactions and defining excitation amplitudes, remains ambiguous.

Despite these difficulties, it is possible to define a product under which QNMs with different frequencies are orthogonal, in analogy to other non-Hermitian physical systems. The product is understood as a complex-valued bilinear form, rather than as an inner product. Based on earlier developments for perturbations of the Schwarzschild family of spacetimes \cite{Ching:1993gt,Ching:1995rt}, Ref.~\cite{GHSS2023} provided a general recipe to construct such a product for perturbations of the Kerr family of spacetimes. The product is built from the symplectic form of the Teukolsky master equation, and exploits the time and azimuthal angular reflection symmetries of the Kerr family via an operator $\mathcal{J}$. At the same time, separate products for the angular and radial solutions of the Teukolsky equation have also been introduced \cite{London:2020uva,London26}, while Ref.~\cite{Motohashi:2024fwt} introduces further bi-orthogonal products inspired by the case of the quantum resonance
states.
Such QNM products have practical applications for the computation of QNM excitation coefficients \cite{GHSS2023} and perturbative frequency shifts~\cite{Leung:1997was,Zimmerman:2014aha,Mark:2014aja,Hussain:2022ins,Li:2023ulk,Cannizzaro:2023jle}, as well as in waveform modelling~\cite{Hamilton:2021pkf,Thompson:2023ase}. 
Beyond the applications in gravitational wave physics, QNM products have also been introduced in asymptotically anti-de Sitter spacetimes motivated by gauge-gravity duality~\cite{Arnaudo:2025bnm}. 

All these products, however, inherit the blow-up behaviour of the QNM eigenfunctions toward the boundaries of the spatial domain, and this issue must be addressed at both foundational and computational levels. On the analytical side, the arguments leading to the definition of these products are initially formulated for functions with compact support. A careful treatment of the full spatial domain, together with a consistent implementation of the boundary conditions, is typically deferred. These issues are then addressed at the computational level through regularisation schemes, such as deforming the integration domain into the complex plane.

From a complementary perspective, it has been understood over the past decades that the ill behaviour of the QNM eigenfunctions reflects a poor choice of coordinate system, not adapted to the radiative nature of wave propagation~\cite{Zenginoglu:2011jz,PanossoMacedo:2024nkw}. Traditionally, the QNM eigenvalue problem arises from a Fourier analysis with respect to Boyer--Lindquist-type coordinates. In this setting, the limits towards the horizon and infinitely far region correspond to the bifurcation sphere $\mathcal{B}$ and spatial infinity $i^0$, respectively, where hypersurfaces of constant time accumulate, leading to the blow-up of the corresponding spatial functions. 

Upon focusing on the geometrical aspects of the problem, the hyperboloidal framework~\cite{Zenginoglu:2011jz,PanossoMacedo:2023qzp,PanossoMacedo:2024nkw} provides a natural resolution to this issue. It prescribes a formulation in which the time foliation intersects the physically relevant regions of spacetime for black-hole and gravitational-wave physics, namely the black-hole horizon ${\cal H}^+$ and future null infinity $\scri^+$. Within this geometrical setting, the corresponding QNM eigenfunctions remain finite, and the QNM eigenfrequencies are properly defined as isolated eigenvalues of the infinitesimal generator of time translations~\cite{Gajic:2024xrn}.

By properly accounting for the radiative boundary conditions at a geometrical level, the hyperboloidal framework not only resolves the difficulties in representing QNM eigenfunctions, but also naturally leads to a formulation of black hole perturbation theory in terms of non-normal physics~\cite{PhysRevX.11.031003,Jaramillo:2022kuv,PanossoMacedo:2025xnf}. In this context, the calculation of QNM excitation coefficients is obtained via finite computations carried out directly across the entire black hole exterior region~\cite{Ansorg:2016ztf,Besson:2024adi,Bourg:2025lpd}.

In a broader perspective, the hyperboloidal framework arises within the context of conformal methods in general relativity, which are particularly suited for the rigorous treatment of spacetime infinities~\cite{Penrose64}. Conformal methods have become central to general relativity, providing fundamental insights into the geometry of black hole horizons, the nature of singularities, and the global causal structure of spacetime~\cite{ValienteKroon2016}. Beyond their mathematical elegance, they also provide powerful geometrical tools for formulating black hole perturbation theory in a fully regular setting~\cite{GasperinPanossoMacedoFeng2026}. In particular, \cite{GasperinJaramillo2022} suggests that the inclusion of fluxes at the conformal boundaries may provide a notion of normal operators in QNMs problems, suggesting a route for formulating the orthogonality project without dismissing the role played by the boundary conditions and non-compact support functions.

Thus, a natural question is whether a geometrical approach to the QNM problem can amend the singular behaviour of the bilinear products and properly incorporate the radiative boundary behaviour into the analysis. Motivated by these developments, the goal of this paper is to analyse the properties of the bilinear forms introduced in \cite{GHSS2023} (see also \cite{Cannizzaro:2023jle} for scalar QNMs) \emph{within the hyperboloidal framework}. Our analysis focuses on the Schwarzschild family of spacetimes, considering conserved currents constructed from different principles starting from the massless wave equation. 

\subsection{Main results}
We extend the general framework for constructing QNM orthogonality products based on \emph{conserved currents} \cite{GHSS2023} to the hyperboloidal framework. For explicit calculations, we focus on scalar perturbations of the Schwarzschild family of spacetimes and on two currents, the \emph{symplectic} and \emph{energy} currents. A crucial role in establishing QNM orthogonality is played by the operator $\mathcal{J}$, which acts on the spacetime as a discrete $t$–$\phi$ symmetry.

An important conceptual outcome of our work is the introduction of \emph{past hyperboloidal coordinates}, which directly encode the action of the operator $\mathcal{J}$ within the hyperboloidal framework. Whilst $\mathcal{J}$ is a discrete symmetry of the spacetime, it does not extend to the wave equation once radiative boundary conditions are imposed. The hyperboloidal approach nevertheless provides a clear \emph{geometric interpretation} of $\mathcal{J}$ within the QNM problem: it maps quasinormal modes on future hyperboloidal coordinates to their \emph{anti-quasinormal modes} (antimodes)\footnote{
Anti-quasinormal modes satisfy outgoing boundary conditions at the white hole horizon and ingoing boundary conditions at past null infinity — the time reversal of the boundary conditions defining QNMs (see Sec. \ref{sec:adv_ret_sols}).
} defined on past hyperboloidal coordinates. 

While a future hyperboloidal foliation yields a regular representation of QNM eigenfunctions, the corresponding anti-QNMs exhibit singular behaviour when evaluated at the black hole horizon $\mathcal{H}^+$ and future null infinity $\mathscr{I}^+$. Conversely, a past hyperboloidal foliation provides a regular representation of the anti-QNM eigenfunctions, but the singular behaviour appears in the QNM solutions as one approaches the white hole horizon $\mathcal{H}^-$ and past null infinity $\mathscr{I}^-$. This shows that the divergence is structural rather than geometric: it is an intrinsic feature that arises in any construction of QNM products based on the operator $\mathcal{J}$.

The conformal treatment of spacetime within the hyperboloidal framework also provides a natural strategy to assess the role of boundary terms in the orthogonality product. In this setting, it becomes more natural to define an extended version of the product that includes the fluxes across the boundaries. At a conceptual level, this approach allows us to show that the Hamiltonian operator is formally normal under the extended product. However, this construction still inherits the divergences associated with the $t$–$\phi$ reflection, and lacks full control over the asymptotic behaviour of the product. As a result, calculations based purely on the (real) radial coordinate have not yet reached the same level of maturity as alternative hyperboloidal strategies to calculate QNMs projections~\cite{Ansorg:2016ztf,Besson:2024adi,Bourg:2025lpd}.

We therefore employ two regularisation methods based on \emph{analytic continuation} to yield finite and well-defined orthogonality products: a \emph{semi-analytic} approach exploiting the properties of special functions~\cite{London26} and a radial \emph{complex contour} integration method~\cite{Leaver1986}. Under these regularization schemes, the fluxes across the boundaries vanish altogether, and the extended product reduces to the product defined purely in the bulk. 

Within these regularisation schemes, both methods produce consistent results for the orthogonality products. Moreover, we are also able to numerically verify QNM orthogonality with the extended product. If we restrict the integration spatial domain far from the asymptotic boundaries, the fluxes are essential for the correct orthogonality result. In the asymptotic regions, an appropriate normalisation is required to avoid numerical roundoff errors arising from ``large/large'' divisions.

Furthermore, we address the calculation of QNM excitation factors and amplitudes \cite{Leaver1986} from initial data. After establishing the relation between orthogonality projections and  Green's functions techniques~\cite{GHSS2023} in the hyperboloidal context, we compute these observables with the regularisation schemes. The agreement of our numerical results confirms the robustness and generality of these strategies. We are, however, unable to compute excitation coefficients with the extended products since, by definition, the calculation of the fluxes requires the knowledge of the full time evolution associated with a given initial data set.

Overall, we establish a flexible and numerically robust framework for computing QNM orthogonality relations and excitation coefficients using hyperboloidal foliations. The construction is valid for any conserved current and provides a consistent, geometrically informed, and physically meaningful description of QNMs in black hole spacetimes.
Our numerical results are restricted to scalar perturbations on the Schwarzschild spacetime. However, the geometrical framework is directly extendable to different geometries, including the Kerr solution. Meanwhile, the incorporation of spin-weighted fields require adapting the ${\cal J}$ operator as described in Ref.~\cite{GHSS2023}.

\subsection{Notation and conventions}
We employ lower case latin letters to represent tensorial quantities in the abstract index notation, whereas Greek letters indicate the components of the tensor in a given coordinate system. Thus, $v^a$ is a vector, whereas $v^{\bar \mu}$ or $v^{\check \mu}$ are its components expressed, for instance, in the coordinate systems $\bar x^\mu$ or $\check x^\mu$, respectively. In particular, we use $x^\mu =(t,r,\theta,\varphi)$ for Schwarzschild-Droste coordinates, whereas $\bar x^\mu =(\tau,\sigma,\theta,\varphi)$ and $\check x^\mu =(\check\tau,\check\sigma,\theta,\check\varphi)$ are, respectively, future and past hyperboloidal coordinates. The scalar field defined over the entire spacetime is denoted by $\Psi$, while $\psi$ represents the conformal scalar field, i.e taking into account the asymptotic decay $\Psi \sim \psi/r$. The subscripts $(\ell,m)$ indicate the harmonic decomposition of the field, so that $\psi_{\ell,m}$ is a function only of the time and radial coordinates. We employ $\phi_{\ell m}$ to denote the field's representation in the frequency domain. The QNM quantities are denoted as $\omega_I$ and $\phi_I$, where the single labels $I, J, \cdots$ represent the particular mode numbers $I=(\ell,m, n)$.

\subsection{Structure of the paper}
The paper is organised as follows. Section \ref{sec:schwarzschild} reviews the Schwarzschild family of spacetimes and the properties of solutions to the massless wave equation. Section \ref{sec:hyper} introduces the hyperboloidal formalism and emphasises the role of the $\mathcal{J}$ symmetry operator. Section \ref{sec:conserved_currents} introduces the symplectic and energy currents, and presents the associated bilinear forms in compactified hyperboloidal coordinates. Section \ref{sec:orthogonality_product} introduces the orthogonality product, and discusses its main properties and the regularisation strategies employed for its computation. The formalism for computing excitation coefficients is presented in Section \ref{sec:excitation_coeffs}. Finally, in Section \ref{sec:discussion} we discuss our results and their applications. Some technical details can be found in the Appendices.

\section{The Schwarzschild family of spacetimes}\label{sec:schwarzschild}
The four dimensional Schwarzschild spacetime is a $1$-parameter family of spacetimes that describes a spherically symmetric black hole. Its line element is traditionally expressed in terms of the Schwarzschild-Droste coordinates $ x^\mu=(t,r,\theta,\varphi)$ as
\begin{equation} \label{eq:Schwarz_BL}
\d s^2 = -f(r) \d t^2  + \dfrac{1}{f(r)} \d r^2 + r^2 {\rm d}\varpi^2
\end{equation}
where $\d \varpi^2$ is the metric on the unit $2$-sphere and
\bea
f(r) =\left( 1 - \dfrac{\rh}{r} \right), \quad \rh = 2M
\eea
where $\rh$ and $M$ describe, respectively, the event horizon radius and the black hole's mass. The tortoise coordinate $r_*$ is defined as
\be
\dfrac{ \d r_*}{\d r} = \dfrac{1}{f(r)}.
\ee

In these coordinates, the line element is symmetric under the transformation
\be
\label{eq:time_reversal}
t = -\check t, \qquad \varphi=-\phip.
\ee
In other words, the metric components satisfy $g_{\mu \nu} = g_{\check\mu \check\nu}$, 
with $\check x{}^\mu = (\check t, r, \theta, \check\varphi)$. 
Since this symmetry plays a central role when defining the QNM bi-orthogonality relation, we will discuss it in the context of wave-like equations in black hole perturbation theory in subsection \ref{sec:J_op_Schwarzschild}. Specifically, we will focus on the case of a massless scalar field propagating on the Schwarzschild spacetime.

\subsection{The massless wave equation: retarded and advanced solutions}\label{sec:adv_ret_sols}
We consider the massless wave equation for the Schwarzschild family of spacetimes. Thus, a massless scalar field $\Psi$ satisfies 
\be \label{eq:KG}
\square \Psi = \dfrac{1}{\sqrt{-\boldsymbol{g}}} \p_{\mu}\bigg( \sqrt{-\boldsymbol{g}} \, g^{\mu\nu} \p_{\nu}  \Psi  \bigg) = 0. \\ 
\ee
In Schwarzschild-Droste coordinates, and taking into account the asymptotic behaviour and a decomposition into spherical harmonics modes via 
\be \label{ansatz}
\Psi(t,r,\theta,\phi) = \dfrac{1}{r} \sum_{\ell m} \psi_{\ell m}(t,r) Y_{\ell m }(\theta, \phi),
\ee
the equation of motion reduces to
\be
\label{eq:scalar_Schwarzschild}
\square_{\ell m} \psi_{\ell m} = 0 
\ee
where
\bea
\label{eq:wave_operator_Schwarzscild}
\square_{\ell m} &=& -\p^2_{tt}  + \p^2_{r_*r_*}  - V_{\ell m}(r), \\
\label{eq:BH_Potential}
V_{\ell m}(r) &=& \dfrac{f(r)}{r^2}\left( \ell (\ell +1) + \dfrac{\rh}{r}\right).
\eea
Eqn.~\eqref{eq:scalar_Schwarzschild} can be solved with outgoing boundary conditions (BCs) associated with energy propagating into the black hole and carried away into the wave zone, i.e. 
\be
\label{eq:BH_QNM_BC_v1}
\psi_{\ell m}(t,r) \sim e^{- i \omega (t \mp r_*)}, \quad r_* \to \pm \infty \, .
\ee
The BCs \eqref{eq:BH_QNM_BC_v1} are formally represented as
\be
\label{eq:BH_QNM_BC_v2}
\lim_{r_* \rightarrow \pm \infty} \bigg( \p_t \psi_{\ell m} \pm \p_{r_*} \psi_{\ell m} \bigg) = 0.
\ee
In terms of structures associated with the light cone, Eqn.~\eqref{eq:BH_QNM_BC_v2} assumes the form
\be
\label{eq:BH_QNM_BC_v3}
\lim_{r_*\to \infty} \nabla_{v} \psi_{\ell m} = 0, \quad \lim_{r_*\to -\infty} \nabla_{u} \psi_{\ell m} = 0,
\ee
with $v^a$ and $u^a$ vectors generating the null cones\footnote{The vectors $v^a$ and $u^a$ are not normalised. The outgoing null tetrad $\ell^a \propto v^a$ and ingoing null tetrad $k^a \propto u^a$ are normalised to $\ell^a k_a = -1$.}
\be
v^a = \dfrac{1}{2} \left( \p_t + \partial_{r_*} \right), \quad u^a = \dfrac{1}{2} \left( \p_t - \partial_{r_*}\right).
\ee
We will refer to solutions of Eqn.~\eqref{eq:scalar_Schwarzschild} satisfying the BCs \eqref{eq:BH_QNM_BC_v3} as {\em retarded} solutions $\psi^{\rm ret}_{\ell m}(t,r)$. Specifically, a Fourier decomposition
\be
\label{eq:FourierDecomp_retSolution_ScalarSchwarzschild}
\psi^{\rm ret}_{\ell m}(t,r) = \dfrac{1}{2\pi} \int_{-\infty}^{+\infty} d\omega \,e^{-i \omega t} \, \phi^{\rm ret}_{\ell m}(r;\omega)
\ee
 yields the radial function $\phi^{\rm ret}_{\ell m}(r;\omega)$ satisfying the ordinary differential equation (ODE)
 \be
 \label{eq:ODE_ret_ScalarSchwarzschild}
\dfrac{d^2}{dr_* ^2}  \phi^{\rm ret}_{\ell m} - V(r) \phi^{\rm ret}_{\ell m} = -\omega^2 \phi^{\rm ret}_{\ell m}
 \ee
with BCs compatible with the physical scenario captured by the formulations \eqref{eq:BH_QNM_BC_v1}-\eqref{eq:BH_QNM_BC_v2}, 
\be
 \label{eq:BC_ret_ScalarSchwarzschild}
\lim_{r_* \rightarrow \pm \infty} \bigg( \dfrac{1}{\phi^{\rm ret}_{\ell m}} \dfrac{d}{d r_*} \phi^{\rm ret}_{\ell m}  \bigg) =  \pm i\,  \omega 
\, .
\ee
Such homogenous retarded solutions $\phi^{\rm ret}_{\ell m}(r;\omega)$ can only be found for a discrete set of QNM frequencies $\omega_{\ell m n}$. A more formal definition of QNMs is given in terms of the poles of the {\em retarded} Green function. These frequencies have ${\rm Im}\left( \omega_{\ell m n}\right) < 0$, which implies that the time evolution $\psi_{\ell m n}(t,r)$ decays in time, but the radial eigenfunctions $\phi_{\ell m n}(r)$ diverge as $r_* \to \pm \infty$. 

\medskip
In principle, one could also solve \eqref{eq:scalar_Schwarzschild} with boundary conditions associated with purely incoming waves from the past null infinity $\mathscr{I}^-$ and energy flowing out of the past horizon $\mathcal{H}^-$. The boundary conditions for this scenario have formulations akin to Eqs.~\eqref{eq:BH_QNM_BC_v2} and \eqref{eq:BH_QNM_BC_v3}, i.e
\bea
\label{eq:WH_QNM_BC_v2}
& \lim_{r_* \rightarrow \pm \infty} \bigg( \p_t \psi_{\ell m} \mp \p_{r_*} \psi_{\ell m} \bigg) = 0, \\
\label{eq:WH_QNM_BC_v3}
&
\displaystyle \lim_{r_*\to \infty} \nabla_{u} \psi_{\ell m} = 0, \quad \lim_{r_*\to -\infty} \nabla_{v} \psi_{\ell m} = 0.
\eea
We will refer to solutions of Eqn.~\eqref{eq:scalar_Schwarzschild} satisfying the BCs  \eqref{eq:WH_QNM_BC_v2} --- or equivalently \eqref{eq:WH_QNM_BC_v3} --- as {\em advanced} solutions $\psi^{\rm adv}_{\ell m}(t,r)$.

Note that the boundary condition \eqref{eq:WH_QNM_BC_v2} arises directly from the operation $t \to -t$ on Eqn.~\eqref{eq:BH_QNM_BC_v2}. In other words, when expressed in the coordinate system $\check x{}^\mu$, the advanced solution $\psi^{\rm adv}_{\ell m}(\check t,r)$ would satisfy the boundary conditions
\be
\label{eq:WH_QNM_BC_J_op}
\lim_{r_* \rightarrow \pm \infty} \bigg( \p_{\check t} \psi_{\ell m} \pm \p_{r_*} \psi_{\ell m} \bigg) = 0,
\ee
i.e. with the same functional form as Eqn.~\eqref{eq:BH_QNM_BC_v2}.

Given the functional symmetry between the equations and the boundary conditions for $\psi^{\rm ret}_{\ell m}(t,r)$ and $\psi^{\rm adv}_{\ell m}(\check t,r)$, the frequency domain decomposition for the advanced field (equivalent to Eqn.~\eqref{eq:FourierDecomp_retSolution_ScalarSchwarzschild}) is given by a Fourier decomposition defined as
\bea
\label{eq:FourierDecomp_advSolution_ScalarSchwarzschild}
\psi^{\rm adv}_{\ell m} &=& \dfrac{1}{2\pi} \int_{-\infty}^{+\infty} d\check\omega \,e^{-i \check\omega \check t} \, \phi^{\rm adv}_{\ell m}(r;\check\omega) \\
&=& \dfrac{1}{2\pi} \int_{-\infty}^{+\infty} d\check\omega \,e^{+i \check\omega t} \, \phi^{\rm adv}_{\ell m}(r;\check\omega),
\eea
with the first and second lines capturing the representation in terms of coordinate system $\check x^\mu$ and $x^\mu$, respectively.

Such a definition implies an ODE defining the advanced radial field that has the same form as Eqn.~\eqref{eq:ODE_ret_ScalarSchwarzschild}, i.e.
 \be
\dfrac{d^2}{dr_* ^2}  \phi^{\rm adv}_{\ell m} - V(r) \phi^{\rm adv}_{\ell m} = -\check\omega{}^2 \phi^{\rm adv}_{\ell m}.
 \ee
Besides, it allows us to express the BCs for the advanced radial field exactly in the same way as the one for the retarded radial solution in Eqn.~\eqref{eq:BC_ret_ScalarSchwarzschild}, i.e.
\be
 \label{eq:BC_adv_ScalarSchwarzschild}
\lim_{r_* \rightarrow \pm \infty} \bigg( \dfrac{1}{\phi^{\rm adv}_{\ell m}} \dfrac{d}{d r_*} \phi^{\rm adv}_{\ell m}  \bigg) =  \pm i\, \check \omega
\ee
In this way, we can identify $\omega=\check\omega$  and the QNM frequencies associated with the {\em advanced} solution have exactly the same values as those for the {\em retarded} one.

Specifically, they satisfy ${\rm Im}\left( \check\omega_{n\ell m}\right) < 0$, implying that  the time evolution shows a field with exponentially growing modes, precisely because the Fourier decomposition goes as $e^{+ i \omega_{\ell m n} t}$, a dynamical behaviour expected due to the nature of the boundary conditions. The advanced radial functions also diverge as $r_* \to \infty$. 

Since the frequency domain formulation is exactly the same for both scenarios, one can employ the same techniques to construct the corresponding solution. In particular, Leaver's approach introduces an auxiliary field 
\be
\label{eq:FreqDomain_LeaverDecomp}
\phi^{\rm ret/adv}_{\ell m}(r;\omega) = Z(r;\omega)\, \bar \phi^{\rm ret/adv}_{\ell m}(r; \omega),
\ee
with $Z(r;\omega)$ a singular function as $r_* \to \pm\infty$ capturing the asymptotic behaviour imposed by the BCs, which can be either derived based on properties of the Eq.~\eqref{eq:ODE_ret_ScalarSchwarzschild}~\cite{Leaver:1985ax}, or inferred geometrically via the hyperboloidal framework \cite{Zenginoglu:2011jz,PanossoMacedo:2023qzp, PanossoMacedo:2024nkw}. The function $\bar \phi^{\rm ret/adv}_{\ell m}(r; \omega)$ is a regular function, with a well-defined Taylor expansion around the horizon \cite{Leaver:1985ax}
\be
\label{eq:TaylorExp_LeaverDecomp}
\bar \phi^{\rm ret/adv}_{\ell m}(r; \omega) = \sum_k a_k(\omega) \left( 1 - \dfrac{\rh}{r} \right)^k.
\ee
The coefficients $a_k(\omega)$ satisfy a three-term recursion relation and the QNMs are those $\omega$-values yielding a minimal solution to the recurrence relation, see \cite{Leaver:1985ax,Dolan:2007mj}. 
Note that, with representations \eqref{eq:FreqDomain_LeaverDecomp} and \eqref{eq:TaylorExp_LeaverDecomp}, one can analytically extend the radial coordinate $r$ to assume complex values, with a procedure equally valid for retarded and advanced solutions.

\subsection{The $\mathcal{J}$ operator in Schwarzschild-Droste coordinates}\label{sec:J_op_Schwarzschild}

Given a QNM solution $\psi_{\ell m n}(t,r)$ satisfying the outgoing boundary conditions Eqn.~\eqref{eq:BH_QNM_BC_v2}, the bi-orthogonality relation of \cite{GHSS2023} involves the function ${\cal J}\psi_{ \ell m n}(t,r)$, where ${\cal J}$ corresponds to the discrete $t-\varphi$ symmetry operator. Specifically, we will denote the result of the operator ${\cal J}$ as 
\be
{\cal J}\psi_{\ell m n} (t,r) = \left. \psi_{\ell m n}(t,r) \right|_{t\to -t}
\ee
Note however, that even though the wave equation Eqn.~\eqref{eq:scalar_Schwarzschild} is symmetric under the $t-\varphi$ symmetry (as it is the spacetime), the boundary conditions Eqn.~\eqref{eq:BH_QNM_BC_v2} --- or equivalently Eqn.~\eqref{eq:BH_QNM_BC_v1} --- are not. Actually, with the ${\cal J}$ operation, the boundary condition Eqn.~\eqref{eq:BH_QNM_BC_v2} is mapped into Eqn.~\eqref{eq:WH_QNM_BC_v2}. Hence, we can interpret the ${\cal J}$ operator as mapping a {\em retarded} $\psi^{\rm ret}_{\ell m}(t,r)$ into an {\em advance} solution $\psi^{\rm adv}_{\ell m}(t,r)$, i.e.,
\be\label{eq:JinBL}
{\cal J}\psi^{\rm ret}_{\ell m}(t,r) = \psi^{\rm adv}_{\ell m}(t,r).
\ee
We remind the reader that $\mathcal{J} Y_{\ell m}(\theta,\varphi)=Y_{\ell m}(\theta,-\varphi)$.

\section{Hyperboloidal Framework}\label{sec:hyper}
In this section, we introduce the hyperboloidal formalism and discuss how the $\mathcal{J}$ operation is adapted to these coordinates. Specifically, we present two hyperboloidal foliations, future and past, which elucidate the time-reversal operation in the hyperboloidal framework.

\subsection{Future foliation}\label{sec:hyp_future}
The future hyperboloidal coordinates $\bar x^\mu = (\tau, \sigma,  \theta, \varphi)$ are related to the Schwarzschild-Droste coordinates $x^\mu=(t,r,\theta,\varphi)$ via~\cite{Zenginoglu:2007jw,Zenginoglu:2011jz, PanossoMacedo:2023qzp}
\bea \label{eq:HypTrasfo}
 && t = \lambda\bigg(\tau - H(\sigma)\bigg), \quad r =  \dfrac{\rh}{\sigma}, 
\eea
where $\lambda$ is a given length scale of the spacetime and 
\be
\label{eq:MinimalGaugeHeightFunction}
H(\sigma) = \dfrac{\rh}{\lambda} \left( -\dfrac{1}{\sigma} + \ln(\sigma) + \ln(1-\sigma) \right)
\ee
is the minimal gauge height function. The radial compactification in Eqn.~\eqref{eq:HypTrasfo} keeps the minimal structure needed to map the infinitely far wave zone to finite coordinate values. Thus, $\sigma = 0$ corresponds to future null infinity, while the black-hole horizon $r = \rh$ is conveniently set to $\sigma_{\rm h} = 1$.

In these coordinates, the Schwarzschild metric naturally rescales as~\cite{PanossoMacedo:2023qzp}
\bea
\label{eq:conf_metric_Schwarzschild}
ds^2 &=& \Omega^{-2} d\s^2, \quad \Omega = \dfrac{\sigma}{\lambda}, \\
d\s^2 &=&\bar g_{\mu\nu} d\bar x^\mu d\bar x^\nu\nn\\
&=&\dfrac{\rh}{\lambda} \Bigg( -p(\sigma) {\rm d}\tau^2 + 2 \gamma(\sigma){\rm d}\tau \d\sigma + w(\sigma) \d\sigma^2 \Bigg) \nonumber \\
&&+ \left(\dfrac{\rh}{\lambda} \right)^2 {\rm d}\varpi^2,
\eea
with 
\bea
\label{eq:metric_funcs}
&& p(\sigma) = \dfrac{\lambda}{\rh} \sigma^2(1-\sigma), \quad \gamma(\sigma) = 1 -2\sigma^2, \\
&& w(\sigma) = \dfrac{4 \rh}{\lambda} (1+\sigma). 
\eea
Some useful relations arising from the definitions of metric functions are given below~\cite{PanossoMacedo:2023qzp}
\be \label{eq:hyper_relations}
\gamma(\sigma) = p(\sigma) \, \p_\sigma H(\sigma), \quad w(\sigma) = \dfrac{1-\gamma(\sigma)^2}{p(\sigma)}.
\ee

Note that Eqn.~\eqref{eq:conf_metric_Schwarzschild} reflects the fact that the hyperboloidal framework is naturally suited to the conformal representation of spacetime; for more details see Appendix \ref{app:conformal}. Given this, it is natural to also conformally rescale the massless scalar field $\Psi$ as
\begin{equation}
\label{Psi_confresc_future}
\Psi(\tau,\sigma, \theta,\varphi) = \Omega \,\psi(\tau,\sigma, \theta,\varphi)\,.
\end{equation}
 This transformation captures the $1/r$ decay of the scalar field towards infinity, cf.~Eqn.~\eqref{ansatz}, thus ensuring $\psi \sim {\cal O}(1)$ as $r \to \infty$. The conformal field $\psi$ satisfies the wave equation $ L \psi= 0$, where $L$ is the conformal wave operator, cf.~\eqref{ConfQuantity3}, explicitly defined in the next sections, according to the particular coordinate system employed.

\subsubsection{Wave equation: retarded field}
 
When expressed in the coordinates $\bar x^\mu$, the conformal wave operator reads
\bea
\label{eq:waveeqHyp}
\bar{L}:= \left.{L}\right|_{\bar x^\mu} = -w(\sigma)\partial_{\tau \tau} + \bar{L}_1 +  \bar{L}_2 \partial_\tau 
\eea
with
\bea
\label{eq:_L1}
&&\bar{L}_1 = \partial_\sigma \bigg( p(\sigma) \partial_\sigma \bigg) - Q  \\
&&\bar{L}_2 = 2\gamma(\sigma)\partial_{\sigma} + \gamma_{,\sigma}(\sigma),
\eea
and the potential operator
\be
Q = -\dfrac{\lambda}{\rh}\Bigg(\dfrac{1}{\sin\theta}\partial_\theta\bigg( \sin\theta \partial_\theta \bigg) + \dfrac{1}{\sin^2\theta}\p^2_{\varphi\varphi} - \sigma\bigg).
\label{eq:Potential_q}
\ee

To look at the boundary conditions for retarded fields in the $\bar x^\mu$ coordinates, we notice that the time and radial derivatives are
\be
\label{eq:hyp_derivative_trasfo}
\rh \, \partial_t = 
+\partial_{\tau},
\qquad
\rh\, \partial_{r_*} = 
-\gamma(\sigma) \partial_{\tau}  - p(\sigma) \p_\sigma.
\ee
The derivatives of $\psi(\tau,\sigma, \theta,\varphi)$ along the null coordinates determining the boundary conditions in Eqs.~\eqref{eq:BH_QNM_BC_v3} and \eqref{eq:WH_QNM_BC_v3} read
\bea
&&\nabla_{v} \psi =\dfrac{1}{2\rh} 
\bigg(1-\gamma(\sigma)\bigg)\p_{\tau} \psi    - p(\sigma) \p_\sigma \psi,  \\
&&\nabla_{u} \psi =\dfrac{1}{2\rh} 
\bigg(1  + \gamma(\sigma)\bigg)\p_{\tau} \psi  + p(\sigma) \p_\sigma \psi.
\eea
Recall that, at future null infinity $\sigma = 0$, we have $\gamma(0)=1$, $p(0)=0$, reducing the above equations trivially to $\left.\nabla_v\psi\right|_{\scri^+}=0$. Similarly, at the black hole horizon $\sigma =1$, we have $\gamma(1)=-1$, $p(1)=0$, leading to $\left.\nabla_u\psi\right|_{{\cal H}^+}=0$. These results rely on the assumption that the functions $\psi$ and its first derivatives are finite at $\sigma = 0$ and $\sigma=1$. Thus, sufficiently regular solutions to the hyperboloidal wave equation satisfy automatically the outgoing boundary conditions, and the question of constructing the retarded solution $\psi^{\rm ret}(\tau, \sigma,\theta,\varphi)$ reduces to imposing regularity conditions fixed by the operator \eqref{eq:waveeqHyp}.

From the wave operator Eqn.~\eqref{eq:waveeqHyp}, we explicit identify the Hamiltonian operator $\bar {\cal H} = i \bar{\rm  L}$, by introducing the state vector $\bar {\boldsymbol u} = ( \psi ,  \p_\tau \psi)^{T}$, and expressing the wave equation as
\bea
\label{eq:H_future_evol}
\p_{\tau} \bar {\boldsymbol u} = \bar {\rm L} \bar {\boldsymbol u}, \quad
\bar {\rm L} =  \left( 
\begin{array}{cc}
0 & 1 \\
w^{-1} \bar L_1 & w^{-1}{\bar L_2}
\end{array}
\right).
\eea

Furthermore, by adopting a frequency-domain decomposition and expanding the perturbation in spherical harmonics as
\be
\psi(\tau, \sigma, \theta, \varphi) = e^{- i \lambda \overline\omega \tau}  \overline\phi_{\ell m}(\sigma) Y_{\ell m}(\theta, \varphi),
\ee
Eqn.~\eqref{eq:Potential_q} yields the eigenvalues of the operator $Q$, which are determined by the rescaled black hole potential $V_{lm}(r)$. Explicitly, one finds
\bea
q_{\ell m}(\sigma) &=& \dfrac{\lambda^2}{p(\sigma)}V_{\ell m}(r) \nn \\
&=& \dfrac{\lambda}{\rh} \bigg(\ell(\ell+1) + \sigma\bigg),
\eea
with $V_{\ell m}$ defined in Eqn.~\eqref{eq:BH_Potential}. 
Then, we obtain the radial equation
\be
\label{eq:Freq_domain_radial_eq_ret}
\bar{\rm A}(\bar \omega)[\bar \phi_{\ell m}] = 0, \quad \bar{\rm A}(\bar \omega) = 
\bar L^{\ell m}_1 -i \lambda \overline\omega \bar L_2 + \lambda^2 \overline\omega^2 w(\sigma).
\ee
In the above expression, the operator $\bar L^{\ell m}_1$ has the same form as $\bar L_1$ in Eqn.~\eqref{eq:_L1}, but with $Q$ replaced by $q_{\ell m}$. In this way, the QNM problem reduces to the eigenvalue system
\be
\label{eq:eigenvalue_future}
 {\bar{\rm L}}_{\ell m}\, \bar {\boldsymbol u}_{\ell m n} = -i\lambda \,\overline\omega_{\ell m n}\bar {\boldsymbol u}_{\ell m n},
\ee
for a discrete set of frequencies labelled by $n$. Regular solutions to the eigenvalue problem Eqn.~\eqref{eq:eigenvalue_future} are therefore the usual QNMs, yielding the dynamics associated with retarded solutions.

\subsection{Past foliation}\label{sec:Hyp_past_foliation}
To obtain a time-reversal transformation similar to Eqn.~\eqref{eq:time_reversal}, we consider the coordinate system $\xp^{\mu} = (\tp, \sigp,  \theta,  \phip)$ which is related to the standard Schwarzschild time via
\bea
\label{PastHypCoordinates}
 &&t = - \lambda \bigg(\tp - H(\sigp) \bigg), \quad r= \frac{ \rh}{\sigp}, \quad
\varphi=-\phip.
\eea
We will refer to these coordinates as the \textit{past hyperboloidal coordinates}, see Appendix \ref{app:TR_hyperboloidal} for a more detailed discussion on derivation and terminology.
The coordinates $\xp^{\mu}$ are related to the future hyperboloidal coordinates $\bar x^{\mu}$ via
\bea \label{pasttofutureHyp}
&& \tp = -\tau + 2 H(\sigma), \quad \sigp=\sigma,
\quad \phip=-\varphi. 
\eea
 One verifies that the metric components Eqn.~\eqref{eq:conf_metric_Schwarzschild} are invariant under the coordinate change Eqn.~\eqref{pasttofutureHyp}, i.e. $\bar g_{ \mu  \nu} = \check g_{ \mu  \nu}.$  Just as above, we also rescale the scalar field $\Psi$ as 
\begin{equation}
\label{Psi_confresc_past}
\Psi(\tp, \sigp,  \theta,  \phip) = \Omega \,\psi(\tp, \sigp,  \theta,  \phip).
\end{equation}

\begin{figure*}[t]
 \includegraphics[width=0.68\columnwidth]{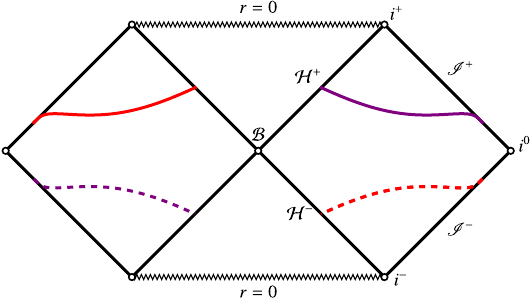}
\includegraphics[width=0.68\columnwidth]{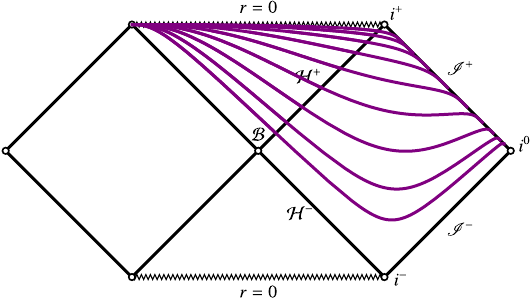}
 \includegraphics[width=0.68\columnwidth]{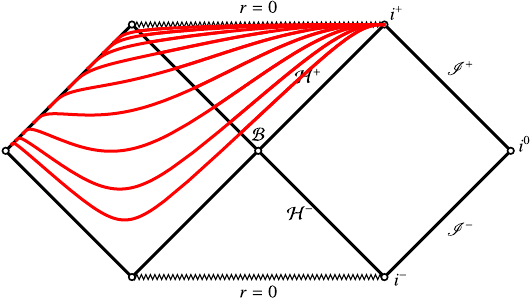}

  \caption{Carter-Penrose diagrams of the Schwarzschild spacetime with future (purple) and past (red) hyperboloidal foliations {\em Left Panel:} On the exterior black hole region represented by the right diamond, future slices extends between the black hole horizon ${\cal H}^+$ and future null infinity ${\scri}^+$ (solid purple line), whereas the past foliation connects the white hole horizon ${\cal H}^-$ and past null infinity ${\scri}^-$ (dashed red line). Visualised on the left diamond's asymptotic flat region, these slices change their roles. {\em Middle Panel:} Future hyperboloidal time flow $\tau > 0$. Quasinormal modes ringdown as $\sim e^{-i \lambda \omega_{\ell m n}\tau}$, and eigenfunctions $\phi_{\ell m n}(\sigma)$ are finite on $\sigma\in[0,1]$. {\em Right Panel:} Past hyperboloidal time flow $\check\tau > 0$. Anti-quasinormal ringdown as $\sim e^{-i \lambda \omega_{\ell m n}\check\tau}$, and eigenfunctions $\phi_{\ell m n}(\check\sigma)$ are finite on $\check\sigma\in[0,1]$. Antimodes expressed back into future hyperboloidal coordinates via the ${\cal J}$ operator grow in time as $\sim e^{i \lambda \omega_{\ell m n}\tau}$ and eigenfunctions diverges as $\sim e^{2 i \lambda \omega_{\ell m n}H(\sigma)}$, cf.~Eq.~\eqref{eq:MinimalGaugeHeightFunction} for the height function $H(\sigma)$.  
  }
    \label{fig:penrose_diagrams}
\end{figure*}

Future (purple) and past (red) hyperboloidal foliations are visualised within the Carter-Penrose diagram in Fig.~$\ref{fig:penrose_diagrams}$.

 \subsubsection{Wave equation: advanced field}
 
 The conformal wave operator is also invariant in the sense that in terms of $\xp{}^\mu$ the equivalent of Eqn.~\eqref{eq:waveeqHyp} reads
\bea
\label{eq:waveeqHypPast}
\check{L} := \left.L\right|_{\xp{}^\mu} = -w(\sigma)\partial_{\tp \tp} + \check{L}_1 +  \check{L}_2 \partial_{\tp} 
\eea
with
\bea
&&\check{L}_1 = \partial_{\sigp} \bigg( p(\sigp) \partial_{\sigp} \bigg) - \check Q \\ 
&&\check{L}_2 = 2\gamma(\sigp)\partial_{\sigp} + \gamma_{,\sigp}(\sigp),
\eea
and the potential operator $\check Q$ as defined in Eqn.~\eqref{eq:Potential_q} but expressed in terms of the $(\check \sigma, \theta, \check \varphi)$ coordinates.

Contrary to future hyperboloidal coordinates, however, the past hyperboloidal coordinates provide a parametrisation best adapted to the {\em advanced} solutions. To appreciate this result, we observe that the coordinate transformation Eqn.~\eqref{PastHypCoordinates} yields
\be
\label{eq:hyp_derivative_trasfo}
\rh \, \partial_t = 
-\partial_{\tp},
\qquad
\rh\, \partial_{r_*} = 
-\gamma(\sigma) \partial_{\tp}  - p(\sigp) \p_{\sigp}.
\ee
and thus $\psi(\tp, \sigp,  \theta,  \phip)$ satisfies
\bea
&&\nabla_{v} \psi =-\dfrac{1}{2\rh} 
\bigg(1+\gamma(\sigp)\bigg)\p_{\tp} \psi    - p(\sigp) \p_{\sigp} \psi,  \\
&&\nabla_{u} \psi = -\dfrac{1}{2\rh} 
\bigg(1  - \gamma(\sigma)\bigg)\p_{\tp} \psi  + p(\sigp) \p_{\sigp} \psi
\eea
Similar to the previous arguments, at null infinity ($\check\sigma=0$) we have $\gamma(0)=1$, $p(0)=0$, which leads to $\nabla_u\psi^{\rm adv}=0$ if $\psi^{\rm adv}(\check t, \check \sigma, \theta,\check \varphi)$ and its first derivatives are {\em finite} at $\sigp = 0$. At the horizon ($\check\sigma=1$), we have $\gamma(1)=-1$, $p(1)=0$ giving $\nabla_v\psi^{\rm adv}=0$ subject to the same regularity conditions at $\sigp =1$. Thus, the ingoing boundary conditions are automatically satisfied through the geometry of the slices, and constructing advanced solutions $\psi^{\rm adv}(\check t, \check \sigma, \theta,\check \varphi)$ to the wave equation Eqn.~\eqref{eq:waveeqHypPast} reduces to imposing regularity conditions in the past hyperboloidal implementation.   

Here again, we also identify the Hamiltonian operator $\check {\cal H} = i \check{\rm  L}$, via
\bea
\label{eq:H_past_evol}
\p_{\check\tau} \check {\boldsymbol u} = \check {\rm L} \check{\boldsymbol u}, \quad
\check {\rm L} =  \left( 
\begin{array}{cc}
0 & 1 \\
w^{-1} \check L_1 & w^{-1}{\check L_2},
\end{array}
\right).
\eea
 with the state vector $\check {\boldsymbol u} = ( \psi ,  \p_{\check\tau} \psi)^{T}$, and we also introduce a frequency domain and harmonic decomposition for the field in the form
\be
\psi(\check\tau, \check\sigma, \theta, \check\varphi) = e^{ -i\lambda\check \omega \check\tau} \check\phi_{\ell m}(\check\sigma) Y_{\ell m}(\theta, \check\varphi).
\ee
Hence, the radial function satisfies an equation with the exact functional form as Eqn.~\eqref{eq:Freq_domain_radial_eq_ret}, i.e. 
\be
\label{eq:Freq_domain_radial_eq_adv}
\check{\rm A}(\check \omega)[\check\phi_{\ell m}]=0, \quad
\check{\rm A}(\check \omega) = \check L^{\ell m}_1 -i\lambda\check \omega\check L_2 +\lambda^2\check \omega^2 w(\check\sigma), 
\ee
with the eigenvalue problem also assuming the same functional form
\be
\label{eq:eigenvalue_past}
 {\check{\rm L}}_{\ell m}\, \check {\boldsymbol u}_{\ell m n} = -i\lambda\,\check \omega_{\ell m n} \check {\boldsymbol u}_{\ell m n}.
\ee
As discussed, regular solutions for the eigenvalue Eqn. ~\eqref{eq:eigenvalue_past} constitute anti-QNMs and yield the dynamics associated with advanced solutions.

\subsection{Hamiltonian evolution}
Seen as a Hamiltonian evolution problem, Eqs.~\eqref{eq:H_future_evol} and \eqref{eq:H_past_evol} provide the same dynamics for their respective fields $\bar {\boldsymbol u}$ and $\check {\boldsymbol u}$ when they are represented in the coordinates best adapted to their radiative properties at the boundaries.

In other words, the Hamiltonian operators $\bar {\cal H}$ and $\check{\cal  H}$ have the same structural form, and therefore the eigenvalue problems Eqs.~\eqref{eq:eigenvalue_future} and \eqref{eq:eigenvalue_past} yield the same spectral values, i.e. $\bar \omega_{\ell m n} = \check \omega_{\ell m n}$, as well as the same eigenvectors $\overline \phi_{\ell m n}(\sigma) = \check \phi_{\ell m n}(\check \sigma)$. 

This property allows us to uniquely identify a radial QNM eigenfunction with a radial anti-QNM eigenfunction with corresponding eigenvalue $\omega_{\ell m n}$, and denote both by functions $\phi_{\ell m n}$. In the same way, we can refer to a single Hamiltonian operator ${\cal H}$. However, the temporal behaviour of these two functions differs significantly. When represented in their adapted coordinates, both fields
\bea
\label{eq:ret_fut}
\psi^{\rm ret}_{\ell m n}(\tau, \sigma, \theta, \varphi) = e^{-i \lambda \omega_{\ell m n}\tau}\phi_{\ell m n}(\sigma) Y_{\ell m }(\theta, \varphi), \\
\label{eq:adv_past}
\psi^{\rm adv}_{\ell m n}(\check\tau, \check \sigma, \theta, \check \varphi) = e^{-i \lambda \omega_{\ell m n} \check \tau}\phi_{\ell m n}(\check \sigma) Y_{\ell m }(\theta, \check \varphi)
\eea
decay in time, respectively, as $\tau$ or $\check \tau$ goes to infinity (because ${\rm Im}(\omega_{\ell m n})<0$). Moreover, they are, respectively, also regular as $\sigma = \check \sigma \to 0$ or $\sigma = \check \sigma \to 1$ along the corresponding surfaces of constant future/past hyperboloidal time. 

Conversely, the representation of an anti-QNM in terms of future hyperboloidal coordinates reads
\be
\label{eq:adv_fut}
\psi^{\rm adv}_{\ell m n}(\tau, \sigma,\theta, \varphi) = e^{i\lambda \omega_{\ell m n} \tau} e^{-2i\lambda \omega_{\ell m n} H(\sigma)}\phi_{\ell m n}(\sigma)Y^\ast_{\ell m }(\theta, \varphi),
\ee
cf.~Eqn.~\eqref{pasttofutureHyp}. Note that the temporal exponent in Eq.~\eqref{eq:adv_fut} carries a flipped sign relative to Eq.~\eqref{eq:ret_fut}, $e^{i\lambda \omega_{\ell m n}\tau}$ rather than $e^{-i\lambda \omega_{\ell m n}\tau}$. Since QNM frequencies satisfy ${\rm Im}(\omega_{\ell m n})<0$, this sign flip converts the decaying factor $e^{-i\lambda \omega_{\ell m n}\tau} = e^{-i\lambda{\rm Re}(\omega_{\ell m n})\tau}e^{\lambda{\rm Im}(\omega_{\ell m n})\tau}$ into the growing factor $e^{i\lambda \omega_{\ell m n}\tau} = e^{i\lambda{\rm Re}(\omega_{\ell m n})\tau}e^{-\lambda{\rm Im}(\omega_{\ell m n})\tau}$, whose real exponential envelope $e^{-\lambda{\rm Im}(\omega_{\ell m n})\tau}$ grows without bound as $\tau\to\infty$. Therefore, the anti-QNM field $\psi^{\rm adv}_{\ell m n}$ will grow in time as $\tau \to \infty$ due to an incoming flux of radiation from the white hole horizon $\mathcal{H}^-$ and past null infinity $\mathscr{I}^-$, and their radial profile will diverge as $e^{-2i\lambda \omega_{\ell m n} H(\sigma)}$ as one approaches future null infinity $\mathscr{I}^+$ at $\sigma =0$ or the black hole horizon $\mathcal{H}^+$ at $\sigma = 1$.

With this analysis, we are now in a position to introduce the ${\cal J}$ operation in the context of the hyperboloidal framework.

\subsection{The $\mathcal{J}$ operator in hyperboloidal coordinates}\label{sec:J_op_Hyp}
As discussed in Sec.~\eqref{sec:J_op_Schwarzschild}, the time reversal property of the ${\cal J}$ operator acts on the complete field, including its behaviour close to the boundaries. For problems where radiative boundary conditions are particularly relevant, the operator maps retarded into advanced solutions, i.e.
\be
\label{eq:J_action}
{\cal J}[\psi^{\rm ret}] = \psi^{\rm adv}, \quad {\cal J}[\psi^{\rm ret}] = \psi^{\rm adv}.
\ee
Considering a specific coordinate system associated with a hyperboloidal foliation, the $\mathcal{J}$ operator is, therefore, effectively realised as a coordinate pullback between future and past hyperboloidal coordinates
\begin{align}
&\mathcal{J}\psi(\tau,\sigma,\theta,\varphi)=\psi(\tp,\sigp,\theta,\check\varphi)\,,\label{eq:J_hyp_map}\\
&\mathcal{J}\psi(\tp,\sigp,\theta,\check\varphi)=\psi(\tau,\sigma,\theta,\varphi)\,,
\end{align}
leading to
\begin{align}
&\partial_\tau \mathcal{J}\psi(\tau,\sigma,\theta,\varphi)=-\partial_{\tp} \psi(\tp,\sigp,\theta,\check\varphi)\,,\label{eq:tau_J}\\
&\partial_\sigma \mathcal{J}\psi(\tau,\sigma,\theta,\varphi)= \partial_{\sigp} \psi (\tp,\sigp,\theta,\check\varphi) + \dfrac{2 \gamma(\check{\sigma})}{p(\check{\sigma})} \partial_{\tp} \psi (\tp,\sigp,\theta,\check\varphi) \nonumber \,\,.
\end{align}
Note that the expressions above are in agreement with the discussion in the previous section. Given a QNM solution $\psi_{\ell m n}(\tau, \sigma, \theta, \varphi)$ as in Eqn.~\eqref{eq:ret_fut}, the operator ${\cal J}$ will map it to its equivalent anti-QNM as in Eqn.~\eqref{eq:adv_fut}.

We conclude this section by reviewing the operator's interaction with the Hamiltonian ${\cal H}$. Applying the ${\cal J}$ operator to the Hamiltonian evolution Eqn.~\eqref{eq:H_future_evol} provides
\bea
{\cal J} \p_\tau \bar u &=& i {\cal J} {\cal H} \bar u, \nn
\eea
but a manipulation of the left-hand side leads to
\bea
{\cal J} \p_\tau \bar u &=& - \p_{\check\tau} \check u  \nn \\
&=& - i{\cal H} \check u \nn \\
&=& - i{\cal H} {\cal J} \bar u.  \nn
\eea
and therefore ${\cal J}{\cal H} = - {\cal H}{\cal J}$.

\subsection{Summary of coordinates and operators}
\label{sec:notation_summary}
Table~\ref{tab:notation} collects the coordinate systems introduced above and
their relation to the Schwarzschild--Droste coordinates $x^\mu$, together
with their respective wave operators.

\begin{table*}[ht!]
\label{tab:notation}
\begin{tabular}{|p{1.7cm}|p{2.3cm}|p{2.6cm}|p{1.8cm}|p{1.8cm}|p{2.6cm}|}
\hline
Coordinates & Coordinate & Coordinate &
\multicolumn{2}{c|}{Radial Boundary Location}  & Associated  \\ \cline{4-5}
System & Notation & Transformation & Infinity & Horizon & Wave-Operator \\
\hline
\hline
Schwarzschild--Droste & $x^\mu=(t,r,\theta,\varphi)$ & --- &
$r\to\infty$ $(i^0)$ & $r\to\rh$ $({\cal B})$ & $\square_{\ell m}$ --- Eq.~\eqref{eq:wave_operator_Schwarzscild} \\
& & & & &\\
Future hyperboloidal & $\bar x^\mu=(\tau,\sigma,\theta,\varphi)$ & $t=\lambda\big(\tau-H(\sigma)\big)$, $r=\dfrac{\rh}{\sigma}$ & $\sigma=0$ $(\scri^+)$ & $\sigma=1$  ($\mathcal H^+$) & $\bar L$ --- Eq.~\eqref{eq:waveeqHyp}  \\
& & & & &\\
Past hyperboloidal & $\check x^\mu=(\check\tau,\check\sigma,\theta,\check\varphi)$ & $\check \tau = -\tau + 2 H(\sigma)$, $\check \sigma = \sigma$, $\check\varphi=-\varphi$ &
$\check\sigma=0$ ($\scri^-$$)$ & $\check\sigma=1$ ($\mathcal H^-$) &
$\check L$ --- Eq.~\eqref{eq:waveeqHypPast} \\
\hline
\end{tabular}
\caption{Coordinate systems employed in this paper. The parameter $\lambda$ is a fixed length scale, $r_{\rm h}=2M$ is the horizon radius, and $H(\sigma)$ denotes the height function, which is determined by Eq.~\eqref{eq:MinimalGaugeHeightFunction} in the minimal gauge.}
\end{table*}


\section{Conserved Currents}
\label{sec:conserved_currents}
In this section, we discuss two examples of conserved currents associated with the wave equation on Schwarzschild background: the \emph{symplectic current} and the \emph{energy current}. Conserved currents represent the cornerstone for the construction of the orthogonality product in the next section.

\subsection{Symplectic current}
The symplectic current is defined as the bilinear current
\begin{align}
\label{eq:J_physical}
J^a(\Psi_1,\Psi_2) =  \Psi_2  \nabla^a  \Psi_1 -  \Psi_1  \nabla^a  \Psi_2,
\end{align}
where $\Psi_1, \Psi_2$ are solutions of the wave equation \eqref{eq:KG} for a massless scalar field. It is straight-forward to show that, on-shell,
\begin{align}
\nabla_a J^a =  \Psi_2  \square  \Psi_1 -  \Psi_1  \square  \Psi_2=0,
\end{align}
and thus the current is conserved.

Then, one can define the integral of the symplectic current $J^a$, on a constant $\tau$ hypersurface $\Sigma_\tau$ as
\be
\label{eq:Pi_prod}
\Pi_S[\Psi_1,\Psi_2]=\int_{\Sigma_\tau} J_a n^a d\Sigma_\tau \,,
\ee
where $n^a$ is the normal vector to $\Sigma_\tau$ and $d\Sigma_\tau$ is the volume form on this hypersurface
\bea \label{eq:hypersurface}
&&n^a=\sqrt{\frac{\sigma^2}{\lambda \,\rh\,w(\sigma)}} \bigg( w(\sigma) (\partial_\tau)^a - \gamma(\sigma) (\partial_\sigma)^a \bigg)\nn\\
&&{\rm d} \Sigma_\tau =\frac{\rh^2}{\sigma^2} \sqrt{\frac{\lambda\,\rh\,w(\sigma)}{\sigma^2}} \sin (\theta) {\rm d} \sigma \; {\rm d} \theta \; {\rm d} \varphi\,.
\eea
Putting everything together, we obtain
\bea
&&\Pi_S[\Psi_1, \Psi_2]=\nn\\
&&\qquad \dfrac{\rh^2}{\lambda^2} \oint {\rm d}\varpi \int\limits_{\substack{\mathcal{C} \\ \tau={\rm const.}}} \d\sigma \Bigg[ w(\sigma) \bigg( \psi_2\p_{\tau} \psi_1 - \psi_1\p_{\tau}\psi_2\bigg) \nn \\
&& \qquad \qquad\qquad  - \gamma(\sigma) \bigg(\psi_2\p_{\sigma} \psi_1- \psi_1\p_{\sigma}\psi_2  \bigg) \Bigg].
\label{eq:Pi_Simpletic}
\eea
In the above expression, we have used the conformal rescaling of the scalar field \eqref{Psi_confresc_future}.  Besides, we allowed for a generic integration contour $\mathcal{C}$ over the $\sigma$ variable to account for upcoming regularisation schemes.

In principle, $\partial_\tau \Pi_S[\Psi_1, \Psi_2]$ is non-zero due to boundary terms at the horizon and at null infinity\footnote{This is particularly relevant as we aim to extend our domain beyond compact support. }. 
Put differently, computing $\Pi_S[\Psi_1, \Psi_2]$ on different $\Sigma_\tau$ surfaces will give different results. These boundary terms correspond to the current's fluxes across infinitesimal hypersurfaces of $\sigma=$ constant, $\Sigma_{\sigma}$, at the endpoints of the $\Sigma_\tau$ hypersurface
\be
\label{eq:Flux}
{\cal F}_S^{\sigma}[\Psi_1, \Psi_2] = \int_{\Sigma_{\sigma}} J_a s^a d\Sigma_{\sigma},
\ee
where $s^a$ is the unit normal to $\Sigma_{\sigma}$ and $d\Sigma_{\sigma}$ its volume element
\bea \label{eq:hypersurface_sigma}
&&s^a=\sqrt{\frac{\sigma^2}{\lambda \,\rh\,p(\sigma)}} \bigg( \gamma(\sigma) (\partial_\tau)^a +p(\sigma) (\partial_\sigma)^a \bigg)\nn\\
&&{\rm d} \Sigma_\sigma =\frac{\rh^2}{\sigma^2} \sqrt{\frac{\lambda\,\rh\,p(\sigma)}{\sigma^2}} \sin (\theta) {\rm d} \tau \; {\rm d} \theta \; {\rm d} \varphi\,.
\eea
Considering the $\Sigma_\sigma$ hypersurfaces as the boundaries $\p{\cal C}$ associated with the integration contour in Eqn.~\eqref{eq:Pi_Simpletic}, the flux reads 
\bea
&&{\cal F}^{\p{\cal C}}_{S}[\Psi_1, \Psi_2]\equiv\int d\tau\, {F}^{\p{\cal C}}_{S} ,\\
&& {F}^{\p{\cal C}}_{S}=  \Bigg\{\dfrac{\rh^2}{\lambda^2} \oint {\rm d}\varpi  \Bigg[ \gamma(\sigma) \bigg( \psi_2\p_{\tau} \psi_1 - \psi_1\p_{\tau}\psi_2\bigg)  \\
&& \left.+ p(\sigma) \bigg(\psi_2\p_{\sigma} \psi_1- \psi_1\p_{\sigma}\psi_2  \bigg) \Bigg]\Bigg\}\right|_{\p\mathcal{C} }. \nn
\label{eq:Flux_Simpletic}
\eea
To clarify the connection between $\Pi_S$ and ${\cal F}^{\p{\cal C}}_{S}$ we point out that according to Gauss' theorem, we have 
\begin{equation}
    \Pi_S|_{\tau_0+\delta \tau_0}-\Pi_S|_{\tau_0}={\cal F}_S^{\partial\mathcal{C}}|_{\tau_0}^{\tau_0+\delta \tau_0} \,,
    \end{equation}
or equivalently
\begin{equation}
 \frac{d\Pi_S}{d\tau}= {F}^{\p{\cal C}}_{S}\Rightarrow\frac{d}{d\tau}\left(\Pi_S-{\cal F}^{\p{\cal C}}_{S}\right)=0 \,.
    \end{equation}
Note that a similar construction can take place on a past hyperboloidal slice. Specifically, on a surface of constant $\check \tau$, $\Sigma_{\check \tau}$, we can define 
\bea
&&\check\Pi_S[\Psi_1, \Psi_2]=\nn\\
&&\qquad \dfrac{\rh^2}{\lambda^2} \oint d\check\varpi \int\limits_{\mathcal{C},\,\check\tau=\check\tau_0} d\sigp \Bigg[ w(\sigp) \bigg( \psi_2\p_{\tp} \psi_1 - \psi_1\p_{\tp}\psi_2\bigg) \nn \\
&&\qquad \qquad - \gamma(\sigp) \bigg(\psi_2\p_{\sigp} \psi_1- \psi_1\p_{\sigp}\psi_2  \bigg) \Bigg],
\eea
where we have also used the conformal re-scaling \eqref{Psi_confresc_past}. In the same way, the flux across its boundaries read
\bea
\check{\cal F}^{\p{\cal C}}_{S}[\Psi_1, \Psi_2]=&&  \Bigg\{\dfrac{\rh^2}{\lambda^2} \oint d\check\varpi  \nn\\
&& \times \int d\check\tau \Bigg[ \gamma(\check\sigma) \bigg( \psi_2\p_{\check\tau} \psi_1 - \psi_1\p_{\check\tau}\psi_2\bigg)  \\
&& \left.+ p(\check\sigma) \bigg(\psi_2\p_{\check\sigma} \psi_1- \psi_1\p_{\check\sigma}\psi_2  \bigg) \Bigg]\Bigg\}\right|_{\p\mathcal{C} }. \nn
\eea

\subsection{Energy current}
Motivated by the role played by the energy norm in the calculation of the QNM pseudospectrum \cite{PhysRevX.11.031003,GasperinJaramillo2022}, here we also consider products derived from the energy current.

Consider a scalar field $\Psi$ that satisfies the massless wave equation \eqref{eq:KG}. The energy momentum tensor associated with this equation is given by
\bea \label{EnergyMomentumTensor}
T_{ab}=  \nabla_a \Psi \nabla_b \Psi - \frac{1}{2}g_{ab}g^{cd}\nabla_c \Psi\nabla_d \Psi \,.
\eea
Given a killing vector field $t^a = \left(\p_t \right)^a$, one can define the current 
\be
\label{eq:J_Energy}
\tilde J_a=T{}_{ab}t^b,
\ee
with the property that
\be
\nabla^a \tilde J_a = \frac{1}{2} T_{ab} \bigg{(} \nabla^a t^b+\nabla^b t^a \bigg{)}=0.
\ee
 
For a spacelike surface with constant time $\tau$, $\Sigma_{\tau}$, with normal vector and volume element given in Eqn.~\eqref{eq:hypersurface}, 
the total energy (on $\Sigma_{\tau}$) is given by 
\bea
&&\tilde E[\Psi]=\frac{1}{2} \oint {\rm d} \varpi \int\limits_{\substack{\mathcal{C} \\ \tau={\rm const.}}} \frac{\rh^2}{\sigma^2} \d\sigma\bigg( w( \sigma) \partial_\tau \Psi \partial_\tau \Psi \nonumber\\
&&\quad + p(\sigma) \partial_\sigma\Psi 
\partial_\sigma \Psi - \frac{\lambda}{\rh} \bigg( \Omega_{AB} \partial_A \Psi \partial_B \Psi \bigg) \bigg) \nonumber\\
&&=\frac{1}{2} \frac{\rh^2}{\lambda^2} \int {\rm d} \varpi \int\limits_{\substack{\mathcal{C} \\ \tau={\rm const.}}} \d\sigma\bigg[ w( \sigma) \partial_\tau \psi \partial_\tau \psi+ \frac{2 p(\sigma)}{\sigma}\psi \partial_\sigma\psi \nonumber\\ 
&& +\frac{p(\sigma)}{\sigma^2}\psi^2 
+ p(\sigma) \partial_\sigma\psi 
\partial_\sigma \psi + \frac{\lambda}{\rh} \bigg( \Omega^{AB} \partial_A \psi \partial_B \psi \bigg) \bigg],
\label{eq:EnergyProd}
\eea
 where $\Omega_{AB}={\rm diag}[1, \sin^2\theta]$ is the metric on the unit sphere $\mathbb{S}^2$ parametrised by $(\theta,\phi)$. As for the sympletic current, we allowed for a generic integration countour ${\cal C}$ for the eventual introduction of regularisation schemes. 

 To re-express Eqn.~\eqref{eq:EnergyProd} in a form akin to the one employed in \cite{PhysRevX.11.031003} one needs to re-arrange some terms \cite{GasperinJaramillo2022}. The angular contribution in Eqn.~\eqref{eq:EnergyProd} reads
\bea
&& \d\varpi\,\Omega^{AB} \partial_A \psi \partial_B \psi = \Bigg[ -\psi\Bigg( \dfrac{\p_\theta\left( \sin\theta \p_\theta\psi\right)}{\sin\theta} + \dfrac{\p^2_{\varphi\varphi}\psi}{\sin^2\theta}\Bigg) \nn \\
&&+\dfrac{\p_\theta\left( \sin\theta\, \psi \, \p_\theta\psi\right)}{\sin\theta} + \dfrac{\p_\varphi\left( \psi \,\psi_\varphi\right)}{\sin^2\theta} \Bigg]\d\varpi.
\label{eq:angular_Pi_E}
\eea
Upon integration on the sphere $(\theta, \varphi)\in [0,\pi]\times[0,2\pi)$, the angular boundary terms vanish assuming regularity of $\psi$ at the poles $\sin\theta = 0$, and periodicity in $\varphi\to\varphi+2\pi$.

We can also re-arrange the radial contribution into
\be
\frac{2 p(\sigma)}{\sigma}\psi \partial_\sigma\psi +\frac{p(\sigma)}{\sigma^2}\psi^2 = \p_\sigma \left( \dfrac{p(\sigma) \psi^2}{\sigma} \right) +\dfrac{\lambda}{\rh}\sigma \psi^2,
\ee
where we used the explicit form of $p(\sigma)$ in Eqn.~\eqref{eq:metric_funcs} to obtain the last term in the above expression. Thus, Eqn.~\eqref{eq:EnergyProd} decomposes into
\bea
&&\tilde E[\Psi] = E[\Psi] + \left.\dfrac{1}{2}\dfrac{\rh^2}{\lambda} \dfrac{p(\sigma) \psi^2}{\sigma}\right|_{\p{\cal C}}, \label{eq:EnergyProd_BoundTerm} \\
&&E[\Psi]=\frac{1}{2} \frac{\rh^2}{\lambda^2}\Bigg\{ \oint {\rm d} \varpi  \times\label{eq:EnergyProd_2} \\
&& \times\int\limits_{\substack{\mathcal{C} \\ \tau={\rm const.}}} \d\sigma\bigg( w( \sigma) (\partial_\tau \psi)^2 + p(\sigma) (\partial_\sigma \psi)^2 + \psi Q[\psi] \bigg) \Bigg\}, \nn
\eea
with the potential operator $Q$ defined in Eqn.~\eqref{eq:Potential_q}.
This is the energy associated with the wave equation expressed as in Eqn.~\eqref{eq:scalar_Schwarzschild}~\cite{GasperinJaramillo2022}. 

The equality $\tilde E[\Psi] = E[\Psi]$ holds if the boundary contribution in Eqn.~\eqref{eq:EnergyProd_BoundTerm} were to vanish.
This is indeed the case, if one considers the integration contour ${\cal C}$ to be the black-hole exterior region  $\sigma\in[0, 1]$. The argument is the same as for the angular boundary term. Here, $p(\sigma)/\sigma \sim \sigma(1-\sigma)$ plays the same role for $\sigma$ as $\sin\theta$ does for $\theta$, particularly since $\psi$ remains finite at the black hole horizon and future null infinity along the hyperboloidal hypersurface $\Sigma_\tau$.

It should be emphasised that the energy $E$ defined above is not conserved, since there is dissipation through the boundaries of the hypersurface $\Sigma_{\tau}$. More specifically, the energy current \eqref{eq:J_Energy} yields the flux contributions
\bea
&&\tilde {\cal F}^{\p{\cal C}}_E[\Psi] =  \dfrac{\rh^2}{\lambda^2} \Bigg\{\oint {\rm d}\varpi  \nn \\
&& \times \left.\int {\rm d}\tau \Bigg[ \gamma(\sigma)  \p_{\tau} \psi ^2  + p(\sigma) \bigg(\p_{\sigma} \psi + \dfrac{\psi}{\sigma}   \bigg)\p_\tau\psi \Bigg]\Bigg\}\right|_{\p\mathcal{C} }.
\label{eq:Flux_Energy}
\eea
The last term can be recast as a total derivative with respect to the hyperboloidal time $\tau$ as
\be
\dfrac{p(\sigma) \psi \p_\tau \psi}{\sigma} = \p_\tau \left( \dfrac{p(\sigma) \psi^2}{2\sigma}\right), \nn
\ee
so that Eqn.~\eqref{eq:Flux_Energy} takes the form
\bea
&&\tilde {\cal F}^{\p{\cal C}}_E[\Psi] =  {\cal F}^{\p{\cal C}}_E[\Psi] + \left.\dfrac{1}{2}\dfrac{\rh^2}{\lambda} \dfrac{p(\sigma) \psi^2}{\sigma}\right|_{\p{\cal C}},\label{eq:Flux_Energy_BoundTerm}  \\
&&{\cal F}^{\p{\cal C}}_E[\Psi]=  \dfrac{\rh^2}{\lambda^2}\Bigg\{ \oint {\rm d}\varpi \times \nn \\
&& \times \int {\rm d}\tau\left. \Bigg[ \gamma(\sigma)  (\p_{\tau} \psi) ^2  + p(\sigma)\p_{\sigma} \psi\p_\tau\psi\Bigg] \Bigg\}\right|_{\p\mathcal{C} }  .
\label{eq:Flux_Energy_2}
\eea
As before, the additional term in Eqn.~\eqref{eq:Flux_Energy_BoundTerm} vanishes if the contour boundaries correspond to $\sigma_0 = 0$ and $\sigma_1 = 1$. For consistency with the notation for a generic integration countour, we will keep these boundary contribution in Eqs.~\eqref{eq:EnergyProd_BoundTerm} and ~\eqref{eq:Flux_Energy_BoundTerm}, and discuss its role when necessary. However, we will define the energy~\cite{PhysRevX.11.031003,GasperinJaramillo2022}---and associated flux---products in terms of Eqs.~\eqref{eq:EnergyProd_2} and \eqref{eq:Flux_Energy_2}.

 Given that the energy current, $\tilde J_a$, is quadratic in the scalar field, we can define the associated bilinear energy current 
 \be
 \label{eq:bilinear_energy_current}
 J_a(\Psi_1,\Psi_2)=\left(\nabla_{(a} \Psi_1 \nabla_{b)} \Psi_2 - \frac{1}{2}g_{ab}g^{cd}\nabla_c \Psi_1\nabla_d \Psi_2 \right)t^b \, .
 \ee 
When $\Psi_1, \Psi_2$ are two solutions of the wave equation for a massless scalar field, it can be shown that $\nabla_a J^a=0$ is also conserved \cite{GasperinJaramillo2022}. As in the previous subsection, we can define the integral of this bilinear current on the $\Sigma_\tau$ hypersurface. Restricting to the effective contribution, as in Eqn.~\eqref{eq:EnergyProd_2}, we define a bilinear product, 
\bea
\label{eq:EnergyPi_2}
\Pi_E[\Psi_1, \Psi_2]&=&\frac{1}{2} \frac{\rh^2}{\lambda^2} \oint {\rm d} \varpi \int\limits_{\substack{\mathcal{C} \\ \tau={\rm const.}}} \d\sigma\bigg( w( \sigma) \partial_\tau \psi_1 \partial_\tau \psi_2 \\ 
&& + p(\sigma) \partial_\sigma\psi_1 
\partial_\sigma \psi_2 + \dfrac{\psi_1 Q[\psi_2] + \psi_2 Q[\psi_1] }{2} \bigg). \nn
\eea
Besides, the corresponding flux contribution reads
\bea
 {\cal F}^{\p{\cal C}}_{E}[\Psi_1, \Psi_2] &=& \Bigg\{ \dfrac{\rh^2}{\lambda^2} \oint {\rm d}\varpi \int {\rm d}\tau \Bigg[ \gamma(\sigma)  \p_{\tau} \psi_1 \p_{\tau} \psi_2  \\
 && \left.+ \dfrac{p(\sigma)}{2} \left( \p_{\sigma} \psi_1\p_\tau\psi_2 +\p_{\sigma} \psi_2\p_\tau\psi_1\right)\Bigg]\Bigg\}\right|_{\p\mathcal{C} }  \nn
\label{eq:Pi_Flux_Energy_2}
\eea
Moreover, similarly to the symplectic current, we can also define $\check\Pi_{E}[\Psi_1, \Psi_2]$, and $\check{\cal F}^{\p{\cal C}}_{E}[\Psi_1, \Psi_2]$ by picking a constant $\check\tau$ hypersurface in the past foliation.

\subsection{Bilinear properties}
The bilinear products defined above are associated with distinct conserved currents and exhibit different behaviour under the interchange of $\Psi_1,\Psi_2$. Specifically,
\begin{align}\label{eq:symm_pis}
\Pi_S[\Psi_1, \Psi_2]&=-\Pi_S[\Psi_2, \Psi_1],\nonumber\\
\Pi_E[\Psi_1, \Psi_2]&=\Pi_E[\Psi_2, \Psi_1]\,.
\end{align}
and also
\begin{align}\label{eq:symm_flux}
{\cal F}^{\p\cal C}_S[\Psi_1, \Psi_2]&=-{\cal F}^{\p\cal C}_S[\Psi_2, \Psi_1],\nonumber\\
{\cal F}^{\p\cal C}_E[\Psi_1, \Psi_2]&={\cal F}^{\p\cal C}_E[\Psi_2, \Psi_1]\,.
\end{align}

\subsubsection{From symplectic to energy}
It is easy to prove that, on shell, the two bilinear forms are related to each other via~\cite{GHSS2023,Hollands:2012sf}
\begin{align}
\label{eq:Pi_S_to_Pi_E}
\Pi_S[\Psi_1, \partial_\tau \Psi_2] &= 2 \Pi_E[\psi_1, \psi_2] \\
&\quad - \dfrac{\rh^2}{\lambda^2} \left. \psi_1 \Big( p(\sigma) \partial_\sigma \psi_2 + \gamma(\sigma) \partial_\tau \psi_2 \Big) \right|_{\p {\cal C}}. \nn
\end{align}
Similarly, the flux contributions at the boundaries are related as
\begin{align}
\label{eq:F_S_to_F_E}
{\cal F}^{\p {\cal C}}_S[\Psi_1, \partial_\tau \Psi_2] &= 2 {\cal F}_E^{\p {\cal C}}[\Psi_1, \Psi_2]  \\
&\quad - \dfrac{\rh^2}{\lambda^2}\left. \psi_1 \Big( p(\sigma) \partial_\sigma \psi_2 + \gamma(\sigma) \partial_\tau \psi_2 \Big) \right|_{\p {\cal C}}.\nn
\end{align}

 In other words, up to boundary terms,  the energy bilinear product can be derived from the bilinear product associated with the symplectic current through the insertion of the symmetry operator corresponding to a time translation invariance. The same is valid for the energy flux. Following \cite{GHSS2023}, this approach enables the identification of bilinear products associated with an infinite tower of conserved currents, obtained from the symplectic current through a recursive action of symmetry operators. In this sense, the bilinear product constructed from the symplectic current appears to play a more fundamental role.

\subsubsection{Hamiltonian action}
With the help of Eqs.~\eqref{eq:symm_pis} and \eqref{eq:Pi_S_to_Pi_E}, it is easy to prove that the action of the Hamiltonian $\mathcal{H}=i\partial_\tau$ on the symplectic product is such that
\bea\label{eq:contour_argument}
&\Pi_S[\mathcal{H}\Psi_1, \Psi_2]+\Pi_S[\Psi_1, \mathcal{H}\Psi_2]= i \dfrac{\rh^2}{\lambda^2} \times \nonumber\\
& \times \bigg( p(\sigma)(\psi_2 \partial_\sigma\psi_1-\psi_1 \partial_\sigma\psi_2) \nn \\
&+ \gamma(\sigma)(\psi_2 \partial_\tau\psi_1-\psi_1 \partial_\tau\psi_2) \bigg)\bigg|_{\p {\cal C}}.\,
\eea
In the same way, the action of the Hamiltonian in the sympletic-fluxes yields
\bea\label{eq:contour_argument_fluex}
&{\cal F}^{\p {\cal C}}_S[\mathcal{H}\Psi_1, \Psi_2]+{\cal F}^{\p {\cal C}}_S[\Psi_1, \mathcal{H}\Psi_2]= i \dfrac{\rh^2}{\lambda^2} \times \nonumber\\
& \times \bigg( p(\sigma)(\psi_2 \partial_\sigma\psi_1-\psi_1 \partial_\sigma\psi_2) \nn \\
&+ \gamma(\sigma)(\psi_2 \partial_\tau\psi_1-\psi_1 \partial_\tau\psi_2) \bigg)\bigg|_{\p {\cal C}}.\,
\eea
Similar expressions also hold for the energy product and energy-flux
\bea
\label{eq:contour_argument2}
&& \Pi_E[\Psi_1, {\cal H} \Psi_2]  + \Pi_E[{\cal H} \Psi_1, \Psi_2] = i\dfrac{\rh^2}{\lambda^2} \times \\
&& \times\left.\bigg[\gamma(\sigma)\p_\tau\psi_1\p_\tau\psi_2 + \dfrac{p(\sigma)}{2}\bigg(\p_\tau\psi_1 \p_\sigma\psi_2 + \p_\tau\psi_2 \p_\sigma\psi_1\bigg)\bigg]\right|_{\p {\cal C}}, \nn \\
&& {\cal F}^{\p {\cal C}}_{E}[\Psi_1,{\cal H} \Psi_2] + {\cal F}^{\p {\cal C}}_{E}[{\cal H} \Psi_1, \Psi_2] = i\dfrac{\rh^2}{\lambda^2} \times \label{eq:contour_argument2_flux} \\
&& 
\left.\bigg[\gamma(\sigma)\p_\tau\psi_1\p_\tau\psi_2 + \dfrac{p(\sigma)}{2}\bigg(\p_\tau\psi_1 \p_\sigma\psi_2 + \p_\tau\psi_2 \p_\sigma\psi_1\bigg)\bigg]\right|_{\p {\cal C}}. \nn
\eea
Note that the right-hand side of Eqs.~\eqref{eq:contour_argument}-\eqref{eq:contour_argument_fluex} and Eqs.~\eqref{eq:contour_argument2}-\eqref{eq:contour_argument2_flux} correspond to the fluxes' integrand in Eqs.~\eqref{eq:Flux_Simpletic} and \eqref{eq:Pi_Flux_Energy_2}, respectively.

If it were not for such boundary terms, the Hamiltonian operator would be formally skew-adjoint, $\mathcal{H}=-\mathcal{H}^\dagger$, under the bilinear products $\Pi[\Psi_1, \Psi_2]$. We emphasise that this notion should not be interpreted in the functional-analytic sense of a Hilbert adjoint, which requires specific assumptions on the domains of ${\cal H}$ and ${\cal H}^\dag$. Here, the adjoint is defined solely with respect to the bilinear product, i.e. via the algebraic expression
\be
\Pi[\Psi_1, \mathcal{H}\Psi_2]=\Pi[\mathcal{H}^\dagger\Psi_1, \Psi_2]\,.
\ee
These results highlight the need to develop strategies to account for the boundary contributions. This can be done either by employing one of the two analytic continuation strategies, discussed in Section~\ref{sec:implementation}, or by introducing an extend notion of the bilinear product, as discussed in the next section.

\subsubsection{Extended bilinear product}

\begin{figure}[t]
\includegraphics[width=1.\columnwidth]{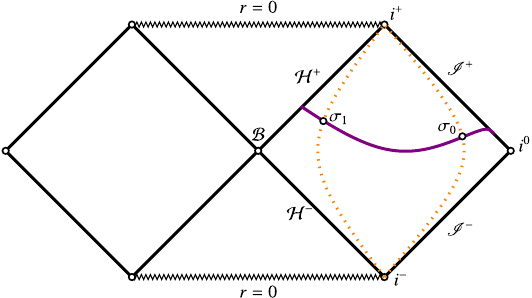}
  \caption{Carter-Penrose diagram representing the domain of the extended product. The purple curve represent a constant $\tau$ hypersurface in the future hyperboloidal foliation $\bar{x}^\mu$. The two dotted orange curves represent constant $\sigma_0$ and $\sigma_1$ hypersurfaces, across which fluxes flow. The extended product \eqref{eq:PiProd_Flux} incorporates the momentary flux at given $\tau=$constant. The limit $\sigma_0\to 0$ and $\sigma_1\to 1$ provides a route to study the interplay between fluxes contributions and regularisation schemes. }
    \label{fig:DomainExtendedProduct}
\end{figure}
To keep explicit control over the behaviour at the domain boundaries, as currently giving rise to the non-normal character of the evolution operator, we extend the definition of the $\Pi$-product in Eqn.~\eqref{eq:Pi_prod} as
\begin{eqnarray}
\label{eq:PiProd_Flux}
\tilde\Pi[\Psi_1, \Psi_2] = \Pi[\Psi_1, \Psi_2] - {\cal F}^{\p {\cal C}}[\Psi_1, \Psi_2],
\end{eqnarray}
which can be used either with the symplectic or the energy currents. Fig.~\ref{fig:DomainExtendedProduct} illustrates the key elements of the extend product, as it incorporates the bulk contribution along the $\tau =$  constant hypersurface parametrised by real coordinates $(\sigma,\theta,\varphi)\in[\sigma_0,\sigma_1]\times[0,\pi]\times[0,2\pi)$, as well as flux contributions across two $\sigma=$ constant boundaries. Then, it follows the conservation law $\p\tau \tilde \Pi[\Psi_1, \Psi_2] = 0$.

Combining Eqs.~\eqref{eq:symm_pis} and \eqref{eq:symm_flux}, the extend products retain the symmetries
\bea
\label{eq:PiFlux_symplectic_antisym}
&&\tilde \Pi_S[\Psi_1, \Psi_2] = -\tilde \Pi_S[\Psi_2, \Psi_1] \nn , \\
&&\tilde \Pi_E[\Psi_1, \Psi_2] = \tilde \Pi_E[\Psi_2, \Psi_1].
\eea
Moreover, the boundary terms in Eqs.~\eqref{eq:Pi_S_to_Pi_E} and \eqref{eq:F_S_to_F_E} cancel each other out exactly, yielding
\begin{equation}
\tilde \Pi_S[\Psi_1, \partial_\tau \Psi_2] = 2 \tilde \Pi_E[\Psi_1, \Psi_2].
\label{eq:PiFlux_symplectic_to_energy}
\end{equation}
In the same way, the boundary terms in Eqs.~\eqref{eq:contour_argument}-\eqref{eq:contour_argument_fluex} for the symplectic current, as well those in \eqref{eq:contour_argument2}-\eqref{eq:contour_argument2_flux} for the energy current also compensate each other, and we obtain the exact relation
\begin{equation}
\label{eq:sympletic_H_relation}
\tilde \Pi[{\cal H} \Psi_1, \Psi_2] + \tilde \Pi[\Psi_1, {\cal H} \Psi_2] = 0
\end{equation}
in both cases. In this way, it follows that the Hamiltonian operator is formally skew-adjoint ${\cal H}^\dag = - {\cal H}$ under the extended product, with the bi-linear adjoint with respect to the extended product Eqn.~\eqref{eq:PiProd_Flux} understood as
\be
\tilde \Pi[\Psi_1,{\cal H} \Psi_2] = \tilde \Pi[{\cal H}^\dag \Psi_1, \Psi_2].
\ee
If we consider pure QNM solutions $\Psi_{I_1}$ and $\Psi_{I_2}$ with respective frequencies $\omega_{I_1}$ and $\omega_{I_2}$, with $I=(
\ell, m, n)$, the bilinear product~\eqref{eq:PiProd_Flux} implies
\be
(\omega_{I_1} + \omega_{I_2})\,\tilde \Pi[\Psi_{I_1},\Psi_{I_2}] = 0.
\ee
In other words, the skew-adjoint structure of the product leads to pairing between frequencies $\omega_{1}$ and $-\omega_{2}$. However, a pairing in the QNM spectrum based on the physical decaying and oscillatory time scales relies, at most, on the mirror symmetry $\omega_{\ell m n} \leftrightarrow -\omega^\ast_{\ell -m n}$. Thus, in general, the factor $(\omega_{I_1}+\omega_{I_2})$ never vanishes, and we conclude
\be
\label{eq:Pi_Flux_Orthogonality}
\tilde \Pi[\Psi_{\rm I_1},\Psi_{\rm I_2}] = 0\quad \forall \, I_1, I_2.
\ee
At this stage, an orthogonality relation holds even for $I_1=I_2$, regardless if $\tilde \Pi$ is calculated in terms of the symplectic or the energy current. It is instructive to look at those cases separately.

Given the antisymmetric structure of the symplectic product, it is immediate that diagonal terms vanish, i.e.~$\tilde \Pi[\Psi_{ I_1},\Psi_{I_1}] = 0$. For $I_1 \neq I_2$ in the symplectic case, or for arbitrary $I_1$ and $I_2$ in the energy product, the result Eqn.~\eqref{eq:Pi_Flux_Orthogonality} is less trivial. 

A direct numerical investigation shows that the individual contributions $\Pi[\Psi_{I_1}, \Psi_{I_2}]$ and ${\cal F}^{\p {\cal C}}[\Psi_{I_1}, \Psi_{I_2}]$ are each non-zero. Remarkably, however, they cancel exactly under Eqn.~\eqref{eq:PiProd_Flux} for any choice of radial boundaries $\sigma_0$ and $\sigma_1$ (including the limits $\sigma_0\to 0$ and $\sigma_1\to 1$), and Eqn.~\eqref{eq:Pi_Flux_Orthogonality} is indeed satisfied.

The vanishing of the $\tilde \Pi$-product for all QNMs indicates that the bilinear form Eqn.~\eqref{eq:PiProd_Flux} is degenerate on the QNM subspace. Thus, to obtain a non-trivial pairing between QNMs, we exploit in the next section the role played by the operator ${\cal J}$~\cite{GHSS2023}, mapping modes into anti-modes—as discussed in Secs.~\ref{sec:J_op_Schwarzschild} and~\ref{sec:J_op_Hyp}.

In summary, both the symplectic and energy currents yield conserved bilinear forms $\Pi_S$ and $\Pi_E$ on constant-$\tau$ (or $\check\tau$) hyperboloidal slices, together with associated boundary fluxes $\mathcal F^{\partial\mathcal C}$ that account for their failure to be strictly conserved on a finite integration domain. These bilinear forms, and their transformation properties under $\mathcal J$, are the building blocks for the orthogonality product introduced in Sec.~\ref{sec:orthogonality_product}.

\section{Orthogonality Product}
\label{sec:orthogonality_product}

Consider a bilinear product in future hyperboloidal coordinates, $\Pi[\Psi_1(x),\Psi_2(x)]$, arising from a conserved current, $J^a[\Psi_1(x),\Psi_2(x)]$. We define the associated orthogonality product 
\bea
\label{eq:J_ProdFuture}
\langle {\Psi}{}_1(x),\Psi_2(x) \rangle_\tau:=\Pi[{\Psi}{}_1(x),\mathcal{J}\Psi_2(x) ]\,.
\eea
Note that here the bilinear product is to be understood either as the one arising from the symplectic current or the energy current. In a similar way, we can define the orthogonality product 
\bea \label{eq:J_ProdPast}
\langle {\Psi}{}_1(\check x),\Psi_2(\check x) \rangle_{\check\tau}:=\check\Pi[{\Psi}{}_1(\check x),\mathcal{J}\Psi_2(\check x) ]\,,
\eea
defined on a constant hyperboloidal slice in the past hyperboloidal parametrisation.

Aside from being $\mathbb{C}$-linear in both entries, the orthogonality product satisfies the following two main properties:
\begin{enumerate}
  \item As we saw in the previous subsection, the action of the Hamiltonian on the bilinear product is symmetric up to a total $\sigma$-derivative, cf.~Eqs.~\eqref{eq:contour_argument} and \eqref{eq:contour_argument2}
    \bea \label{eq:hamiltonaction}
\Pi[\mathcal{H}\Psi_1, \Psi_2]+\Pi[\Psi_1, \mathcal{H}\Psi_2] \nn \\
  =0+\text{total deriv. term}\,. 
  \eea
  Rewriting Eqn.~\eqref{eq:hamiltonaction} in terms of the orthogonality product and using $\mathcal{J} \mathcal{H}=-\mathcal{H} \mathcal{J} $, we obtain
  \bea
  \label{eq:hamiltonaction_ortho_prod}
  \langle\mathcal{H}\Psi_1, \Psi_2\rangle-\langle\Psi_1, \mathcal{H}\Psi_2\rangle\nonumber\\
  =0+\text{total deriv. term}
  \eea
The boundary terms in Eqs.~\eqref{eq:hamiltonaction_ortho_prod} correspond to those of Eqs.~\eqref{eq:contour_argument} and \eqref{eq:contour_argument2} under the action of ${\cal J}$ in $\Psi_2$.

Setting the boundary terms aside for the moment, Eq.~\eqref{eq:hamiltonaction_ortho_prod} suggests that
\be
\langle  \mathcal{H}\Psi_1, \Psi_2\rangle= \langle \Psi_1, \mathcal{H}\Psi_2\rangle=\langle \mathcal{H}^\dagger\Psi_1, \Psi_2\rangle\,.
\ee
i.e., indicates a  formally self-adjoint Hamiltonian operator under the orthogonality product Eqn.~\eqref{eq:J_ProdFuture}, as opposed to the skew-symmetric structure from the previous section.

Specialising on QNMs functions $\Psi_{I_1}$ and $\Psi_{I_2}$, we have
\be
(\omega_{I_1}-\omega_{I_2})\,\langle\Psi_{I_1},\Psi_{I_2}\rangle = 0,
\ee
which is a necessary condition for the orthogonality relation (up to normalisation)
\be
\langle\Psi_{I_1},\Psi_{I_2}\rangle=\delta_{\ell_1\ell_2}\delta_{m_1m_2}\delta_{n_1n_2}\,.
\ee
In other words, QNMs are orthogonal to each other under the product Eqn.~\eqref{eq:J_ProdFuture}. As emphasised earlier, this is subject to an appropriate integration strategy that removes boundary terms.

\item It is easy to show that 
\begin{eqnarray}
&&\langle \Psi_1(x), \mathcal{J} \Psi_2(x)\rangle_\tau =c \,\langle \Psi_2(x), \mathcal{J} \Psi_1(x)\rangle_\tau, \nn \\
&&\langle \Psi_1(x),\Psi_2(x)\rangle_\tau =c \,\langle \mathcal{J} \Psi_2(x), \mathcal{J} \Psi_1(x)\rangle_\tau\,.
\end{eqnarray}
where $c=-1$ for the symplectic current and $c=1$ for the energy current. Taking $\Psi_1, \Psi_2$ as QNMs, and remembering that the action of $\mathcal{J}$ on a QNM gives an anti-QNM, the above expressions imply that the orthogonality product between two anti-QNMs is essentially equivalent to that of two QNMs with the same frequencies. 
Equivalently, the same statement can be written as
\be 
\langle{\Psi}{}_1(x),\Psi_2(x) \rangle_\tau= \mathcal{J}\langle{(\mathcal{J}\Psi}{}_1)(\check x),(\mathcal{J}{\Psi}{}_2)(\check x)\rangle_{\check\tau}, \nn
\ee
making explicit that the orthogonality product of two QNMs on a future hyperboloidal foliation coincides to that of two anti-QNMs on a past hyperboloidal foliation.     
\end{enumerate}

In what follows, we will only consider the orthogonality product in future hyperboloidal coordinates. As such, we will suppress the index $\tau$ of the orthogonality product.

\subsection{Extended Orthogonality properties}\label{sec:extended_ortho_prod}

To account for the boundary terms, we extend the orthogonality product from Eqn.~\eqref{eq:J_ProdFuture} into its equivalent definition based on extending the bilinear $\tilde \Pi[\Psi_1, \Psi_2]$ via
\be
\label{eq:J_Prod_Flux}
(\Psi_1, \Psi_2) = \tilde \Pi[\Psi_1, {\cal J}\Psi_2].
\ee
Combining Eq.~\eqref{eq:sympletic_H_relation} with the identity ${\cal H}{\cal J} = -{\cal J}{\cal H}$, we compute exactly
\bea
(\Psi_1, {\cal H}\Psi_2)
&=& ({\cal H}\Psi_1, \Psi_2).
\eea

Thus, if we define the bi-linear adjoint with respect to the product Eqn.~\eqref{eq:J_Prod_Flux} by
\be
(\Psi_1, {\cal H}\Psi_2) = ({\cal H}^\dag\Psi_1, \Psi_2),
\ee
we obtain ${\cal H}^\dag = {\cal H}$, i.e. the Hamiltonian operator is formally self-adjoint under the product Eqn.~\eqref{eq:J_Prod_Flux} (in the same purely algebraic sense as discussed previously).

From the above results, it follows that, under the ${\cal J}$-bilinear product~\eqref{eq:J_Prod_Flux}
\be
(\omega_{1}-\omega_{2})\,( \Psi_{\rm 1},\Psi_{2} )= 0.
\ee
for QNMs functions $\Psi_{1}$ and $\Psi_{2}$.

Thus, by incorporating fluxes into the definition of the product, the cancellations of boundary terms occur identically for arbitrary choices of a radial integration contour ${\cal C}$ with boundary $\p {\cal C}$.

\subsection{Explicit expressions}
Upon specialising Eqn.~\eqref{eq:J_ProdFuture} to the symplectic current, we obtain 
\bea
&& \langle \Psi_1(x^\mu),\Psi_2(x^{\mu})\rangle:=\Pi_S[\Psi_1, \mathcal{J}\Psi_2]=\nn\\
&&\dfrac{\rh^2}{\lambda^2} \oint {\rm d}\varpi \int\limits_{\substack{\mathcal{C} \\ \tau={\rm const.}}} \d\sigma \Bigg[ w(\sigma) \bigg( \mathcal{J}\psi_2\p_{\tau} \psi_1 - \psi_1\p_{\tau} \mathcal{J}\psi_2\bigg) \nn \\
&&-  \gamma(\sigma) \bigg(\mathcal{J}\psi_2\p_{\sigma} \psi_1- \psi_1\p_{\sigma}\mathcal{J}\psi_2  \bigg) \Bigg].
\label{eq:S_prod_J}
\eea
In the case of QNMs $\Psi_1$, $\Psi_{2}$, this reduces to
\bea
&& \langle \Psi_1(x^\mu),\Psi_2(x^{\mu})\rangle:= e^{-i(\omega_{I_1}-\omega_{I_2}) \lambda \tau_o}\, \dfrac{\rh^2}{\lambda^2} \times \nn\\
&& \times \left( \oint {\rm d}\varpi Y_{\ell_1 m_1}(\theta, \varphi)Y^*_{\ell_2 m_2}(\theta, \varphi) \right)\int\limits_{\substack{\mathcal{C} \\ \tau={\rm const.}}} \d\sigma   \times \nn \\
&&\times e^{-i2 \lambda \omega_{I_2} H(\sigma)} \Bigg[\bigg(-i w(\sigma)\lambda(\omega_{I_1} - \omega_{I_2}) - \frac{2 i \lambda \omega_{I_2}}{p(\sigma)}\bigg) \phi_{I_1} \phi_{I_2} \nn \\
&& 
- \gamma(\sigma) \bigg( \phi_{I_2} \p_{\sigma}\phi_{I_1} - \phi_{I_1} \p_{\sigma}\phi_{I_2} \bigg) \Bigg]\,,
\label{eq:S_prod_explicit}
\eea
where we have also made explicit the decomposition in spherical harmonics. Equivalent expression results for the symplectic fluxes upon the action of the ${\cal J}$ operator, and specialising to a QNM eigenfunctions
\bea
&&{\cal F}^{\p {\cal C}}_S[\Psi_1, {\cal J}\Psi_2]
=
\dfrac{\rh^2}{\lambda^2}
\delta_{\ell_1, \ell_2} \delta_{m_1, m_2}\,
e^{-i\lambda(\omega_{I_1}-\omega_{I_2}) \tau_0}
\times \nn  \\
&&\times
e^{-2i\lambda \omega_{I_2} H(\sigma)} \Bigg[
\dfrac{p(\sigma)}{-i\lambda(\omega_{I_1}-\omega_{I_2})}
\bigg(
\partial_\sigma \phi_{I_1}(\sigma)\phi_{I_2}(\sigma) \nn \\
&& -
\partial_\sigma \phi_{I_2}(\sigma)\phi_{I_1}(\sigma)
\bigg)
+
\gamma(\sigma_i)\phi_{I_1}(\sigma_i)\phi_{I_2}(\sigma_i)
\Bigg] \Bigg|_{\p {\cal C}}.
\label{eq:Pi_Sympletic_QNM_Flux}
\eea
For simplicity, the above expression already performs the integral over the spherical harmonics.

Similarly, we provide explicit expressions for the energy orthogonality product. Given fields $\Psi_1$ and $\Psi_2$ on a constant $\tau=\tau_0$ hypersurface, the energy product is 
\bea
&&\langle \Psi_1,\Psi_2 \rangle_E
:=\Pi_E[\Psi_1, \mathcal{J}\Psi_2]=\nn  \\
&&\frac{1}{2}\frac{\rh^2}{\lambda^2}\oint {\rm d}\varpi \int\limits_{\substack{\mathcal{C} \\ \tau={\rm const.}}} (w(\sigma) \partial_\tau \psi{}_1\partial_\tau\mathcal{J}\psi_{2} +p(\sigma)\partial_\sigma \psi_1\partial_\sigma\mathcal{J}\psi_2 \nn \\
&&+ \dfrac{q_{\ell_1 m_1}(
\sigma) + q_{\ell_2 m_2}(
\sigma)}{2}\psi{}_1\mathcal{J}\psi_2 )\; {\rm d}\sigma\,.  \eea
When $\Psi_1$ and $\Psi_2$ are two QNMs, the product takes the form 
\bea
&&\langle \Psi_{1},\Psi_{2} \rangle_E= \frac{1}{2}\frac{\rh^2}{\lambda^2}e^{-i \lambda (\omega_{I_1}-\omega_{I_2})\tau_0} \times \nn \\
&& \times\left( \oint {\rm d}\varpi Y_{\ell_1 m_1}(\theta, \varphi)Y^*_{\ell_2 m_2}(\theta, \varphi) \right) \int\limits_{\substack{\mathcal{C} \\ \tau={\rm const.}}} e^{-i\,2 \omega_{I_2} \lambda H(\sigma)} \times \nn\\
&&\times \bigg(\lambda^2 \omega_{I_1}\omega_{I_2}w(\sigma) \phi_{I_1}  \phi_{I_2} +p(\sigma)\partial_\sigma \phi{}_{I_1} \partial_\sigma \phi_{I_2} \label{eq:E_prod_explicit} \\
&& -2i \lambda \omega_{I_2}\gamma(\sigma)\p_\sigma \phi{}_{I_1} \phi_{I_2} + \dfrac{q_{\ell_1\,m_1}(\sigma) + q_{\ell_2\,m_2}(\sigma)}{2}\phi{}_{I_1}  \phi_{I_2} \bigg)\; {\rm d}\sigma. \nn 
\eea
The equivalent expression for the fluxes associated with the energy current is
\bea
&&
{\cal F}_E^{\p {\cal C}}[\Psi_{1}, {\cal J}\Psi_{2}]
=
\dfrac{\rh^2}{2 \lambda^2}
\delta_{\ell_1, \ell_2} \delta_{m_1, m_2}\,
e^{-i\lambda(\omega_{I_1}-\omega_{I_2}) \tau_0}
\times \nn \\
&&
\label{eq:Pi_Energy_QNM_Flux}
\times
e^{-2i\lambda\omega_{I_2} H(\sigma)}\dfrac{p(\sigma)}{\omega_{I_1}-\omega_{I_2}}
\bigg(
\omega_{I_1}\partial_\sigma \phi_{I_2}(\sigma)\phi_{I_1}(\sigma) \nn \\
&& -
\omega_{I_2}\partial_\sigma \phi_{I_1}(\sigma)\phi_{I_2}(\sigma)
\bigg)\Bigg|_{\p {\cal C}}.
\eea

We finish this section drawing attention to the term $e^{-i\,2 \omega_{I_2} \lambda H(\sigma)}$ present in the integrand of both products \eqref{eq:S_prod_explicit} and \eqref{eq:E_prod_explicit}, as well as in the respective fluxes \eqref{eq:Pi_Sympletic_QNM_Flux} and \eqref{eq:Pi_Energy_QNM_Flux}.
The height function is explicitly given in Eqn.~\eqref{eq:MinimalGaugeHeightFunction}, which implies that $e^{-i\,2 \omega_{I_2} \lambda H(\sigma)} \to \infty$ for $\sigma \to 0$ and $\sigma \to 1$, whenever ${\rm Im}(\omega_{I_2})<0$. As discussed in Sec.~\ref{sec:J_op_Hyp}, this term is a direct consequence of applying the ${\cal J}$ operation in the product's definition \eqref{eq:J_ProdFuture}, which maps a QNM into an anti-QNM. The integrand's divergence, therefore, reflects the singular 
behaviour of anti-QNMs at the black-hole horizon and future null infinity. Hence, regularisation schemes are required to perform the integrals \eqref{eq:S_prod_explicit} and \eqref{eq:E_prod_explicit}.

\subsection{Regularisation strategies}\label{sec:implementation}
In this Section, we scrutinise the singular behaviour towards the black-hole hole horizon and future null infinity introduced by the evaluation of the anti-QNMs thereon. We begin with a direct analysis of the blow-ups in terms of the extended orthogonality product, which points out to the limitations of this scheme in practical terms. Then, we describe two alternative methods to regularise the orthogonality product, both relying on the analytic continuation of the relevant parameters. 

The \emph{semi-analytic approach} first restricts the frequency parameter in a region where the radial integral converges and it has a closed form expression in terms of special function. Then, the parameter $\omega$ is analytically extended for the QNM values. The second strategy relies on the introduction of a \emph{complex integration contour} for the radial variable, choosing the radial integration path such that the integral is also convergent. Crucially, we verify that the underlying fluxes vanish in both strategies.

\subsubsection{Boundary behaviour}
We expand the expressions \eqref{eq:S_prod_explicit}-\eqref{eq:Pi_Energy_QNM_Flux} around future null infinity $\sigma_0 = \epsilon$ and the black-hole horizon $\sigma_1 = 1 - \epsilon$, for $\epsilon\gtrsim 0$. For that purpose, we make use of the regularity condition for the radial fields provided by the radial equation \eqref{eq:Freq_domain_radial_eq_ret} 
\bea
\p_\sigma \phi_{I_1}(\sigma_0) &=&\Bigg( -2i \rh \omega_{I_1} + i\dfrac{\ell_1(\ell_1+1)}{2\rh \omega_{I_1}} \Bigg) \times  \phi_{I_1}(\sigma_0) \\
&& + {\cal O}(\epsilon) \nn \\
\p_\sigma \phi_{I_1}(\sigma_1) &=&\Bigg(1 -2i\rh \omega_{I_1} + \dfrac{\ell_1(\ell_1+1) -(2\rh \omega_{I_1})^2}{1 - 2i\rh \omega_{I_1}} \Bigg) \times \nn \\
&& \times \phi_{I_1}(\sigma_1) + {\cal O}(\epsilon).
\eea
These conditions are directly applicable to the fluxes expressions yielding for the symplectic current Eqn.~\eqref{eq:Pi_Sympletic_QNM_Flux}  
\bea
\label{eq:flux_symp_asymp_scri}
{\cal F}_S^{\sigma 0} &=& \dfrac{\rh^2}{\lambda^2}\,\delta_{\ell_1, \ell_2} \delta_{m_1, m_2}\,
e^{-i\lambda(\omega_{I_1}-\omega_{I_2}) \tau_0}  \, \phi_{I_1}(0)\phi_{I_2}(0) \times \nn\\
&&\times e^{{2i \rh \omega_{I_2}}/\epsilon} \, \epsilon^{-2i \rh \omega_{I_2}}  \Bigg[ 1 +{\cal O}(\epsilon) \Bigg] \\
\label{eq:flux_symp_asymp_hrzn}
{\cal F}_S^{\sigma 1} &=& -\dfrac{\rh^2}{\lambda^2}\,\delta_{\ell_1, \ell_2} \delta_{m_1, m_2}\,
e^{-i\lambda(\omega_{I_1}-\omega_{I_2}) \tau_0} \, \phi_{I_1}(1)\phi_{I_2}(1)  \times \nn \\
&& \times \epsilon^{-2i\rh \omega_2}\Bigg[ 1 +{\cal O}(\epsilon)\Bigg],
\eea
and for the energy current Eqn.~\eqref{eq:Pi_Energy_QNM_Flux}
\bea
\label{eq:flux_energy_asymp_scri}
&&{\cal F}_E^{\sigma 0} =\delta_{\ell_1, \ell_2} \delta_{m_1, m_2}\,
e^{-i\lambda(\omega_{I_1}-\omega_{I_2}) \tau_0} \,\dfrac{\ell_1(\ell_1+1)}{4}\,     \times \nn \\
&&\times \dfrac{i}{\lambda}\dfrac{\omega_{I_1} + \omega_{I_2}}{\omega_{I_1}\omega_{I_2}}  \phi_{I_1}(0)\phi_{I_2}(0)\,  e^{{2i \rh \omega_{I_2}}/ \epsilon} \, \epsilon^{2 - 2i \rh \omega_{I_2}}  \nn \\
&& \times \Bigg[1 +{\cal O}(\epsilon) \Bigg], \\
\label{eq:flux_energy_asymp_hrzn}
&&{\cal F}_E^{\sigma 1} = \dfrac{\rh}{2\lambda}\,\delta_{\ell_1, \ell_2} \delta_{m_1, m_2}\,
e^{-i\lambda(\omega_{I_1}-\omega_{I_2})\tau_0} \, \phi_{I_1}(1)\phi_{I_2}(1) \times \nn \\
&&\times   \bigg(1+\ell_1(\ell_1 +1)\bigg)  \Bigg( \dfrac{1 - 2i\rh (\omega_{I_1} +\omega_{I_2})}{(1-2i\rh \omega_{I_1})(1-2i\rh \omega_{I_2})}\Bigg) \times\nn \\
&&  \times \epsilon^{1-2i\rh \omega_{I_2}}  \Bigg[ 1 +{\cal O}(\epsilon)\Bigg].
\eea
Close to these boundaries, the integral for the bulk contributions goes for the symplectic current Eqn.~\eqref{eq:S_prod_explicit}, as
\bea
&& \Pi_S \underset{\sigma\sim0}{\approx}  \dfrac{\rh^3}{\lambda^3}\,\delta_{\ell_1, \ell_2} \delta_{m_1, m_2}\,
e^{-i\lambda(\omega_{I_1}-\omega_{I_2}) \tau_0} \,  2i\lambda \omega_{I_2}  \times  \label{eq:PI_sym_0} \\ 
&&\times \phi_{I_1}(0)\phi_{I_2}(0) \int \d\sigma\, e^{-2 i \rh  \omega_{I_2}/\sigma} \, \sigma^{-2 - 2i \rh \omega_{I_2}} \Bigg( 1+ {\cal O}(\sigma) \Bigg), \nn \\
&& \Pi_S \underset{\sigma\sim 1}{\approx} \dfrac{\rh^3}{\lambda^3}\,\delta_{\ell_1, \ell_2} \delta_{m_1, m_2}\,
e^{-i\lambda(\omega_{I_1}-\omega_{I_2})\tau_0} \,  2i\lambda \omega_{I_2}  \times  \\
&& \phi_{I_1}(1)\phi_{I_2}(1) \times \int \d\sigma\, (1- \sigma)^{-1 - 2i \rh \omega_{I_2}}\Bigg(1 + {\cal O}(1-\sigma)\Bigg), \nn
\eea
while for the energy current they read, cf.~Eq.~\eqref{eq:E_prod_explicit}
\bea
&& \Pi_E \underset{\sigma\sim0}{\approx}  \dfrac{\rh}{\lambda}\,\delta_{\ell_1, \ell_2} \delta_{m_1, m_2}\,
e^{-i\lambda(\omega_{I_1}-\omega_{I_2}) \tau_0} \,  \dfrac{\ell_1(\ell_1+1)}{2}  \times \nn \\ 
&&\times \dfrac{\omega_{I_1}+\omega_{I_2}}{\omega_{I_{1}}} \phi_{I_{1}}(0)\phi_{I_{2}}(0)  \int \d\sigma\, e^{-2i \rh \omega_{I_{2}}/\sigma} \sigma^{-2i \rh \omega_{I_{2}}}  \nn \times \\
&& \times \Bigg( 1+ {\cal O}(\sigma) \Bigg), \label{eq:PI_en_0} \\
&& \Pi_E \underset{\sigma\sim 1}{\approx} \dfrac{\rh}{2\lambda}\,\delta_{\ell_1, \ell_2} \delta_{m_1, m_2}\,
e^{-i\lambda(\omega_{I_{1}}-\omega_{I_{2}}) \tau_0} \, \phi_{I_{1}}(1)\phi_{I_{2}}(1) \times \nn \\
&&\times   \bigg(1+\ell_1(\ell_1 +1)\bigg)   \dfrac{1 - 2i\rh (\omega_{I_{1}} +\omega_{I_{2}})}{1-2i\rh \omega_{I_{1}}} \times\nn \\
&&  \times \int \d\sigma\, (1-\sigma)^{-2i\rh \omega_{I_{2}}}  \Bigg[ 1 +{\cal O}(1-\sigma)\Bigg].
\eea
Thus, the asymptotic structure of the bulk and flux contributions suggests a nontrivial interplay between the divergent terms arising at the boundaries. In particular, the leading-order behaviour indicates that the singular factors appearing in the bulk integrals cancel those present in the flux terms. However, the exponential behaviour of the integrands near $\sigma=0$ (cf.~Eqs.~\eqref{eq:PI_sym_0} and \eqref{eq:PI_en_0}) prevents a straightforward term-by-term cancellation at the level of the real-line integral.

This observation suggests that a more refined treatment of the boundary contributions is required. In practice, performing the calculation directly on the real line is numerically challenging, as we discuss in Sec.~\ref{sec:num_direct_integration}. 

Instead, we explore the regularisation strategies discussed in the literature \cite{GHSS2023,London26}. By exploiting analytical extensions one can control these divergences and extract finite, well-defined results. 
It is natural to expect that a detailed treatment of such manipulations in the complex plane may be understood and reinterpreted in terms of the bulk integrals and the physical flux terms. This expectation is motivated by standard results in complex analysis regarding contour deformation and analytic continuation (see, e.g.,~\cite{BleisteinHandelsman1986,ChurchillBrown2014}). Establishing this connection rigorously, however, requires a detailed analysis of the analytic structure of the integrand and its behaviour near the boundaries, which we leave for future work.

\subsubsection{Semi-analytic integration}\label{sec:semianal-int}
One route to integrating orthogonality products is through a semi-analytic approach inspired by \cite{London26}. The QNM field $\phi_I(\sigma)$ solves the radial equation \eqref{eq:Freq_domain_radial_eq_ret}. This is a confluent Heun equation with regular singular points at $\sigma=\{1, \infty\}$ and an irregular singular point at $\sigma=0$ ---see \cite{Minucci:2024qrn}.
Then, we use Frobenius method and expand $\psi_I(\sigma)$ around the regular singular point $\sigma=1$ (black hole event horizon) via
\bea
\label{eq:FrobeniusSeries}
&&\phi_I(\sigma) = \sum_{p=0}^{\infty} a_p (\omega_{I}) \, (1-\sigma)^{r_I+p}
\eea
which, when replaced in the radial equation, allows us to determine $r_I$ to be $\{0, 2 i \rh \omega_{I} \}$. We restrict ourselves to the value $r_{I}=0$ corresponds to the regular expansions around the black-hole horizon $\mathcal{H}^+$. We recall that, in the minimal gauge, the expansion of the hyperboloidal field \eqref{eq:FrobeniusSeries} coincides with the one suggested by Leaver~\cite{Leaver:1985ax}. Hence, the coefficients $a_p$ satisfy a three terms recurrence relation as detailed in \cite{Leaver:1985ax} and the QNM values $\omega_{I}$ are those for which the sequence $\{a_p(\omega_{\rm I})\}$ is a minimal solution.

When we replace the resulting series Eqn.~\eqref{eq:FrobeniusSeries} in
the bilinear form, the latter splits into  a sum of distinct integrals, each of which can be recast as 
\bea
\label{eq:TricomiIntegral}
U(a,b,z)=\dfrac{1}{\Gamma(a)}\int^{\infty}\limits_{\substack{{0}}}  e^{-zt} t^{a-1}(1+t)^{b-a-1}\, d t,
\eea
 using $t=\dfrac{(1-\sigma)}{\sigma}$, $z=-2 i \rh \omega_I $ and integrating on the real line. The values of $a$ and $b$ are positive integers, depending on the particular contribution of $\sigma$ and $(1-\sigma)$ from the products and the series expansion \eqref{eq:FrobeniusSeries}.
 
 The function $U(a,b,z)$ is the Tricomi hypergeometric function of the second kind and is only convergent for ${\rm Re}(z)>0$ and ${\rm Re}(a)>0$.  In the case of QNMs, ${\rm Im}(\omega_I)<0 \Rightarrow {\rm Re}(z)<0$. Thus, we assume the validity of the integral \eqref{eq:TricomiIntegral} in its domain of convergence, and then we analytically continue the function $U(a,b,z)$ over ${\rm Re}(z)$. 
 
 To do this, we can either use  the functions inbuilt in $\textsc{Mathematica}$ as a black box to compute directly $U(a,b,z)$ for $Re(z)<0$,  or expressing $U(a,b,z)$ in terms of the Kummer function $M(a,b,z)$
\bea
U(a,b,z)&=&\dfrac{\pi}{\sin(\pi b)}\bigg[ \dfrac{M(a,b,z)}{\Gamma(1+a-b)\Gamma(b)} \nn \\
&&-z^{1-b}\dfrac{M(1+a-b,2-b,z)}{\Gamma(a)\Gamma(2-b)}\bigg]. \nn
\eea
which is defined for all complex $a$, $b$, $z$. In the above expression we have used that
\bea
\Gamma(b)\Gamma(1-b)=\dfrac{\pi}{\sin(\pi b)}, \qquad {\rm for} \quad b \notin \mathbb{Z}. \nn
\eea

In doing so, the fluxes are set to zero at both boundaries. Specifically, upon analysing the fluxes close to $\sigma=0$, we see that the exponential factor $e^{-2i\lambda\omega_2 \,H(\sigma)}$ gives $e^{-z/\sigma} \sigma^{z}$, which for ${\rm Re}(z)>0$, kills all the fluxes at null infinity. Now, when looking at the fluxes at the event horizon ($\sigma\to1$), the integrand gives $(1-\sigma)^{z}$, which once again vanishes for ${\rm Re}(z)>0$ and removes the flux contributions. 

We conclude that for ${\rm Im}(\omega)>0 \Rightarrow {\rm Re}(z)>0$, the orthogonality products coincide $\braket{\Psi_1,\Psi_2} = (\Psi_1, \Psi_2)$, and the bulk integral can be analytically extended into the QNM region ${\rm Im}(\omega)<0$.

\subsubsection{Complex contour}\label{sec:complexcontour-int} 
An alternative approach to computing the orthogonality product relies on choosing a complex integration contour, $\mathcal{C}$, for numerical integration. The complex contour is more easily defined in the un-compactified coordinate $r$, rather than the compactified $\sigma$, as the singularity at $r=\infty$ ($\sigma=0$) can be more easily approached from a direction of convergence. 
\begin{figure*}[t!]
\centering
\includegraphics[width=0.49\textwidth]{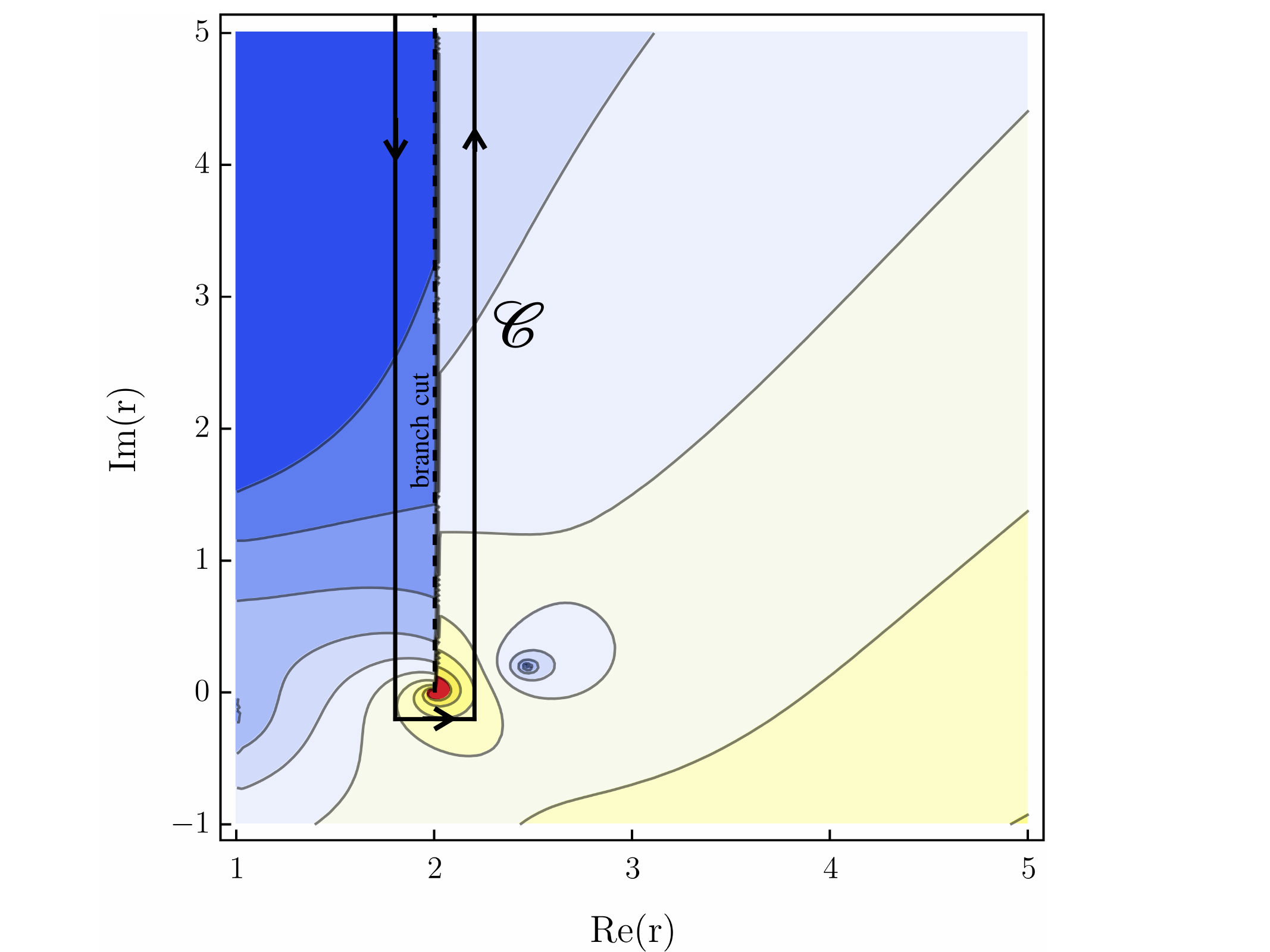}\hspace{-10pt}
\includegraphics[width=0.49\textwidth]{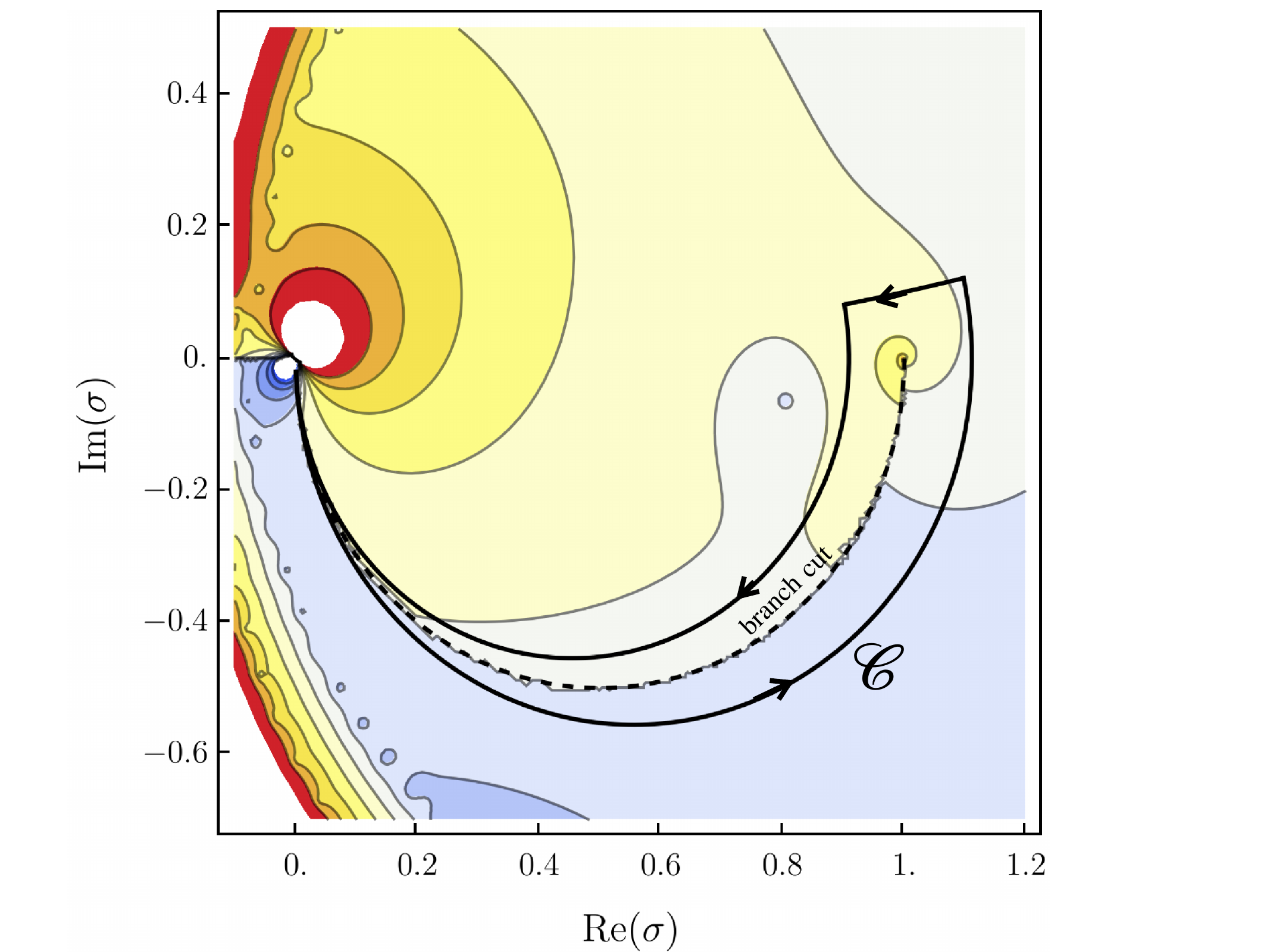}
\caption{Complex contour (solid black line) for the definition of the QNM products in the coordinate $r$ (left panel) and the compactified coordinate $\sigma$ (right panel). 
  The colours show where the integrand decays (blue) or grows (red) exponentially. The branch cut (dashed black line) of the integrand has been rotated to lie inside the contour.}
    \label{fig:contour_complex}
\end{figure*}

We therefore analytically extend the integrand in $r$, placing the branch cut of the integrand emanating from the horizon $\rh$ parallel to the positive imaginary axis, as shown in Fig.~\ref{fig:contour_complex}. At large $r\gg \rh$, the integrand behaves as $\sim e^{-z \,r/ \rh}$, with $z=-2\,i\omega_I\rh$ as in the previous regularisation stategy. Then, for any mode with ${\rm Re}(i\,z)>0$, the integral is convergent if the contour approaches complex infinity parallel to the positive imaginary axis, on either side of the branch cut. 
We use a contour $\mathcal{C}$ that runs along and around the branch cut and ends at complex infinity at both ends, as shown in Fig.~\ref{fig:contour_complex}.  For modes with ${\rm Re}(i\,z)<0$, we would instead place the branch cut parallel to the negative imaginary axis, and reflect the contour accordingly.

Although the contour was more easily defined in terms of the coordinate $r$, it can also be expressed in terms of the compactified coordinate $\sigma$. This is shown in the right panel of Fig.~\ref{fig:contour_complex}. We obtain quantitatively similar values for the QNM product when evaluating it in terms of the compactified coordinate. 

Finally, we observe that the same strategy regularising the bulk integral, is also responsible for setting the fluxes to zero, and therefore the products also coincide  $\braket{\Psi_1,\Psi_2} = (\Psi_1, \Psi_2)$ when evaluated along the complex contour ${\cal C}$.  

To summarize, the orthogonality product pairs a QNM with an equivalent anti-mode, i.e., the $\mathcal J$-image of another QNM, which is necessarily singular at the boundaries adapted to the first; this divergence is regularized either by a semi-analytic continuation via Tricomi functions or by complex-contour deformation of the radial integral, and the two methods agree. Including the boundary fluxes explicitly (Sec.~\ref{sec:extended_ortho_prod}) renders the Hamiltonian formally normal, at the cost of requiring full time-domain information rather than initial data alone.

\section{Excitation Coefficients}
\label{sec:excitation_coeffs}
One of the main applications of QNM orthogonality products is the calculation of QNM excitation amplitudes $c_{I}$ directly from a given initial data set $\{ {\Psi}_{\rm ID}(\sigma, \theta, \varphi), \dot{\Psi}_{\rm ID}(\sigma, \theta, \varphi) \}$, i.e., without having to infer them from the time evolution $\Psi(\tau, \sigma, \theta, \varphi)$ uniquely determined by the wave equation~\eqref{eq:waveeqHyp}. In this section, provide a detailed comparison between the determination of QNM excitation amplitudes via  Green's function and the orthogonality projections previously introduced. 

In particular, we focus initially on the QNM amplitudes arising from the projection~\cite{GHSS2023}
\be\label{eq:exc_coeff_product}
c_I = \frac{ \langle {\Psi},\Psi_I \rangle}{\langle {\Psi}{}_I,\Psi_I \rangle} \, ,
\ee
and then discuss the formalism under the extended producted introduced in Sec.~\ref{sec:extended_ortho_prod}.

\subsection{Projections versus Green's function}
To show that Eq.~\eqref{eq:exc_coeff_product} does correspond to the QNM excitation amplitude, we follow Ref.~\cite{GHSS2023} and compare this expression with the original definition of the excitation coefficients based on a Laplace transform explicitly calculated within the hyperboloidal framework. 

We start from the wave equation in hyperboloidal coordinates, Eqn.~\eqref{eq:waveeqHyp}. On a Schwarzschild background, we fix the angular component and suppress its $\ell, m$ indices\footnote{By omitting $(\ell, m)$, we restrict the notation to the QNM overtone number $n$, as opposed to the collective index $I=(\ell, m, n)$.}. We also denote by $\{ {\phi}_{\rm ID}(\sigma), \dot{\phi}_{\rm ID}(\sigma) \}$ the  initial data set associated with the particular $\ell, m$ contribution.
We then perform a Laplace transform in $\tau$, 
\bea
\tilde{\phi}(\omega; \sigma) = \int_0^\infty {\rm d}\tau \left[ e^{i \lambda\omega \tau} \psi(\tau, \sigma) \right],
\eea
with the inverse transform defined as 
\bea \label{eq:inverse_Laplace}
\psi(\tau,\sigma) = \frac{1}{2\pi}\int_{-\infty+i\epsilon}^{\infty+i\epsilon} d\omega \left[ e^{-i\lambda\omega \tau} \tilde\phi(\omega; \sigma) \right] \, .
\eea
The Laplace transform $\tilde{\phi}$ must satisfy
\bea\label{eq:laplace_transformed_eq}
\bar{\rm A}( \omega)[\tilde{\phi}] = \mathcal{I}(\omega; \sigma),
\eea
with the operator $\bar{\rm A}( \omega)$ defined in Eqn.~\eqref{eq:Freq_domain_radial_eq_ret} and $\mathcal{I}(\omega;\sigma)$ a functional of the initial data $\{ {\phi}_{\rm ID}(\sigma), \dot{\phi}_{\rm ID}(\sigma)\}$
\be
\mathcal{I}(\omega; \sigma) = -w(\sigma) \dot{\phi}_{\rm ID}(\sigma) + \left[ \bar{L}_2+ i \lambda \omega w(\sigma) \right] {\phi}_{\rm ID}(\sigma) \, .
\ee
Eqn.~\eqref{eq:laplace_transformed_eq} can be solved in terms the Green's function $G(\omega;\sigma, \sigma')$ defined via
\bea
\bar{\rm A}(\bar \omega)[ G(\omega;\sigma, \sigma') ] =  \delta(\sigma-\sigma').
\eea
To construct the Green's function, we first restrict ourselves in the inverse Laplace transformation's domain of convergence ${\rm Im}(\omega)>0$, and consider two linearly independent solutions of the homogenous equation $\bar{\rm A}( \omega)[\phi_i] = 0$ ($i=0,1$). In this region, the homogenous solution $\phi_0$ and $\phi_1$ are regular, respectively, at the boundaries $\sigma = 0$ and $\sigma=1$, corresponding therefore to hyperboloidal representation of the ``up'' and ``in'' radial solutions.

Leveraging Abel identity, their corresponding Wronskian reads
\bea
W(\omega; \sigma)= W_o(\omega)\dfrac{e^{2\i \lambda \omega H(\sigma)}}{p(\sigma)}.
\eea
The explicit form of $\sigma$-independent term $W_o(\omega)$ is not relevant for our arguments. It suffices to recall that, at the QNM frequencies, $W_o(\omega_n) = 0$, yielding 
\be
\label{eq:in_at_QNM}
\phi_1(\omega_n; \sigma) = A_+(\omega_n) \phi_0(\omega_n; \sigma).
\ee

From the Wronskian, we obtain the Green's function
\bea
G(\omega;\sigma, \sigma') = \frac{\phi_0(\sigma_<) \phi_1(\sigma_>)}{p(\sigma') W(\omega; \sigma')}
\eea
where $\sigma_<={\rm min}(\sigma,\sigma')$ and $\sigma_>={\rm max}(\sigma,\sigma')$. 

The inhomogeneous solution is then given by
\be
\tilde{\phi}(\omega;\sigma) = \int_0^1 \d\sigma' \left[ \frac{\phi_0(\omega;\sigma_<) \, \phi_1(\omega;\sigma_>)}{ p(\sigma') W(\omega;\sigma')} \mathcal{I}(\omega;\sigma') \right],
\ee
which can be substituted back into Eqn.~\eqref{eq:inverse_Laplace}, to obtain the full time evolution $\psi(\tau,\sigma)$.

To single out the QNMs' contribution in the solution, we close the $\omega$ contour in Eqn.~\eqref{eq:inverse_Laplace} in the lower half plane. 
For an observer at $\mathscr{I}^+$, $\sigma = 0$, thus we consider $\sigma < \sigma'$ and approximate the QNM contribution by
\bea
\psi(\tau,\sigma) \simeq \sum_n e^{-i \lambda \omega_n \tau} c_n \phi_n(\sigma),
\eea
where we have defined $\phi_n(\sigma) = \phi_0(\omega_n;\sigma)$ as QNM function. The excitation coefficients then follow as
\bea\label{eq:exc_coeff_hyper}
c_n = -i \lambda A_+(\omega_n) \int_0^1 \d\sigma' \left[ \frac{\phi_n(\sigma') \mathcal{I}(\omega_n;\sigma')}{ p(\sigma')\partial_\omega W(\omega;\sigma')|_{\omega=\omega_n}} \right].
\eea
The Wronskian's $\omega$-derivative reads, at a QNM frequency,
\bea
\label{eq:Wronskian_omega_der}
\p_\omega W(\omega; \sigma')|_{\omega=\omega_n}&
=&  \p_\omega W_o(\omega)|_{\omega=\omega_n}\dfrac{e^{2\i \lambda \omega_n H(\sigma')}}{p(\sigma')}.
\eea
This allows us to separate, in the excitation coefficients, an intrinsic excitation factor and an overlap integral dependent on the initial data 
\bea\label{eq:exc_coeff_hyper_factors}
c_n = b_n I_n \, ,
\eea
with
\bea\label{eq:exc_factors_hyper}
b_n &:=&  \frac{i\, \lambda^3 A_+(\omega_n)}{\rh^2 \p_\omega W_o(\omega)|_{\omega=\omega_n}}, \\
\label{eq:overlap_hyper}
I_n &:=& - \dfrac{\rh^2}{\lambda^2} \int_{\cal C} \d\sigma' e^{-2\i \lambda \omega_n H(\sigma')} \phi_n(\sigma') \mathcal{I}(\omega_n;\sigma'),
\eea
where we introduced a factor of ${\rh^2}/{\lambda^2}$ in both to simplify the correspondence with the orthogonality product. For the same reason, we have also adapted the integration domain $\sigma'\in[0,1]$ to a generic contour ${\cal C}$, so that we can employ the regularisation schemes from the previous sections.

One can easily verify that the overlap integral Eqn.~\eqref{eq:overlap_hyper} is the same as the one given by the symplectic product between the QNM and the initial data. This only requires integrating by parts, discarding  boundary terms of the form $\propto\gamma(\sigma) e^{-2 i \lambda \omega_n H(\sigma) }\phi_n(\sigma) \phi_{\rm ID}(\sigma) $ that vanish under regularisation schemes discussed in the previous sections. Specifically, we find 
\bea\label{eq:exc_coeff_hyperproof}
I_n = \left.\bigg( \langle \Psi, \Psi_n \rangle -\left.\dfrac{\rh^2}{\lambda^2} \gamma\,  \Psi \left( {\cal J}\Psi_n\right) \right|_{\p {\cal C}}\bigg)\right|_{\tau=0}.
\eea

A comparison between Eqs.~\eqref{eq:exc_coeff_hyper_factors} and \eqref{eq:exc_coeff_hyperproof} against Eq.~\eqref{eq:exc_coeff_product} indicates that we need to identify $b_n$ with the inverse QNM norm $\langle \Psi_n, \Psi_n \rangle$. 
Following~\cite{GHSS2023}, we consider the product $ \langle \Psi_1, \Psi_n \rangle$, starting with a generic frequency $\omega$ for ``in'' solution $\Psi_1$. we recall that Eqn.~\eqref{eq:S_prod_explicit} directly yields
\bea
\label{eq:0=0}
\mathcal{L}_\tau   \langle \Psi_1, \Psi_n \rangle = -i (\omega - \omega_n) \lambda \langle \Psi_1, \Psi_n \rangle  \,,
\eea
and therefore, we may recover $\langle \Psi_{n}, \Psi_n \rangle$ via
\bea
\label{eq:norm_psi_n}
\langle \Psi_{n}, \Psi_n \rangle = \lim_{\omega \to \omega_n} \dfrac{\mathcal{L}_\tau   \langle \Psi_{1}, \Psi_n \rangle}{-i \lambda (\omega - \omega_n)\, A_+(\omega_n)}
\eea
where we have used Eqn.~\eqref{eq:in_at_QNM} to express $\Psi_1|_{\omega=\omega_n}$.

While it is clear that the r.h.s of Eqn.~\eqref{eq:0=0}, or equivalently denominator of \eqref{eq:norm_psi_n}, vanishes at $\omega=\omega_n$, the respective l.h.s (numerator) directly relates to the fluxes across the boundaries, cf.~Eqn.~\eqref{eq:Flux_Simpletic} 
\bea
&&\mathcal{L}_\tau   \langle \Psi_{1}, \Psi_n \rangle=\dfrac{d}{d\tau} {\cal F}^{\p{\cal C}}_S[\Psi_{1}, {\cal J}\Psi_n] \nn\\
&&= \dfrac{\rh^2}{\lambda^2}e^{-i \lambda (\omega-\omega_{n})\tau} e^{-2 i \lambda \omega_{n}H(\sigma)} \times \nn \\
\label{eq:flux_in_QNM}
&&\times\Bigg[ i \lambda (\omega_{n}- \omega)\gamma(\sigma) \phi_{1}(\omega;\sigma) \phi_0(\omega_n;\sigma)   \\
&&+ p(\sigma) \bigg( \phi'_{1}(\omega;\sigma) \phi_0(\omega_n;\sigma) - \phi_{1}(\omega;\sigma) \phi'_0(\omega;\sigma)\bigg) \left.\Bigg]\right|_{\p {\cal C}}. \nn
\eea
One can verify that these fluxes vanish identically when $\omega = \omega_n$. This result is expected because, when $\omega = \omega_n$, the amount of information from the QNM $\Psi_n$ crossing into ${\cal H}^+$ and $\scri^+$ is exactly the same as the one coming from ${\cal H}^-$ and $\scri^-$ from the anti-QNM ${\cal J}\Psi_n$. Thus, Eq.~\eqref{eq:norm_psi_n} reduces to a $0/0$ limit, and its evaluation follows by using l'Hopital rule with derivatives applied to the limiting variable $\omega$.

The denominator contribution follows from the right-hand side of Eqn.~\eqref{eq:0=0} via
\bea
\left.\dfrac{\d}{\d \omega}{\rm r.h.s.}\right|_{\omega_n}= - i \lambda A_+(\omega_n)\langle \Psi_n, \Psi_n  \rangle \, .
\eea
To perform the $\omega$-derivative of the numerator, one first notices the term proportional to $p(\sigma)$ in Eqn.~\eqref{eq:flux_in_QNM} relates to the Wronskian between the homogenous solutions $\phi_0$ and $\phi_1$. Hence,
\bea
\left.\dfrac{\d}{\d \omega}{\rm l.h.s.}\right|_{\omega_n}
=&& \dfrac{\rh^2}{\lambda^2}  e^{- 2 i \lambda \omega_n H(\sigma)}  
\bigg[ i \lambda \gamma(\sigma)    \phi_1(\omega; \sigma) \phi_0(\omega_n;\sigma) \nn \\
&&+ p(\sigma)  \frac{\d}{\d \omega}  W[\phi_0(\omega_n;\sigma), \phi_1(\omega;\sigma)]\left. \bigg] \right|_{\substack{\p{\cal C}, \\ \omega=\omega_n} }   \nn
\eea
A careful manipulation of the Wronskian's limit~\cite{GHSS2023} allows us to leverage Eqn.~\eqref{eq:Wronskian_omega_der} and reduce Eqn.~\eqref{eq:norm_psi_n} into
\be
\langle \Psi_{n}, \Psi_n \rangle =  -i \dfrac{\rh^2}{\lambda^3} \dfrac{\p_{\omega}W_0(\omega)|_{\omega=\omega_n}}{A_+(\omega_n)}  + \left.\dfrac{\rh^2}{\lambda^2}\gamma \Psi_n ({\cal J} \Psi_n)\right|_{\p{\cal C}}, 
\ee
and hence, identify from Eq.~\eqref{eq:exc_factors_hyper}
\be
\label{eq:exc_factor_prod}
b_n = \Bigg( \langle \Psi_{n}, \Psi_n \rangle - \left.\dfrac{\rh^2}{\lambda^2}\gamma \Psi_n ({\cal J} \Psi_n)\right|_{\p{\cal C}} \Bigg)^{-1}.
\ee
Employing the regularisation schemes in Eqs.~\eqref{eq:exc_coeff_hyperproof} and \eqref{eq:exc_factor_prod}, the boundary terms vanish, and we recover the correspondence between Eqs.~\eqref{eq:exc_coeff_product} and \eqref{eq:exc_coeff_hyper_factors}.

Thanks to the relation between the energy and symplectic currents, Eqn.~\eqref{eq:Pi_S_to_Pi_E}, it is easy to see that the QNM excitation coefficients can also be written in terms of the energy product as
\be\label{eq:exc_coeff_product_E}
c_n = \frac{ \langle {\Psi}{}_{\rm ID},\Psi_n \rangle_E}{\langle {\Psi}{}_n,\Psi_n \rangle_E} \, .
\ee
It is natural to define the excitation factors in terms of the energy product as
\bea \label{eq:excitationfactor_energyproduct}
b^{E}_n = \langle \Psi_n, \Psi_n \rangle_E^{-1} \, ,
\eea
such that $b^{E}_n =  \frac{2}{i \lambda \omega_n} b_n$. 

\subsubsection{Initial data regularisation schemes}

The comparison between the Green's function approach against the orthogonality projection led to boundary terms, vanishing under the regularisation scheme. 

Implementing the {\em complex contour} strategy is, therefore, straightforward once an explicit form of the initial data set $\{ {\phi}_{\rm ID}(\sigma), \dot{\phi}_{\rm ID}(\sigma) \}$ is available to be evaluated at complex values of $\sigma$.

The {\em semi-analytic integration} requires an assumption that the functions $\{ {\phi}_{\rm ID}(\sigma), \dot{\phi}_{\rm ID}(\sigma) \}$ are analytic in the interval $\sigma\in(0,1]$, thus allowing for a Taylor expansion around the horizon in the form
\bea
&&\phi_{\rm ID}(\sigma) = \sum_{q=0}^{\infty} b_q \, (1-\sigma)^{r_1+q}, \\
&&\dot \phi_{\rm ID}(\sigma)=\sum_{k=0}^{\infty} c_k \, (1-\sigma)^{r_3+k},
\eea
which allows us to rewrite the bilinear form $ \langle \Psi_{\rm ID},\Psi_{I} \rangle$ as sums of $\Gamma(a)$ and hypergeometric functions $U(a,b,z)$ in analogy to the description in Sec.~\ref{sec:semianal-int}.

\subsection{Limitation of extended orthogonality product}\label{sec:project_limitation}

While comparing the the QNM excitation coefficients obtained via Green's function technique and the orthogonality product, one is faced once again with boundary terms that only vanish when regularisation schemes are employed. Hence, a natural question is whether the extended product defined in Sec.~\ref{sec:extended_ortho_prod} may formally account for these boundary terms.

Although the extended orthogonality product~\eqref{eq:J_Prod_Flux} provides a formal structure under which the Hamiltonian operator is formally self-adjoint, the presence of flux terms in its definition limits its direct applicability to the calculation of QNM excitation amplitudes.
Specifically, QNM excitation amplitudes $c_{I}$ should result from projecting the solution $\Psi(\tau, \sigma, \theta, \varphi)$ to the initial value problem into a given QNM state $\Psi_I(\tau, \sigma, \theta, \varphi)$ via
\be\label{eq:exc_coeff_exteded_product}
c_I = \frac{ ( {\Psi},\Psi_I )}{( {\Psi}{}_I,\Psi_I )} \, .
\ee
By its definition, the extended product includes a bulk and flux contributions as
\bea
( \Psi,\Psi_I ) &=& \Pi[ \Psi(0,\sigma, \theta,\varphi) , \Psi_I(0,\sigma, \theta,\varphi) ] \\ 
&-&{\cal F}^{\p {\cal C}}[\Psi(\tau,\sigma, \theta,\varphi), \Psi_I(\tau,\sigma, \theta,\varphi)]\bigg|_{\tau=0}. \nn
\eea
The above expression brings an explicit dependence on the coordinate to emphasise that the bulk contribution can be taken at the initial data slice, and therefore be calculated solely from initial data set $\{ {\Psi}_{\rm ID}(\sigma, \theta, \varphi), \dot{\Psi}_{\rm ID}(\sigma, \theta, \varphi) \}$. The flux contributions, on the other hand, are notions defined globally in time, and therefore, they depend on the entire solution $\Psi(\tau, \sigma, \theta, \varphi)$. 

This property prevents a direct practical implementation, and further fundamental analysis are required to explore the extended product's applicability in {\em predicting} QNM excitation amplitudes solely from the initial data. Initial simplistic attempts based on series expansion for the solution proved ineffective in addressing these issues, and we believe the matter requires a more fundamental input from complex analysis to understand the relation between the physical fluxes and the singular terms.

\section{Results}

\begin{table}[t!]
\begin{center}
\begin{tabular}{|l|c|} 
 \hline
 $(\ell m n)$ & Frequency $M\omega_{\ell m n}$  \\ \hline 
 \hline
 000 & $0.1104549390801739 - 0.1048957170561837 i$ \\ \hline
 001 & $0.0861169182774843 - 0.3480524468204372 i$ \\ \hline
 002 & $0.0757419355330204 - 0.6010785901043246 i$ \\ \hline
 100 & $ 0.2929361332672827 - 0.0976599889135786 i$ \\ \hline
 101 & $ 0.2644486506048260- 0.3062573915590371 i$ \\ \hline
\end{tabular}
    \caption{QNM frequencies used in this work, extracted from the Python package \texttt{qnm} \cite{Stein:2019mop}. 
    }
    \label{tab:qnm_frequencies}
\end{center}
\end{table}

\subsection{Orthogonality product}
\begin{table*}[t!]
\begin{center}
\begin{tabular}{ p{1.5cm}|p{3.1cm}p{3.2cm}|p{2.8cm}p{3.2cm} }
\hline
Mode pair
& \multicolumn{2}{c|}{Symplectic current} 
& \multicolumn{2}{c}{Energy current} \\ \cline{2-5}
$I_1,I_2$
& Complex contour 
& Semi-analytic method 
& Complex contour 
& Semi-analytic method \\ 
\hline
   000, 000 &  $1.0$ & $1.0$ &  $1.0$ & $1.0$\\
   000, 001 &  $(1.2 - 0.6i)\times 10^{-6}$ &  $(11 - 3.8i)\times 10^{-6}$ &  $(-3.1 + 5.9i)\times 10^{-6}$ &  $-(6.3 - 11i) \times 10^{-6}$ \\
    001, 000 &  $-(1.2+2.6 i) \times 10^{-8}$ & $-(4.5 + 2.02i)\times 10^{-8}$ & $(0.7+2.4 i) \times 10^{-8}$  & $(2.7 + 1.9i)\times 10^{-8}$ \\
     001, 001 &  $1.0$ & $1.0$ &  $1.0$ & $1.0$\\
     000, 002 &  $(-7.3 + 8.3 i)\times 10^{-5}$ &  $(-1.4 + 0.32i)\times 10^{-4}$ &  $(-2.1 + 1.3 i)\times 10^{-4}$ &  $(-3.2 +2.9i) \times 10^{-4}$ \\
     002, 000 & $(-0.1+1.1 i) \times 10^{-6}$ &  $-(0.60 + 1.2i) \times 10^{-8} $ & $(-2.4+4.9 i) \times 10^{-7}$ &  $-(1.5+6.3i) \times 10^{-9}$ \\
      002, 002 &  $1.0$ & $1.0$ &  $1.0$ & $1.0$\\
       100, 100 &  $1.0$ & $1.0$  &  $1.0$ & $1.0$\\
       100, 101 & $(5.3 +8.5 i)\times10^{-12}$ &  $-(2.4 + 4.2i)\times10^{-12}$  &  $(2.2 + 6.7 i)\times10^{-12}$ & $-(3.5 + 2.6 i) \times 10^{-12}$  \\
     101, 100 &   $(1.6-3.1 i) \times 10^{-14}$ &   $(4.6 + 4.3i)\times 10^{-15}$ & $(-1.4+ 3.9 i) \times 10^{-14}$ & $( 3.8 + 5.3 i) \times10^{-15}$\\
      101, 101 &  $1.0$ &$1.0$ &  $1.0$ & $1.0$\\
      \hline
\end{tabular}
    \caption{Normalised orthogonal products between scalar QNMs for hyperboloidal foliations of Schwarzschild spacetime, with $I=\ell m n$. We report results obtained with both the orthogonal symplectic and the orthogonal energy products, evaluated with both the complex contour and the semi-analytic  methods. In all cases, the products are consistent with zero (i.e., the modes are orthogonal). However, the precision depends on the ordering of the two modes in question. Similar results can be obtained when applying the $\mathcal{J}$-symmetry to $\Psi_1(\bar{x}^\mu)$ and computing the products between two anti-modes according to Property 2.
   }
    \label{tab:results_ortho}
\end{center}
\end{table*}

We begin this section by explicitly stating the frequencies of the QNMs used to testing our results; these have been extracted from the Python package \texttt{qnm} \cite{Stein:2019mop} and are shown in Table \ref{tab:qnm_frequencies}. 

We then explicitly implement the orthogonality product \eqref{eq:S_prod_explicit} via the regularisation strategies devised in sec.~\eqref{sec:implementation}. In doing so, we carry out a comparison of the two approaches described in the previous subsection (namely the complex contour approach and the semi-analytic approach) and the two orthogonality products (namely the symplectic orthogonality product and the energy orthogonality product), with results summarised in Table \ref{tab:results_ortho}. Besides, we also discuss the practicality and limitations of the extended product \eqref{eq:J_Prod_Flux} for numerical integration direct on the real line.

We finish with a proof-of-principle calculation of QNMs excitation factor and subsequentially excitation amplitudes via projection of initial data in the QNM eigenfunctions, as summarised in Tables \ref{tab:exc_factors} and \ref{tab:results_exc_coeffs}.

\subsubsection{Analytic continuation approaches}

\begin{figure}[t]
\includegraphics[width=\columnwidth]{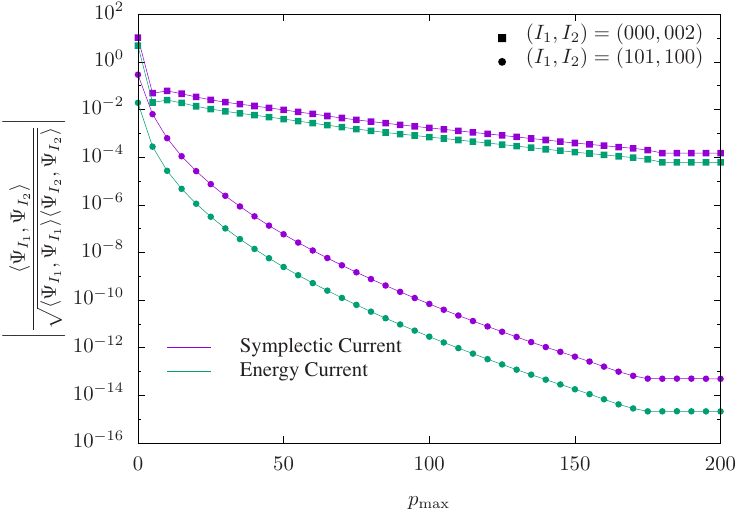}
  \caption{Convergence of the normalised orthogonality products computed within the semi-analytic regularisation scheme as a function of the truncation parameter $p_{\rm max}$. Results for two representative pairs of QNMs: $(I_1,I_2)=(000,002)$ (squares) and $(I_1,I_2)=(101,100)$ (circles), and for both the symplectic current (purple) and the energy current (green). In both cases, the curves saturate around $p_{\rm max}\sim175$, beyond which including further terms in the sum no longer improves the accuracy of the result. The convergence rate, equal for both currents, is neither purely exponential nor algebraic and depends on the specific pair of mode.}
\label{fig:QNMOrtho_SemiAnal_Convergence}
    \label{fig:QNMOrtho_SemiAnal_Convergence}
\end{figure}

For both regularisation schemes, we compute QNM eigenfunctions $\psi_{I}$ using Leaver's method, i.e. we consider the Frobenius expansion \eqref{eq:FrobeniusSeries} and select the solution to the indicial equation with $r_I=0$~\cite{Leaver:1985ax}. In particular, we iterate the 3-term recursion relation to a truncated value $p_{\rm max}$.

In the semi-analytic regularisation scheme, we consider $p_{\rm max}=400$ terms and the precision of $16$ digits. Without code optmisation, the computation time ranges from a few minutes for $p_{\rm max}=100$ to roughly half an hour for $p_{\rm max}=400$. Within the current numerical implementation, including additional terms in the sum significantly increases the computational cost without contributing to an enhancement in accuracy. Indeed, extending the calculation to $p_{\rm max}=700$ does not produce a noticeable change in the main results within the significant digits displayed here.

In particular, Fig.~\ref{fig:QNMOrtho_SemiAnal_Convergence} displays the convergence of the normalised product with respect to $p_{\rm max}$, where we observe a saturation around $p_{\rm max}\sim 175$. It also shows how different products converge with different rates (which is neither exponential nor algebraical). The overall convergence behaviour is the same for the symplectic and energy currents, but the efficiency depends significantly on the particular pair of modes. A rigorous estimate for the expected convergence behaviour according to the frequency values goes beyond the scope of this section.

When evaluating the products with the complex contour method, we include up $p_{\rm max}=700$ terms in the sum. The integration time is less than a minute for low-$n$ modes, but it increases quickly for modes with small real frequencies ($n\gtrsim 3$ in Schwarzschild), for which the integrand decays slowly and the integration contour must be extended further in the imaginary direction to achieve convergence.

In Table~\ref{tab:results_ortho}, we summarise our results for the (normalised) orthogonality of QNMs, finding 
\be
\label{eq:QNMortho}
\frac{\langle\Psi_{I_1},\Psi_{I_2} \rangle}{\sqrt{\langle\Psi_{I_1},\Psi_{I_1}\rangle   \langle\Psi_{I_2},\Psi_{I_2}\rangle }}=\delta_{12}\,.
\ee
Specifically, we see that the overlap of two distinct QNMs is vanishing within our machine roundoff error. 

When comparing the results for the two different conserved currents for a given integration strategy, we see that the energy current is slightly better behaved because of the extra derivative in the integrand (at infinity). It is also interesting to compare the two integration approaches.  The complex integration contour approach, which relies on the real part of the QNM frequency to damp the integrand, is less effective when the latter is small. We see the effect of this on higher overtone modes, for which orthogonality becomes less precise. Whether the numerical efficiency can be improved by following an optimal integration strategy, adapting the contour to a particular frequency, as in Ref. ~\cite{London26}, requires further analysis. The semi-analytic approach, relying on the integration via Tricomi hypergeometric functions, shows that the results for the two currents are comparable for the given number of terms in the sum. However, the energy current behaves better when considering higher overtones, for which more terms in the sums are required.

\subsubsection{Extended orthogonality product}
\label{sec:num_direct_integration}
\begin{figure}[t]
\includegraphics[width=\columnwidth]{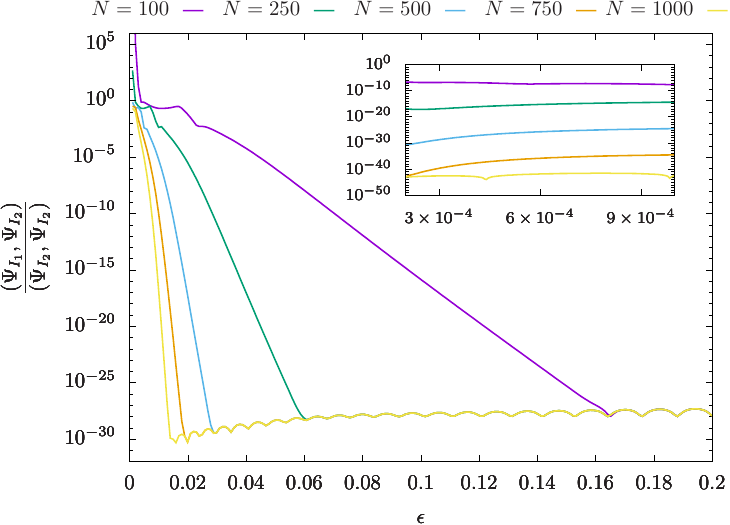}
  \caption{Numerical convergence of the normalised extended product $(\psi_{I_1},\psi_{I_2})/(\psi_{I_2},\psi_{I_2})$ as a function $\epsilon$ parametrising the boundary offset $\sigma\in[\epsilon, 1-\epsilon]$. Inside the bulk, orthogonality is precisely obtained. Even though the increase of numerical resolution $N$ allows extending the orthogonality towards $\epsilon \sim 0$. numerical roundoff error accumulate as $\epsilon \to 0$ in a failure to perform a ``large/large" limit. By adapting the integration scheme (multi-domain spectral methods + Analytical mesh refinement), we recover the orthogonlaty towards $\epsilon \to 0$. Results obtained with the sympletic current for $I_1=(\ell_1, m_1, n_1)=(0,0,0)$ and $I_2=(\ell_2, m_2, n_2)=(0,0,1)$}
    \label{fig:QNMOrtho_noAC}
\end{figure}

We integrate the extended product \eqref{eq:J_Prod_Flux} numerically for the QNM modes $I_1 = (\ell_1, m_1, n_1) = (0,0,0)$ and $I_2 = (\ell_2, m_2, n_2) = (0,0,1)$. Here, the eigenfunctions result from the eigenvalue problem \eqref{eq:eigenvalue_future}, with the operator $\overline{\rm L}$ discretized via a Chebyshev spectral method using a grid based on Chebyshev--Lobatto collocation points~\cite{PhysRevX.11.031003}. We employ a total of $N_{\rm QNM} = 150$ grid points for this calculation.

The radial integral in Eq.~\eqref{eq:J_Prod_Flux} is then performed over the interval $\sigma \in [\epsilon, 1-\epsilon]$. This interval is also discretised using Chebyshev--Lobatto collocation points, and we vary the total number of grid points $N$ in the numerical integration to study its convergence. We refer to this set of points as the $(\epsilon, N)$-grid. Our goal is twofold: (i) to confirm QNM orthogonality for any finite region $\epsilon \neq 0$ in the exterior black hole region, and (ii) to explore its validity as the asymptotic boundaries are included in the limit $\epsilon \to 0$.

For this purpose, the original QNM eigenfunctions obtained on $\sigma \in [0,1]$ with $N_{\rm QNM} = 150$ must be spectrally interpolated onto the new $(\epsilon, N)$-grid. Fig.~\ref{fig:QNMOrtho_noAC} displays the results for the normalized extended product
${(\Psi_{I_1},\Psi_{I_2})}/{(\Psi_{I_2},\Psi_{I_2})}$ as a function of the cutoff parameter $\epsilon$, for different resolution $N=100, 250, 500, 750$ and $1000$.

The orthogonality  is clearly recovered inside the bulk $\epsilon \gtrsim 0$, and increasing the numerical resolution allows us to recover it up to $\epsilon \sim 2 \times 10^{-2}$. As $\epsilon \to 0$, however, round off errors prevent an accurate cancellation of ``large/large'' division, and the achieving orthogonality requires a massive increase in the numerical resolution $N$. 

To ensure that the loss of convergence is a numerical artifact, and not a flaw in the analysis of the extended product, we employ a more sophisticated integration scheme. The $(\epsilon, N)$-grid is divided into two subdomains, $\sigma^{(0)} \in [\epsilon, 1/2]$ and $\sigma^{(1)} \in [1/2, 1-\epsilon]$, each with $N^{(0)} = N^{(1)} = N/2$ grid points. In this way, the strong gradients near the boundaries are isolated within each subdomain, where we employ analytical mesh refinement techniques~\cite{PanossoMacedo:2022fdi,Zhou:2025xta} with parameter $\kappa^{(0)} = \kappa^{(1)} = |\ln(\epsilon)|$ to increase the number of grid points around the problematic boundaries. The results are displayed in the inset of Fig.~\eqref{fig:QNMOrtho_noAC}, where we observe a very accurate orthogonality extending all the way towards $\epsilon\sim 10^{-4}$.

Note that Table ~\ref{tab:results_ortho} normalises the product with respect to the geometric mean between the modes $\psi_{I_1}$ and $\psi_{I_2}$. Here instead, the normalisation is in accordance with the projection of a generic function onto the mode $\psi_{I_2}$, cf.~\eqref{eq:exc_coeff_product}. This normalisation is preferable for the integration along the real line, because the behaviour near the boundaries is dominated by $e^{- 2 i \lambda \omega_{I_2} H(\sigma)}$. We have verified that our conclusion holds also for a real line integration normalised by the geometric mean. However, an accurate extension to the boundary $\epsilon \sim 0$ requires a prohibitive increase in the numerical resources.

\subsection{Excitation factors and amplitudes}
We finish this section with an explicit calculation of the QNMs excitation factors and amplitudes.

Table \ref{tab:exc_factors} presents the excitation factors $b_n$ and $b^E_n$ associated with the symplectic and energy products, cf.~Eqs.~\eqref{eq:exc_factor_prod} and \eqref{eq:excitationfactor_energyproduct}. Apart from confirming their relation via $b^{E}_n = \frac{2}{i \lambda \omega_n} b_n$, these results also make evident the role played by the regularisation schemes in obtaining a finite value for the norm $\langle \Psi_{I}, \Psi_{I}\rangle$. 

For the QNM excitation amplitudes, we consider a proof-of-principle initial data\footnote{Note that analytic continuation can be used for other choices of initial data that can be expanded as Taylor series.}
\be
\label{eq:ID}
\phi_{\rm ID}= 1, \quad \dot\phi_{\rm ID}= 0,
\ee
usually used to benchmark studies in the hyperboloidal literature. Apart from modelling an evolution stage in which the wave is already spread all the way from the black-hole horizon to future null infinity, the spectral representation of the solution to this initial data in terms of the QNMs function is known to converge \cite{Ansorg:2016ztf}.

Table \ref{tab:results_exc_coeffs} shows the excitation coefficients $c_{I}$ for the initial data \eqref{eq:ID}. We observe an internal consistency between the two regularisation scheme, and also across projections defined by both currents. Minor discrepancies between implementations, mostly for higher overtones, are associated to numerical truncation errors. Most importantly, these values coincide with those obtained for completely alternative methods in the hyperboloidal framework, for which no regularisation strategy is required, e.g.~\cite{Ansorg:2016ztf,Bourg:2025lpd,Besson:2024adi}.

Employing the extended orthogonality product to numerically compute QNM excitation factors and amplitudes requires a deeper understanding of its analytical foundations. Since the excitation factors are directly related to the QNM norm $\langle \Psi_{I}, \Psi_{I}\rangle$, this quantity diverges when evaluated via direct integration along the real line. Although the divergence rate can be used to counterbalance the numerical growth arising from the projection $\langle \Psi_{I_1}, \Psi_{I_2}\rangle$ (cf.~Fig.~\ref{fig:QNMOrtho_noAC}), we are currently unable to establish a direct numerical connection to $b_n$. 

Moreover, as discussed in Sec.~\ref{sec:project_limitation}, the flux terms appearing in the definition of the extended product require global knowledge of the time evolution. Consequently, projecting directly onto the initial data set is not feasible at the current stage of our understanding.

\begin{table*}[t!]
\begin{center}
\begin{tabular}{c|cc|cc}
\hline
mode 
& \multicolumn{2}{c|}{$b_n$} 
& \multicolumn{2}{c}{$b_n^E$} \\ \cline{2-5}
$(\ell m n)$
& Complex contour 
& Semi-analytic method 
& Complex contour 
& Semi-analytic method \\ 
\hline
  000 & $ 0.811 - 0.201 i$ &$0.8492 - 0.2370 i $&  $1.35 - 2.39 i$ &  $1.355 - 2.557 i$ \\
001 & $ -0.216 - 0.257 i$ & $-0.2541 - 0.2645  i$ &  $-0.378 - 0.275 i$ & $-0.4325 - 0.2730 i$ \\
002 &  $ 0.0726 + 0.1604 i$ & $0.09516 + 0.1741 i$ & $0.076 + 0.124i$ &$0.09588 + 0.1327i$ \\
100 & $ -0.6023 + 0.0909 i$ & $-0.6026 + 0.09120 i$ & $-0.169 + 0.972i$ & $-0.1685 + 0.9723 i$ \\
101 &  $0.115 + 0.755 i$ & $0.1158 + 0.7550 i$ &  $0.717 + 0.614 i$ & $0.7180 + 0.6125 i$ \\
 \hline
\end{tabular}
    \caption{Excitation factors obtained via the symplectic product \eqref{eq:exc_factor_prod} and energy product \eqref{eq:excitationfactor_energyproduct}.
    We have set $\lambda=2 \rh $. 
 }
    \label{tab:exc_factors}
\end{center}
\end{table*}

\begin{table*}[t!]
\begin{center}
\begin{tabular}{ p{0.45cm}|p{3.32cm}p{3.32cm}|p{3.32cm}p{3.32cm} }
\hline
mode 
& \multicolumn{2}{c|}{Symplectic current} 
& \multicolumn{2}{c}{Energy current} \\ \cline{2-5}
$(\ell m n)$
& Complex contour 
& Semi-analytic method 
& Complex contour 
& Semi-analytic method \\ 
\hline
  000 & $0.548761 + 0.396585 i$ &  $0.548761 + 0.396585 i$ &  $0.548761 + 0.396585 i$ &  $0.548761 + 0.396585 i$ \\
001 &  $-0.011943 + 0.079987 i$ &  $-0.011939 + 0.079984 i$ &  $-0.011943 + 0.079987 i$ & $-0.011941 +0.079984 i$ \\
002 &  $(8.8386 - 3.7824 i) \times 10^{-4}$ &  $(8.8594 - 3.7351 i) \times 10^{-4}$ &  $(8.8398 - 3.7347 i) \times 10^{-4}$ & $(8.8371 - 3.7416i) \times 10^{-4} $ \\
100 &  $0.447111 + 0.184345 i$ &  $0.447111 + 0.184345 i$ &  $0.447111 + 0.184345 i$ & $0.447111 + 0.184345 i$ \\
101 &  $0.0521156 + 0.0235427 i$ &  $0.0521156 + 0.0235427 i$ &  $0.0521156 + 0.0235427 i$ &  $0.0521156 + 0.0235427 i$ \\
 \hline
\end{tabular}
    \caption{Projection of constant initial data on scalar QNM for hyperboloidal foliations of Schwarzschild spacetime, divided by the normalisation constant $\langle \Psi_I, \Psi_I \rangle$. We report results obtained with the symplectic and the energy products, evaluated with both the complex contour and the semi-analytic methods. In all cases, the projection coefficients are consistent within $6$ digits. However, it must be noted that the excitation coefficient corresponding to the $\omega_{002}$ overtone being a zero of order $10^{-4}$ is more sensitive to the precision used.}
    \label{tab:results_exc_coeffs}
\end{center}
\end{table*}

\section{Discussion}\label{sec:discussion}
In this work, we have extended the formalism introduced in Ref.~\cite{GHSS2023} to compute QNM orthogonality products and excitation factors using \emph{conserved currents} within the hyperboloidal framework. The main motivation was twofold: (i) the hyperboloidal framework is known to geometrically address issues related to the divergence of QNM eigenfunctions toward asymptotic boundaries~\cite{Zenginoglu:2011jz, PanossoMacedo:2024nkw}, a feature present in the current orthogonality product formalism; and (ii) it provides a natural setting for conformal methods in black hole perturbation theory, paving the way for the systematic inclusion of asymptotic boundary terms in the definition of the product~\cite{GasperinJaramillo2022}.

In addressing point (i), we found that the divergence is an intrinsic structural feature of the formalism underlying the orthogonality product. In particular, the starting point for constructing the orthogonality property is the existence of a conserved current $J^a$, which naturally induces a bilinear relation between two solutions $\Psi_1$ and $\Psi_2$ of a wave-like equation. A key element in the theory is the use of the spacetime discrete $t$--$\phi$ symmetry, which acts on one of the solutions via the operator $\mathcal{J}$.

This work develops the $t$--$\phi$ symmetry within the hyperboloidal framework, and it implements the action of the ${\cal J}$ operator by identifying two sets of hyperboloidal hypersurfaces: a \emph{future hyperboloidal foliation} $\bar{x}^\mu$ and a \emph{past hyperboloidal foliation} $\check{x}^a$. While the former provides a natural parametrisation for functions satisfying purely outgoing boundary conditions at the horizon and null infinity, the latter is adapted to radiative problems with purely ingoing boundary conditions in the asymptotic regions. Interpreting the ${\cal J}$ operator from this perspective clarifies the structure of the orthogonality product: QNM projections pair a given QNM function with a corresponding anti-QNM state. As a consequence, divergent contributions in the calculation become unavoidable. In particular, evaluating an anti-QNM at the black hole horizon and future null infinity—due to its representation in terms of future hyperboloidal coordinates—leads to divergences.

Turning now to point (ii), we have initiated an exploratory route to extend the product beyond its original definition. Specifically, we have carefully kept track of boundary terms that naturally arise when manipulating the definition and properties of the orthogonality product. These boundary terms are directly related to the fluxes associated with the conserved current $J^a$. As previously suggested~\cite{GasperinJaramillo2022}, incorporating such fluxes into the product definition allows us to demonstrate that the Hamiltonian operator is formally normal under these bilinear relations\footnote{Recall that this notion is understood in a purely algebraic sense and does not rely on assumptions over the operator's domain, required for a more rigorous functional-analytic definition.}. However, this strategy presents challenges at the implementation level, and further work is required to develop it fully.

We employed two currents to construct the orthogonality relation: a symplectic current, conserved by construction in terms of the wave operator, and an energy current, conserved due to the existence of a timelike Killing vector. While the former was the natural choice in previous works~\cite{GHSS2023}, the latter has played a central role in the introduction of the pseudospectrum into gravitational physics~\cite{PhysRevX.11.031003,GasperinJaramillo2022}. 

The two currents emphasise complementary aspects of the field: the symplectic product reflects the intrinsic algebraic structure of QNMs, while the energy product encodes the physical energy content along hyperboloidal slices. Building on the original formalism~\cite{GHSS2023}, this work explicitly establishes the equivalence between the two products, thereby bridging a gap in the literature.

Specifically, we applied the formalism to a scalar field on a Schwarzschild background. This setting already contains all the essential conceptual ingredients, and extending the framework to the Teukolsky equation in Kerr spacetime is therefore primarily a technical, rather than conceptual, step. This would involve, for instance, including the azimuthal component $t-\phi$ transformation associated with the ${\cal J}$ operator, as well as adapting it to the spin-weighted master functions appearing the Teukolsky formalism~\cite{GHSS2023}.

Regarding the \emph{implementation}, we employed two complementary strategies to regularise the divergences~\cite{GHSS2023, London26}. In a \emph{semi-analytic approach}, the product is first solved in a region of the complex frequency domain where the integral is well-defined. The result is then analytically extended into all frequencies. In particular, the QNM radial equations are represented via a Frobenius expansion around the horizon. Substituting the series into the bilinear forms leads to integrals expressed in terms of \emph{Tricomi hypergeometric functions} $U(a,b,z)$. This approach is robust for higher overtones and systematically accommodates differences in behaviour between symplectic and energy products.

Alternatively, a \emph{complex contour integration method} extends the radial coordinate into the complex plane, deforming the integration contour around branch cuts from the horizon while approaching complex infinity. Damping along the contour ensures convergence of the integrand without relying on analyticity assumptions required when performing series expansions. This method is particularly efficient for low-overtone modes. 

At an investigative level, we have observed that leading-order divergences in bulk integrals at the boundaries of the physical spatial domain, along a constant-time hypersurface, cancel out exactly with the corresponding fluxes. Moreover, the fluxes vanish altogether when employing regularisation schemes based on analytic continuation. Clarifying the connection between bulk and flux contributions through complex contour methods remains an interesting direction for future work. Yet, by including the fluxes in the definitions of the the bi-linear products, we found strong evidence for the orthogonality relations by a direct integration over the real line, once the numerics are adjusted to cope with the cancellation of large numbers near the domain boundaries.

One of the key applications of the formalism is the calculation of QNM excitation factors and, most importantly, excitation amplitudes directly from the initial data. Building on Ref.~\cite{GHSS2023}, we have explicitly established the connection between orthogonality products and Green function techniques for computing QNM factors and excitation amplitudes within the hyperboloidal framework. While our numerical results show consistency and robustness when employing regularisation schemes for the explicit calculation of excitation factors and amplitudes, further studies are required to fully understand the role played by boundary terms and fluxes at a more fundamental level.

The incorporation of fluxes in the definition of the orthogonality product plays a dual role: on the one hand, it allows for a proper control of the properties of the Hamiltonian operator; on the other hand, it requires global-in-time knowledge of the fields. This feature prevents a straightforward formulation in terms of the initial data set, as the fluxes also depend on the corresponding time evolution. 

The next steps in our research program include extending the present analysis to rotating (Kerr) black holes and exploring a broader class of perturbations, including scalar, vector, and tensor modes. At a more fundamental level, a tension remains between the proposed formalism and alternative hyperboloidal approaches that achieve QNM projections without encountering divergences or requiring regularization schemes~\cite{Ansorg:2016ztf,Besson:2024adi,Bourg:2025lpd}. 

Further investigation is therefore needed to fully understand the interplay between physical fluxes, regularisation methods, and Green’s function techniques, as well as a coherent description across all QNM bi-orthogonality products discussed in the literature~\cite{Ansorg:2016ztf,GHSS2023,Besson:2024adi,Motohashi:2024fwt,Bourg:2025lpd,London26}.

Particularly promising settings include asymptotically de Sitter and anti-de Sitter spacetimes, which introduce additional elements for understanding the impact of boundary conditions and cosmological constants on perturbation dynamics.

\begin{acknowledgments}
The authors gratefully acknowledge S. R.~Green, J.~Lestingi, L. T.~London, and S. A.~Maenaut for helpful discussions. 
M. M. would like to thank The University of Texas at Austin for their hospitality during the final stages of this work.
M. M. is supported by the European Union’s Horizon Europe 2024 research and innovation programme under the Marie Sk{\l}odowska-Curie grant agreement No. 101154525. 
R.P.M. acknowledges support from the Villum Investigator program supported by the VILLUM Foundation (grant no.\ VIL37766) and the DNRF Chair program (grant no.\ DNRF162) by the Danish National Research Foundation. The Center of Gravity is a Center of Excellence funded by the Danish National Research Foundation under grant No. DNRF184. 
L. S. acknowledges support from the UKRI Horizon guarantee funding (project no. EP/Y023706/1). L. S. is also supported by a University of Nottingham Anne McLaren Fellowship. 
C.P.\ acknowledges support from a Royal Society -- Science Foundation Ireland University Research Fellowship via grant URF/R1/211027.  
\end{acknowledgments}

\bibliography{bibitems.bib}

\begin{thebibliography}{54}%
\makeatletter
\providecommand \@ifxundefined [1]{%
 \@ifx{#1\undefined}
}%
\providecommand \@ifnum [1]{%
 \ifnum #1\expandafter \@firstoftwo
 \else \expandafter \@secondoftwo
 \fi
}%
\providecommand \@ifx [1]{%
 \ifx #1\expandafter \@firstoftwo
 \else \expandafter \@secondoftwo
 \fi
}%
\providecommand \natexlab [1]{#1}%
\providecommand \enquote  [1]{``#1''}%
\providecommand \bibnamefont  [1]{#1}%
\providecommand \bibfnamefont [1]{#1}%
\providecommand \citenamefont [1]{#1}%
\providecommand \href@noop [0]{\@secondoftwo}%
\providecommand \href [0]{\begingroup \@sanitize@url \@href}%
\providecommand \@href[1]{\@@startlink{#1}\@@href}%
\providecommand \@@href[1]{\endgroup#1\@@endlink}%
\providecommand \@sanitize@url [0]{\catcode `\\12\catcode `\$12\catcode
  `\&12\catcode `\#12\catcode `\^12\catcode `\_12\catcode `\%12\relax}%
\providecommand \@@startlink[1]{}%
\providecommand \@@endlink[0]{}%
\providecommand \url  [0]{\begingroup\@sanitize@url \@url }%
\providecommand \@url [1]{\endgroup\@href {#1}{\urlprefix }}%
\providecommand \urlprefix  [0]{URL }%
\providecommand \Eprint [0]{\href }%
\providecommand \doibase [0]{https://doi.org/}%
\providecommand \selectlanguage [0]{\@gobble}%
\providecommand \bibinfo  [0]{\@secondoftwo}%
\providecommand \bibfield  [0]{\@secondoftwo}%
\providecommand \translation [1]{[#1]}%
\providecommand \BibitemOpen [0]{}%
\providecommand \bibitemStop [0]{}%
\providecommand \bibitemNoStop [0]{.\EOS\space}%
\providecommand \EOS [0]{\spacefactor3000\relax}%
\providecommand \BibitemShut  [1]{\csname bibitem#1\endcsname}%
\let\auto@bib@innerbib\@empty
\bibitem [{\citenamefont {Abedi}\ \emph {et~al.}(2025)\citenamefont {Abedi}
  \emph {et~al.}}]{Berti:2025hly}%
  \BibitemOpen
  \bibfield  {author} {\bibinfo {author} {\bibfnamefont {J.}~\bibnamefont
  {Abedi}} \emph {et~al.},\ }\href@noop {} {\bibinfo {title} {{Black hole
  spectroscopy: from theory to experiment}}} (\bibinfo {year} {2025}),\ \Eprint
  {https://arxiv.org/abs/2505.23895} {arXiv:2505.23895 [gr-qc]} \BibitemShut
  {NoStop}%
\bibitem [{\citenamefont {Abac}\ and\ \citenamefont {et~al}(2025)}]{LIGO2025}%
  \BibitemOpen
  \bibfield  {author} {\bibinfo {author} {\bibfnamefont {A.~G.}\ \bibnamefont
  {Abac}}\ and\ \bibinfo {author} {\bibnamefont {et~al}} (\bibinfo
  {collaboration} {LIGO Scientific, Virgo, and KAGRA Collaborations}),\ }\href
  {https://doi.org/10.1103/kw5g-d732} {\bibfield  {journal} {\bibinfo
  {journal} {Phys. Rev. Lett.}\ }\textbf {\bibinfo {volume} {135}},\ \bibinfo
  {pages} {111403} (\bibinfo {year} {2025})}\BibitemShut {NoStop}%
\bibitem [{\citenamefont {Abac}\ \emph {et~al.}(2026)\citenamefont {Abac} \emph
  {et~al.}}]{LIGOScientific:2026wpt}%
  \BibitemOpen
  \bibfield  {author} {\bibinfo {author} {\bibfnamefont {A.~G.}\ \bibnamefont
  {Abac}} \emph {et~al.} (\bibinfo {collaboration} {LIGO Scientific, VIRGO,
  KAGRA}),\ }\href@noop {} {\  (\bibinfo {year} {2026})},\ \Eprint
  {https://arxiv.org/abs/2603.19021} {arXiv:2603.19021 [gr-qc]} \BibitemShut
  {NoStop}%
\bibitem [{\citenamefont {Berti}\ \emph {et~al.}(2016)\citenamefont {Berti},
  \citenamefont {Sesana}, \citenamefont {Barausse}, \citenamefont {Cardoso},\
  and\ \citenamefont {Belczynski}}]{Berti:2016lat}%
  \BibitemOpen
  \bibfield  {author} {\bibinfo {author} {\bibfnamefont {E.}~\bibnamefont
  {Berti}}, \bibinfo {author} {\bibfnamefont {A.}~\bibnamefont {Sesana}},
  \bibinfo {author} {\bibfnamefont {E.}~\bibnamefont {Barausse}}, \bibinfo
  {author} {\bibfnamefont {V.}~\bibnamefont {Cardoso}},\ and\ \bibinfo {author}
  {\bibfnamefont {K.}~\bibnamefont {Belczynski}},\ }\href
  {https://doi.org/10.1103/PhysRevLett.117.101102} {\bibfield  {journal}
  {\bibinfo  {journal} {Phys. Rev. Lett.}\ }\textbf {\bibinfo {volume} {117}},\
  \bibinfo {pages} {101102} (\bibinfo {year} {2016})},\ \Eprint
  {https://arxiv.org/abs/1605.09286} {arXiv:1605.09286 [gr-qc]} \BibitemShut
  {NoStop}%
\bibitem [{\citenamefont {Maggiore}\ \emph {et~al.}(2020)\citenamefont
  {Maggiore} \emph {et~al.}}]{Maggiore:2019uih}%
  \BibitemOpen
  \bibfield  {author} {\bibinfo {author} {\bibfnamefont {M.}~\bibnamefont
  {Maggiore}} \emph {et~al.},\ }\href
  {https://doi.org/10.1088/1475-7516/2020/03/050} {\bibfield  {journal}
  {\bibinfo  {journal} {JCAP}\ }\textbf {\bibinfo {volume} {03}},\ \bibinfo
  {pages} {050}},\ \Eprint {https://arxiv.org/abs/1912.02622} {arXiv:1912.02622
  [astro-ph.CO]} \BibitemShut {NoStop}%
\bibitem [{\citenamefont {Cabero}\ \emph {et~al.}(2020)\citenamefont {Cabero},
  \citenamefont {Westerweck}, \citenamefont {Capano}, \citenamefont {Kumar},
  \citenamefont {Nielsen},\ and\ \citenamefont {Krishnan}}]{Cabero:2019zyt}%
  \BibitemOpen
  \bibfield  {author} {\bibinfo {author} {\bibfnamefont {M.}~\bibnamefont
  {Cabero}}, \bibinfo {author} {\bibfnamefont {J.}~\bibnamefont {Westerweck}},
  \bibinfo {author} {\bibfnamefont {C.~D.}\ \bibnamefont {Capano}}, \bibinfo
  {author} {\bibfnamefont {S.}~\bibnamefont {Kumar}}, \bibinfo {author}
  {\bibfnamefont {A.~B.}\ \bibnamefont {Nielsen}},\ and\ \bibinfo {author}
  {\bibfnamefont {B.}~\bibnamefont {Krishnan}},\ }\href
  {https://doi.org/10.1103/PhysRevD.101.064044} {\bibfield  {journal} {\bibinfo
   {journal} {Phys. Rev. D}\ }\textbf {\bibinfo {volume} {101}},\ \bibinfo
  {pages} {064044} (\bibinfo {year} {2020})},\ \Eprint
  {https://arxiv.org/abs/1911.01361} {arXiv:1911.01361 [gr-qc]} \BibitemShut
  {NoStop}%
\bibitem [{\citenamefont {Toubiana}\ \emph {et~al.}(2024)\citenamefont
  {Toubiana}, \citenamefont {Pompili}, \citenamefont {Buonanno}, \citenamefont
  {Gair},\ and\ \citenamefont {Katz}}]{Toubiana:2023cwr}%
  \BibitemOpen
  \bibfield  {author} {\bibinfo {author} {\bibfnamefont {A.}~\bibnamefont
  {Toubiana}}, \bibinfo {author} {\bibfnamefont {L.}~\bibnamefont {Pompili}},
  \bibinfo {author} {\bibfnamefont {A.}~\bibnamefont {Buonanno}}, \bibinfo
  {author} {\bibfnamefont {J.~R.}\ \bibnamefont {Gair}},\ and\ \bibinfo
  {author} {\bibfnamefont {M.~L.}\ \bibnamefont {Katz}},\ }\href
  {https://doi.org/10.1103/PhysRevD.109.104019} {\bibfield  {journal} {\bibinfo
   {journal} {Phys. Rev. D}\ }\textbf {\bibinfo {volume} {109}},\ \bibinfo
  {pages} {104019} (\bibinfo {year} {2024})},\ \Eprint
  {https://arxiv.org/abs/2307.15086} {arXiv:2307.15086 [gr-qc]} \BibitemShut
  {NoStop}%
\bibitem [{\citenamefont {Leaver}(1986)}]{Leaver1986}%
  \BibitemOpen
  \bibfield  {author} {\bibinfo {author} {\bibfnamefont {E.~W.}\ \bibnamefont
  {Leaver}},\ }\href {https://doi.org/10.1103/PhysRevD.34.384} {\bibfield
  {journal} {\bibinfo  {journal} {Phys. Rev. D}\ }\textbf {\bibinfo {volume}
  {34}},\ \bibinfo {pages} {384} (\bibinfo {year} {1986})}\BibitemShut
  {NoStop}%
\bibitem [{\citenamefont {Price}(1972)}]{Price:1971fb}%
  \BibitemOpen
  \bibfield  {author} {\bibinfo {author} {\bibfnamefont {R.~H.}\ \bibnamefont
  {Price}},\ }\href {https://doi.org/10.1103/PhysRevD.5.2419} {\bibfield
  {journal} {\bibinfo  {journal} {Phys. Rev. D}\ }\textbf {\bibinfo {volume}
  {5}},\ \bibinfo {pages} {2419} (\bibinfo {year} {1972})}\BibitemShut
  {NoStop}%
\bibitem [{\citenamefont {Gundlach}\ \emph {et~al.}(1994)\citenamefont
  {Gundlach}, \citenamefont {Price},\ and\ \citenamefont
  {Pullin}}]{Gundlach:1993tp}%
  \BibitemOpen
  \bibfield  {author} {\bibinfo {author} {\bibfnamefont {C.}~\bibnamefont
  {Gundlach}}, \bibinfo {author} {\bibfnamefont {R.~H.}\ \bibnamefont
  {Price}},\ and\ \bibinfo {author} {\bibfnamefont {J.}~\bibnamefont
  {Pullin}},\ }\href {https://doi.org/10.1103/PhysRevD.49.883} {\bibfield
  {journal} {\bibinfo  {journal} {Phys. Rev. D}\ }\textbf {\bibinfo {volume}
  {49}},\ \bibinfo {pages} {883} (\bibinfo {year} {1994})},\ \Eprint
  {https://arxiv.org/abs/gr-qc/9307009} {arXiv:gr-qc/9307009} \BibitemShut
  {NoStop}%
\bibitem [{\citenamefont {De~Amicis}\ \emph {et~al.}(2024)\citenamefont
  {De~Amicis}, \citenamefont {Albanesi},\ and\ \citenamefont
  {Carullo}}]{DeAmicis:2024not}%
  \BibitemOpen
  \bibfield  {author} {\bibinfo {author} {\bibfnamefont {M.}~\bibnamefont
  {De~Amicis}}, \bibinfo {author} {\bibfnamefont {S.}~\bibnamefont
  {Albanesi}},\ and\ \bibinfo {author} {\bibfnamefont {G.}~\bibnamefont
  {Carullo}},\ }\href {https://doi.org/10.1103/PhysRevD.110.104005} {\bibfield
  {journal} {\bibinfo  {journal} {Phys. Rev. D}\ }\textbf {\bibinfo {volume}
  {110}},\ \bibinfo {pages} {104005} (\bibinfo {year} {2024})},\ \Eprint
  {https://arxiv.org/abs/2406.17018} {arXiv:2406.17018 [gr-qc]} \BibitemShut
  {NoStop}%
\bibitem [{\citenamefont {Arnaudo}\ \emph
  {et~al.}(2025{\natexlab{a}})\citenamefont {Arnaudo}, \citenamefont
  {Carballo},\ and\ \citenamefont {Withers}}]{Arnaudo:2025uos}%
  \BibitemOpen
  \bibfield  {author} {\bibinfo {author} {\bibfnamefont {P.}~\bibnamefont
  {Arnaudo}}, \bibinfo {author} {\bibfnamefont {J.}~\bibnamefont {Carballo}},\
  and\ \bibinfo {author} {\bibfnamefont {B.}~\bibnamefont {Withers}},\
  }\href@noop {} {\  (\bibinfo {year} {2025}{\natexlab{a}})},\ \Eprint
  {https://arxiv.org/abs/2510.18956} {arXiv:2510.18956 [gr-qc]} \BibitemShut
  {NoStop}%
\bibitem [{\citenamefont {Arnaudo}\ and\ \citenamefont
  {Withers}(2025)}]{Arnaudo:2025kit}%
  \BibitemOpen
  \bibfield  {author} {\bibinfo {author} {\bibfnamefont {P.}~\bibnamefont
  {Arnaudo}}\ and\ \bibinfo {author} {\bibfnamefont {B.}~\bibnamefont
  {Withers}},\ }\href@noop {} {\  (\bibinfo {year} {2025})},\ \Eprint
  {https://arxiv.org/abs/2511.17703} {arXiv:2511.17703 [gr-qc]} \BibitemShut
  {NoStop}%
\bibitem [{\citenamefont {De~Amicis}\ \emph {et~al.}(2026)\citenamefont
  {De~Amicis}, \citenamefont {Cannizzaro}, \citenamefont {Carullo},\ and\
  \citenamefont {Sberna}}]{DeAmicis:2025xuh}%
  \BibitemOpen
  \bibfield  {author} {\bibinfo {author} {\bibfnamefont {M.}~\bibnamefont
  {De~Amicis}}, \bibinfo {author} {\bibfnamefont {E.}~\bibnamefont
  {Cannizzaro}}, \bibinfo {author} {\bibfnamefont {G.}~\bibnamefont
  {Carullo}},\ and\ \bibinfo {author} {\bibfnamefont {L.}~\bibnamefont
  {Sberna}},\ }\href {https://doi.org/10.1103/bgjv-lvyd} {\bibfield  {journal}
  {\bibinfo  {journal} {Phys. Rev. D}\ }\textbf {\bibinfo {volume} {113}},\
  \bibinfo {pages} {024048} (\bibinfo {year} {2026})},\ \Eprint
  {https://arxiv.org/abs/2506.21668} {arXiv:2506.21668 [gr-qc]} \BibitemShut
  {NoStop}%
\bibitem [{\citenamefont {Su}\ \emph {et~al.}(2026)\citenamefont {Su},
  \citenamefont {Khera}, \citenamefont {Casals}, \citenamefont {Ma},
  \citenamefont {Chowdhuri},\ and\ \citenamefont {Yang}}]{Su:2026fvj}%
  \BibitemOpen
  \bibfield  {author} {\bibinfo {author} {\bibfnamefont {J.}~\bibnamefont
  {Su}}, \bibinfo {author} {\bibfnamefont {N.}~\bibnamefont {Khera}}, \bibinfo
  {author} {\bibfnamefont {M.}~\bibnamefont {Casals}}, \bibinfo {author}
  {\bibfnamefont {S.}~\bibnamefont {Ma}}, \bibinfo {author} {\bibfnamefont
  {A.}~\bibnamefont {Chowdhuri}},\ and\ \bibinfo {author} {\bibfnamefont
  {H.}~\bibnamefont {Yang}},\ }\href@noop {} {\  (\bibinfo {year} {2026})},\
  \Eprint {https://arxiv.org/abs/2601.22015} {arXiv:2601.22015 [gr-qc]}
  \BibitemShut {NoStop}%
\bibitem [{\citenamefont {Ma}(2026)}]{Ma:2026qbq}%
  \BibitemOpen
  \bibfield  {author} {\bibinfo {author} {\bibfnamefont {S.}~\bibnamefont
  {Ma}},\ }\href@noop {} {\  (\bibinfo {year} {2026})},\ \Eprint
  {https://arxiv.org/abs/2604.08680} {arXiv:2604.08680 [gr-qc]} \BibitemShut
  {NoStop}%
\bibitem [{\citenamefont {Ching}\ \emph {et~al.}(1995)\citenamefont {Ching},
  \citenamefont {Leung}, \citenamefont {Suen},\ and\ \citenamefont
  {Young}}]{Ching:1993gt}%
  \BibitemOpen
  \bibfield  {author} {\bibinfo {author} {\bibfnamefont {E.~S.~C.}\
  \bibnamefont {Ching}}, \bibinfo {author} {\bibfnamefont {P.~T.}\ \bibnamefont
  {Leung}}, \bibinfo {author} {\bibfnamefont {W.~M.}\ \bibnamefont {Suen}},\
  and\ \bibinfo {author} {\bibfnamefont {K.}~\bibnamefont {Young}},\ }\href
  {https://doi.org/10.1103/PhysRevLett.74.4588} {\bibfield  {journal} {\bibinfo
   {journal} {Phys. Rev. Lett.}\ }\textbf {\bibinfo {volume} {74}},\ \bibinfo
  {pages} {4588} (\bibinfo {year} {1995})},\ \Eprint
  {https://arxiv.org/abs/gr-qc/9408043} {arXiv:gr-qc/9408043} \BibitemShut
  {NoStop}%
\bibitem [{\citenamefont {Ching}\ \emph {et~al.}(1996)\citenamefont {Ching},
  \citenamefont {Leung}, \citenamefont {Suen},\ and\ \citenamefont
  {Young}}]{Ching:1995rt}%
  \BibitemOpen
  \bibfield  {author} {\bibinfo {author} {\bibfnamefont {E.~S.~C.}\
  \bibnamefont {Ching}}, \bibinfo {author} {\bibfnamefont {P.~T.}\ \bibnamefont
  {Leung}}, \bibinfo {author} {\bibfnamefont {W.~M.}\ \bibnamefont {Suen}},\
  and\ \bibinfo {author} {\bibfnamefont {K.}~\bibnamefont {Young}},\ }\href
  {https://doi.org/10.1103/PhysRevD.54.3778} {\bibfield  {journal} {\bibinfo
  {journal} {Phys. Rev. D}\ }\textbf {\bibinfo {volume} {54}},\ \bibinfo
  {pages} {3778} (\bibinfo {year} {1996})},\ \Eprint
  {https://arxiv.org/abs/gr-qc/9507034} {arXiv:gr-qc/9507034} \BibitemShut
  {NoStop}%
\bibitem [{\citenamefont {Green}\ \emph {et~al.}(2023)\citenamefont {Green},
  \citenamefont {Hollands}, \citenamefont {Sberna}, \citenamefont {Toomani},\
  and\ \citenamefont {Zimmerman}}]{GHSS2023}%
  \BibitemOpen
  \bibfield  {author} {\bibinfo {author} {\bibfnamefont {S.~R.}\ \bibnamefont
  {Green}}, \bibinfo {author} {\bibfnamefont {S.}~\bibnamefont {Hollands}},
  \bibinfo {author} {\bibfnamefont {L.}~\bibnamefont {Sberna}}, \bibinfo
  {author} {\bibfnamefont {V.}~\bibnamefont {Toomani}},\ and\ \bibinfo {author}
  {\bibfnamefont {P.}~\bibnamefont {Zimmerman}},\ }\href
  {https://doi.org/10.1103/PhysRevD.107.064030} {\bibfield  {journal} {\bibinfo
   {journal} {Phys. Rev. D}\ }\textbf {\bibinfo {volume} {107}},\ \bibinfo
  {pages} {064030} (\bibinfo {year} {2023})}\BibitemShut {NoStop}%
\bibitem [{\citenamefont {London}(2023)}]{London:2020uva}%
  \BibitemOpen
  \bibfield  {author} {\bibinfo {author} {\bibfnamefont {L.~T.}\ \bibnamefont
  {London}},\ }\href {https://doi.org/10.1103/PhysRevD.107.044056} {\bibfield
  {journal} {\bibinfo  {journal} {Phys. Rev. D}\ }\textbf {\bibinfo {volume}
  {107}},\ \bibinfo {pages} {044056} (\bibinfo {year} {2023})},\ \Eprint
  {https://arxiv.org/abs/2006.11449} {arXiv:2006.11449 [gr-qc]} \BibitemShut
  {NoStop}%
\bibitem [{\citenamefont {London}(2026)}]{London26}%
  \BibitemOpen
  \bibfield  {author} {\bibinfo {author} {\bibfnamefont {L.~T.}\ \bibnamefont
  {London}},\ }\href {https://doi.org/10.1103/vy3q-nx3w} {\bibfield  {journal}
  {\bibinfo  {journal} {Phys. Rev. D}\ }\textbf {\bibinfo {volume} {113}},\
  \bibinfo {pages} {044008} (\bibinfo {year} {2026})}\BibitemShut {NoStop}%
\bibitem [{\citenamefont {Leung}\ \emph {et~al.}(1997)\citenamefont {Leung},
  \citenamefont {Liu}, \citenamefont {Suen}, \citenamefont {Tam},\ and\
  \citenamefont {Young}}]{Leung:1997was}%
  \BibitemOpen
  \bibfield  {author} {\bibinfo {author} {\bibfnamefont {P.~T.}\ \bibnamefont
  {Leung}}, \bibinfo {author} {\bibfnamefont {Y.~T.}\ \bibnamefont {Liu}},
  \bibinfo {author} {\bibfnamefont {W.~M.}\ \bibnamefont {Suen}}, \bibinfo
  {author} {\bibfnamefont {C.~Y.}\ \bibnamefont {Tam}},\ and\ \bibinfo {author}
  {\bibfnamefont {K.}~\bibnamefont {Young}},\ }\href
  {https://doi.org/10.1103/PhysRevLett.78.2894} {\bibfield  {journal} {\bibinfo
   {journal} {Phys. Rev. Lett.}\ }\textbf {\bibinfo {volume} {78}},\ \bibinfo
  {pages} {2894} (\bibinfo {year} {1997})},\ \Eprint
  {https://arxiv.org/abs/gr-qc/9903031} {arXiv:gr-qc/9903031} \BibitemShut
  {NoStop}%
\bibitem [{\citenamefont {Zimmerman}\ \emph {et~al.}(2015)\citenamefont
  {Zimmerman}, \citenamefont {Yang}, \citenamefont {Mark}, \citenamefont
  {Chen},\ and\ \citenamefont {Lehner}}]{Zimmerman:2014aha}%
  \BibitemOpen
  \bibfield  {author} {\bibinfo {author} {\bibfnamefont {A.}~\bibnamefont
  {Zimmerman}}, \bibinfo {author} {\bibfnamefont {H.}~\bibnamefont {Yang}},
  \bibinfo {author} {\bibfnamefont {Z.}~\bibnamefont {Mark}}, \bibinfo {author}
  {\bibfnamefont {Y.}~\bibnamefont {Chen}},\ and\ \bibinfo {author}
  {\bibfnamefont {L.}~\bibnamefont {Lehner}},\ }\href
  {https://doi.org/10.1007/978-3-319-10488-1_19} {\bibfield  {journal}
  {\bibinfo  {journal} {Astrophys. Space Sci. Proc.}\ }\textbf {\bibinfo
  {volume} {40}},\ \bibinfo {pages} {217} (\bibinfo {year} {2015})},\ \Eprint
  {https://arxiv.org/abs/1406.4206} {arXiv:1406.4206 [gr-qc]} \BibitemShut
  {NoStop}%
\bibitem [{\citenamefont {Mark}\ \emph {et~al.}(2015)\citenamefont {Mark},
  \citenamefont {Yang}, \citenamefont {Zimmerman},\ and\ \citenamefont
  {Chen}}]{Mark:2014aja}%
  \BibitemOpen
  \bibfield  {author} {\bibinfo {author} {\bibfnamefont {Z.}~\bibnamefont
  {Mark}}, \bibinfo {author} {\bibfnamefont {H.}~\bibnamefont {Yang}}, \bibinfo
  {author} {\bibfnamefont {A.}~\bibnamefont {Zimmerman}},\ and\ \bibinfo
  {author} {\bibfnamefont {Y.}~\bibnamefont {Chen}},\ }\href
  {https://doi.org/10.1103/PhysRevD.91.044025} {\bibfield  {journal} {\bibinfo
  {journal} {Phys. Rev. D}\ }\textbf {\bibinfo {volume} {91}},\ \bibinfo
  {pages} {044025} (\bibinfo {year} {2015})},\ \Eprint
  {https://arxiv.org/abs/1409.5800} {arXiv:1409.5800 [gr-qc]} \BibitemShut
  {NoStop}%
\bibitem [{\citenamefont {Hussain}\ and\ \citenamefont
  {Zimmerman}(2022)}]{Hussain:2022ins}%
  \BibitemOpen
  \bibfield  {author} {\bibinfo {author} {\bibfnamefont {A.}~\bibnamefont
  {Hussain}}\ and\ \bibinfo {author} {\bibfnamefont {A.}~\bibnamefont
  {Zimmerman}},\ }\href {https://doi.org/10.1103/PhysRevD.106.104018}
  {\bibfield  {journal} {\bibinfo  {journal} {Phys. Rev. D}\ }\textbf {\bibinfo
  {volume} {106}},\ \bibinfo {pages} {104018} (\bibinfo {year} {2022})},\
  \Eprint {https://arxiv.org/abs/2206.10653} {arXiv:2206.10653 [gr-qc]}
  \BibitemShut {NoStop}%
\bibitem [{\citenamefont {Li}\ \emph {et~al.}(2024)\citenamefont {Li},
  \citenamefont {Hussain}, \citenamefont {Wagle}, \citenamefont {Chen},
  \citenamefont {Yunes},\ and\ \citenamefont {Zimmerman}}]{Li:2023ulk}%
  \BibitemOpen
  \bibfield  {author} {\bibinfo {author} {\bibfnamefont {D.}~\bibnamefont
  {Li}}, \bibinfo {author} {\bibfnamefont {A.}~\bibnamefont {Hussain}},
  \bibinfo {author} {\bibfnamefont {P.}~\bibnamefont {Wagle}}, \bibinfo
  {author} {\bibfnamefont {Y.}~\bibnamefont {Chen}}, \bibinfo {author}
  {\bibfnamefont {N.}~\bibnamefont {Yunes}},\ and\ \bibinfo {author}
  {\bibfnamefont {A.}~\bibnamefont {Zimmerman}},\ }\href
  {https://doi.org/10.1103/PhysRevD.109.104026} {\bibfield  {journal} {\bibinfo
   {journal} {Phys. Rev. D}\ }\textbf {\bibinfo {volume} {109}},\ \bibinfo
  {pages} {104026} (\bibinfo {year} {2024})},\ \Eprint
  {https://arxiv.org/abs/2310.06033} {arXiv:2310.06033 [gr-qc]} \BibitemShut
  {NoStop}%
\bibitem [{\citenamefont {Cannizzaro}\ \emph {et~al.}(2024)\citenamefont
  {Cannizzaro}, \citenamefont {Sberna}, \citenamefont {Green},\ and\
  \citenamefont {Hollands}}]{Cannizzaro:2023jle}%
  \BibitemOpen
  \bibfield  {author} {\bibinfo {author} {\bibfnamefont {E.}~\bibnamefont
  {Cannizzaro}}, \bibinfo {author} {\bibfnamefont {L.}~\bibnamefont {Sberna}},
  \bibinfo {author} {\bibfnamefont {S.~R.}\ \bibnamefont {Green}},\ and\
  \bibinfo {author} {\bibfnamefont {S.}~\bibnamefont {Hollands}},\ }\href
  {https://doi.org/10.1103/PhysRevLett.132.051401} {\bibfield  {journal}
  {\bibinfo  {journal} {Phys. Rev. Lett.}\ }\textbf {\bibinfo {volume} {132}},\
  \bibinfo {pages} {051401} (\bibinfo {year} {2024})},\ \Eprint
  {https://arxiv.org/abs/2309.10021} {arXiv:2309.10021 [gr-qc]} \BibitemShut
  {NoStop}%
\bibitem [{\citenamefont {Hamilton}\ \emph {et~al.}(2021)\citenamefont
  {Hamilton}, \citenamefont {London}, \citenamefont {Thompson}, \citenamefont
  {Fauchon-Jones}, \citenamefont {Hannam}, \citenamefont {Kalaghatgi},
  \citenamefont {Khan}, \citenamefont {Pannarale},\ and\ \citenamefont
  {Vano-Vinuales}}]{Hamilton:2021pkf}%
  \BibitemOpen
  \bibfield  {author} {\bibinfo {author} {\bibfnamefont {E.}~\bibnamefont
  {Hamilton}}, \bibinfo {author} {\bibfnamefont {L.}~\bibnamefont {London}},
  \bibinfo {author} {\bibfnamefont {J.~E.}\ \bibnamefont {Thompson}}, \bibinfo
  {author} {\bibfnamefont {E.}~\bibnamefont {Fauchon-Jones}}, \bibinfo {author}
  {\bibfnamefont {M.}~\bibnamefont {Hannam}}, \bibinfo {author} {\bibfnamefont
  {C.}~\bibnamefont {Kalaghatgi}}, \bibinfo {author} {\bibfnamefont
  {S.}~\bibnamefont {Khan}}, \bibinfo {author} {\bibfnamefont {F.}~\bibnamefont
  {Pannarale}},\ and\ \bibinfo {author} {\bibfnamefont {A.}~\bibnamefont
  {Vano-Vinuales}},\ }\href {https://doi.org/10.1103/PhysRevD.104.124027}
  {\bibfield  {journal} {\bibinfo  {journal} {Phys. Rev. D}\ }\textbf {\bibinfo
  {volume} {104}},\ \bibinfo {pages} {124027} (\bibinfo {year} {2021})},\
  \Eprint {https://arxiv.org/abs/2107.08876} {arXiv:2107.08876 [gr-qc]}
  \BibitemShut {NoStop}%
\bibitem [{\citenamefont {Thompson}\ \emph {et~al.}(2024)\citenamefont
  {Thompson}, \citenamefont {Hamilton}, \citenamefont {London}, \citenamefont
  {Ghosh}, \citenamefont {Kolitsidou}, \citenamefont {Hoy},\ and\ \citenamefont
  {Hannam}}]{Thompson:2023ase}%
  \BibitemOpen
  \bibfield  {author} {\bibinfo {author} {\bibfnamefont {J.~E.}\ \bibnamefont
  {Thompson}}, \bibinfo {author} {\bibfnamefont {E.}~\bibnamefont {Hamilton}},
  \bibinfo {author} {\bibfnamefont {L.}~\bibnamefont {London}}, \bibinfo
  {author} {\bibfnamefont {S.}~\bibnamefont {Ghosh}}, \bibinfo {author}
  {\bibfnamefont {P.}~\bibnamefont {Kolitsidou}}, \bibinfo {author}
  {\bibfnamefont {C.}~\bibnamefont {Hoy}},\ and\ \bibinfo {author}
  {\bibfnamefont {M.}~\bibnamefont {Hannam}},\ }\href
  {https://doi.org/10.1103/PhysRevD.109.063012} {\bibfield  {journal} {\bibinfo
   {journal} {Phys. Rev. D}\ }\textbf {\bibinfo {volume} {109}},\ \bibinfo
  {pages} {063012} (\bibinfo {year} {2024})},\ \Eprint
  {https://arxiv.org/abs/2312.10025} {arXiv:2312.10025 [gr-qc]} \BibitemShut
  {NoStop}%
\bibitem [{\citenamefont {Arnaudo}\ \emph
  {et~al.}(2025{\natexlab{b}})\citenamefont {Arnaudo}, \citenamefont
  {Carballo},\ and\ \citenamefont {Withers}}]{Arnaudo:2025bnm}%
  \BibitemOpen
  \bibfield  {author} {\bibinfo {author} {\bibfnamefont {P.}~\bibnamefont
  {Arnaudo}}, \bibinfo {author} {\bibfnamefont {J.}~\bibnamefont {Carballo}},\
  and\ \bibinfo {author} {\bibfnamefont {B.}~\bibnamefont {Withers}},\ }\href
  {https://doi.org/10.1007/JHEP09(2025)010} {\bibfield  {journal} {\bibinfo
  {journal} {JHEP}\ }\textbf {\bibinfo {volume} {09}},\ \bibinfo {pages}
  {010}},\ \Eprint {https://arxiv.org/abs/2505.04696} {arXiv:2505.04696
  [hep-th]} \BibitemShut {NoStop}%
\bibitem [{\citenamefont {Zenginoglu}(2011)}]{Zenginoglu:2011jz}%
  \BibitemOpen
  \bibfield  {author} {\bibinfo {author} {\bibfnamefont {A.}~\bibnamefont
  {Zenginoglu}},\ }\href {https://doi.org/10.1103/PhysRevD.83.127502}
  {\bibfield  {journal} {\bibinfo  {journal} {Phys. Rev. D}\ }\textbf {\bibinfo
  {volume} {83}},\ \bibinfo {pages} {127502} (\bibinfo {year} {2011})},\
  \Eprint {https://arxiv.org/abs/1102.2451} {arXiv:1102.2451 [gr-qc]}
  \BibitemShut {NoStop}%
\bibitem [{\citenamefont {Panosso~Macedo}\ and\ \citenamefont
  {Zenginoglu}(2024)}]{PanossoMacedo:2024nkw}%
  \BibitemOpen
  \bibfield  {author} {\bibinfo {author} {\bibfnamefont {R.}~\bibnamefont
  {Panosso~Macedo}}\ and\ \bibinfo {author} {\bibfnamefont {A.}~\bibnamefont
  {Zenginoglu}},\ }\href {https://doi.org/10.3389/fphy.2024.1497601} {\bibfield
   {journal} {\bibinfo  {journal} {Front. in Phys.}\ }\textbf {\bibinfo
  {volume} {12}},\ \bibinfo {pages} {1497601} (\bibinfo {year} {2024})},\
  \Eprint {https://arxiv.org/abs/2409.11478} {arXiv:2409.11478 [gr-qc]}
  \BibitemShut {NoStop}%
\bibitem [{\citenamefont {Panosso~Macedo}(2024)}]{PanossoMacedo:2023qzp}%
  \BibitemOpen
  \bibfield  {author} {\bibinfo {author} {\bibfnamefont {R.}~\bibnamefont
  {Panosso~Macedo}},\ }\href {https://doi.org/10.1098/rsta.2023.0046}
  {\bibfield  {journal} {\bibinfo  {journal} {Phil. Trans. Roy. Soc. Lond. A}\
  }\textbf {\bibinfo {volume} {382}},\ \bibinfo {pages} {20230046} (\bibinfo
  {year} {2024})},\ \Eprint {https://arxiv.org/abs/2307.15735}
  {arXiv:2307.15735 [gr-qc]} \BibitemShut {NoStop}%
\bibitem [{\citenamefont {Gajic}\ and\ \citenamefont
  {Warnick}(2024)}]{Gajic:2024xrn}%
  \BibitemOpen
  \bibfield  {author} {\bibinfo {author} {\bibfnamefont {D.}~\bibnamefont
  {Gajic}}\ and\ \bibinfo {author} {\bibfnamefont {C.~M.}\ \bibnamefont
  {Warnick}},\ }\href@noop {} {\bibinfo {title} {{Quasinormal modes on Kerr
  spacetimes}}} (\bibinfo {year} {2024}),\ \Eprint
  {https://arxiv.org/abs/2407.04098} {arXiv:2407.04098 [gr-qc]} \BibitemShut
  {NoStop}%
\bibitem [{\citenamefont {Jaramillo}\ \emph {et~al.}(2021)\citenamefont
  {Jaramillo}, \citenamefont {Macedo},\ and\ \citenamefont
  {Sheikh}}]{PhysRevX.11.031003}%
  \BibitemOpen
  \bibfield  {author} {\bibinfo {author} {\bibfnamefont {J.~L.}\ \bibnamefont
  {Jaramillo}}, \bibinfo {author} {\bibfnamefont {R.~P.}\ \bibnamefont
  {Macedo}},\ and\ \bibinfo {author} {\bibfnamefont {L.~A.}\ \bibnamefont
  {Sheikh}},\ }\href {https://doi.org/10.1103/PhysRevX.11.031003} {\bibfield
  {journal} {\bibinfo  {journal} {Phys. Rev. X}\ }\textbf {\bibinfo {volume}
  {11}},\ \bibinfo {pages} {031003} (\bibinfo {year} {2021})}\BibitemShut
  {NoStop}%
\bibitem [{\citenamefont {Jaramillo}(2022)}]{Jaramillo:2022kuv}%
  \BibitemOpen
  \bibfield  {author} {\bibinfo {author} {\bibfnamefont {J.~L.}\ \bibnamefont
  {Jaramillo}},\ }\href {https://doi.org/10.1088/1361-6382/ac8ddc} {\bibfield
  {journal} {\bibinfo  {journal} {Class. Quant. Grav.}\ }\textbf {\bibinfo
  {volume} {39}},\ \bibinfo {pages} {217002} (\bibinfo {year} {2022})},\
  \Eprint {https://arxiv.org/abs/2206.08025} {arXiv:2206.08025 [gr-qc]}
  \BibitemShut {NoStop}%
\bibitem [{\citenamefont {Panosso~Macedo}\ \emph {et~al.}(2025)\citenamefont
  {Panosso~Macedo}, \citenamefont {Katagiri}, \citenamefont {Kubota},\ and\
  \citenamefont {Motohashi}}]{PanossoMacedo:2025xnf}%
  \BibitemOpen
  \bibfield  {author} {\bibinfo {author} {\bibfnamefont {R.}~\bibnamefont
  {Panosso~Macedo}}, \bibinfo {author} {\bibfnamefont {T.}~\bibnamefont
  {Katagiri}}, \bibinfo {author} {\bibfnamefont {K.-i.}\ \bibnamefont
  {Kubota}},\ and\ \bibinfo {author} {\bibfnamefont {H.}~\bibnamefont
  {Motohashi}},\ }\href@noop {} {\  (\bibinfo {year} {2025})},\ \Eprint
  {https://arxiv.org/abs/2512.02110} {arXiv:2512.02110 [gr-qc]} \BibitemShut
  {NoStop}%
\bibitem [{\citenamefont {Ansorg}\ and\ \citenamefont
  {Panosso~Macedo}(2016)}]{Ansorg:2016ztf}%
  \BibitemOpen
  \bibfield  {author} {\bibinfo {author} {\bibfnamefont {M.}~\bibnamefont
  {Ansorg}}\ and\ \bibinfo {author} {\bibfnamefont {R.}~\bibnamefont
  {Panosso~Macedo}},\ }\href {https://doi.org/10.1103/PhysRevD.93.124016}
  {\bibfield  {journal} {\bibinfo  {journal} {Phys. Rev. D}\ }\textbf {\bibinfo
  {volume} {93}},\ \bibinfo {pages} {124016} (\bibinfo {year} {2016})},\
  \Eprint {https://arxiv.org/abs/1604.02261} {arXiv:1604.02261 [gr-qc]}
  \BibitemShut {NoStop}%
\bibitem [{\citenamefont {Besson}\ and\ \citenamefont
  {Jaramillo}(2025)}]{Besson:2024adi}%
  \BibitemOpen
  \bibfield  {author} {\bibinfo {author} {\bibfnamefont {J.}~\bibnamefont
  {Besson}}\ and\ \bibinfo {author} {\bibfnamefont {J.~L.}\ \bibnamefont
  {Jaramillo}},\ }\href {https://doi.org/10.1007/s10714-025-03438-6} {\bibfield
   {journal} {\bibinfo  {journal} {Gen. Rel. Grav.}\ }\textbf {\bibinfo
  {volume} {57}},\ \bibinfo {pages} {110} (\bibinfo {year} {2025})},\ \Eprint
  {https://arxiv.org/abs/2412.02793} {arXiv:2412.02793 [gr-qc]} \BibitemShut
  {NoStop}%
\bibitem [{\citenamefont {Bourg}\ \emph {et~al.}(2025)\citenamefont {Bourg},
  \citenamefont {Panosso~Macedo}, \citenamefont {Spiers}, \citenamefont
  {Leather}, \citenamefont {B{\'e}atrice},\ and\ \citenamefont
  {Pound}}]{Bourg:2025lpd}%
  \BibitemOpen
  \bibfield  {author} {\bibinfo {author} {\bibfnamefont {P.}~\bibnamefont
  {Bourg}}, \bibinfo {author} {\bibfnamefont {R.}~\bibnamefont
  {Panosso~Macedo}}, \bibinfo {author} {\bibfnamefont {A.}~\bibnamefont
  {Spiers}}, \bibinfo {author} {\bibfnamefont {B.}~\bibnamefont {Leather}},
  \bibinfo {author} {\bibfnamefont {B.}~\bibnamefont {B{\'e}atrice}},\ and\
  \bibinfo {author} {\bibfnamefont {A.}~\bibnamefont {Pound}},\ }\href
  {https://doi.org/10.1103/fbz4-qsvn} {\bibfield  {journal} {\bibinfo
  {journal} {Phys. Rev. D}\ }\textbf {\bibinfo {volume} {112}},\ \bibinfo
  {pages} {044049} (\bibinfo {year} {2025})},\ \Eprint
  {https://arxiv.org/abs/2503.07432} {arXiv:2503.07432 [gr-qc]} \BibitemShut
  {NoStop}%
\bibitem [{\citenamefont {Penrose}(2011)}]{Penrose64}%
  \BibitemOpen
  \bibfield  {author} {\bibinfo {author} {\bibfnamefont {R.}~\bibnamefont
  {Penrose}},\ }\href {https://doi.org/10.1007/s10714-010-1110-5} {\bibfield
  {journal} {\bibinfo  {journal} {General Relativity and Gravitation}\ }\textbf
  {\bibinfo {volume} {43}},\ \bibinfo {pages} {901} (\bibinfo {year}
  {2011})}\BibitemShut {NoStop}%
\bibitem [{\citenamefont {Valiente~Kroon}(2016)}]{ValienteKroon2016}%
  \BibitemOpen
  \bibfield  {author} {\bibinfo {author} {\bibfnamefont {J.~A.}\ \bibnamefont
  {Valiente~Kroon}},\ }\href@noop {} {\emph {\bibinfo {title} {Conformal
  Methods in General Relativity}}}\ (\bibinfo  {publisher} {Cambridge
  University Press},\ \bibinfo {year} {2016})\BibitemShut {NoStop}%
\bibitem [{\citenamefont {Gasperin}\ \emph {et~al.}(2026)\citenamefont
  {Gasperin}, \citenamefont {Panosso~Macedo},\ and\ \citenamefont
  {Feng}}]{GasperinPanossoMacedoFeng2026}%
  \BibitemOpen
  \bibfield  {author} {\bibinfo {author} {\bibfnamefont {E.}~\bibnamefont
  {Gasperin}}, \bibinfo {author} {\bibfnamefont {R.}~\bibnamefont
  {Panosso~Macedo}},\ and\ \bibinfo {author} {\bibfnamefont {J.}~\bibnamefont
  {Feng}}\ }\href {https://doi.org/arXiv:2602.19245} {arXiv:2602.19245}
  (\bibinfo {year} {2026})\BibitemShut {NoStop}%
\bibitem [{\citenamefont {Gasperín}\ and\ \citenamefont
  {Jaramillo}(2022)}]{GasperinJaramillo2022}%
  \BibitemOpen
  \bibfield  {author} {\bibinfo {author} {\bibfnamefont {E.}~\bibnamefont
  {Gasperín}}\ and\ \bibinfo {author} {\bibfnamefont {J.~L.}\ \bibnamefont
  {Jaramillo}},\ }\href {https://doi.org/10.1088/1361-6382/ac5054} {\bibfield
  {journal} {\bibinfo  {journal} {Classical and Quantum Gravity}\ }\textbf
  {\bibinfo {volume} {39}},\ \bibinfo {pages} {115010} (\bibinfo {year}
  {2022})}\BibitemShut {NoStop}%
\bibitem [{\citenamefont {Leaver}(1985)}]{Leaver:1985ax}%
  \BibitemOpen
  \bibfield  {author} {\bibinfo {author} {\bibfnamefont {E.~W.}\ \bibnamefont
  {Leaver}},\ }\href {https://doi.org/10.1098/rspa.1985.0119} {\bibfield
  {journal} {\bibinfo  {journal} {Proc. Roy. Soc. Lond. A}\ }\textbf {\bibinfo
  {volume} {402}},\ \bibinfo {pages} {285} (\bibinfo {year}
  {1985})}\BibitemShut {NoStop}%
\bibitem [{\citenamefont {Dolan}(2007)}]{Dolan:2007mj}%
  \BibitemOpen
  \bibfield  {author} {\bibinfo {author} {\bibfnamefont {S.~R.}\ \bibnamefont
  {Dolan}},\ }\href {https://doi.org/10.1103/PhysRevD.76.084001} {\bibfield
  {journal} {\bibinfo  {journal} {Phys. Rev. D}\ }\textbf {\bibinfo {volume}
  {76}},\ \bibinfo {pages} {084001} (\bibinfo {year} {2007})},\ \Eprint
  {https://arxiv.org/abs/0705.2880} {arXiv:0705.2880 [gr-qc]} \BibitemShut
  {NoStop}%
\bibitem [{\citenamefont {Zenginoglu}(2008)}]{Zenginoglu:2007jw}%
  \BibitemOpen
  \bibfield  {author} {\bibinfo {author} {\bibfnamefont {A.}~\bibnamefont
  {Zenginoglu}},\ }\href {https://doi.org/10.1088/0264-9381/25/14/145002}
  {\bibfield  {journal} {\bibinfo  {journal} {Class. Quant. Grav.}\ }\textbf
  {\bibinfo {volume} {25}},\ \bibinfo {pages} {145002} (\bibinfo {year}
  {2008})},\ \Eprint {https://arxiv.org/abs/0712.4333} {arXiv:0712.4333
  [gr-qc]} \BibitemShut {NoStop}%
\bibitem [{\citenamefont {Hollands}\ and\ \citenamefont
  {Wald}(2013)}]{Hollands:2012sf}%
  \BibitemOpen
  \bibfield  {author} {\bibinfo {author} {\bibfnamefont {S.}~\bibnamefont
  {Hollands}}\ and\ \bibinfo {author} {\bibfnamefont {R.~M.}\ \bibnamefont
  {Wald}},\ }\href {https://doi.org/10.1007/s00220-012-1638-1} {\bibfield
  {journal} {\bibinfo  {journal} {Commun. Math. Phys.}\ }\textbf {\bibinfo
  {volume} {321}},\ \bibinfo {pages} {629} (\bibinfo {year} {2013})},\ \Eprint
  {https://arxiv.org/abs/1201.0463} {arXiv:1201.0463 [gr-qc]} \BibitemShut
  {NoStop}%
\bibitem [{\citenamefont {Bleistein}\ and\ \citenamefont
  {Handelsman}(1986)}]{BleisteinHandelsman1986}%
  \BibitemOpen
  \bibfield  {author} {\bibinfo {author} {\bibfnamefont {N.}~\bibnamefont
  {Bleistein}}\ and\ \bibinfo {author} {\bibfnamefont {R.~A.}\ \bibnamefont
  {Handelsman}},\ }\href@noop {} {\emph {\bibinfo {title} {Asymptotic
  Expansions of Integrals}}}\ (\bibinfo  {publisher} {Dover Publications},\
  \bibinfo {address} {New York},\ \bibinfo {year} {1986})\BibitemShut {NoStop}%
\bibitem [{\citenamefont {Brown}\ and\ \citenamefont
  {Churchill}(2014)}]{ChurchillBrown2014}%
  \BibitemOpen
  \bibfield  {author} {\bibinfo {author} {\bibfnamefont {J.~W.}\ \bibnamefont
  {Brown}}\ and\ \bibinfo {author} {\bibfnamefont {R.~V.}\ \bibnamefont
  {Churchill}},\ }\href@noop {} {\emph {\bibinfo {title} {Complex Variables and
  Applications}}},\ \bibinfo {edition} {9th}\ ed.\ (\bibinfo  {publisher}
  {McGraw-Hill Education},\ \bibinfo {address} {New York},\ \bibinfo {year}
  {2014})\BibitemShut {NoStop}%
\bibitem [{\citenamefont {Minucci}\ and\ \citenamefont
  {Panosso~Macedo}(2025)}]{Minucci:2024qrn}%
  \BibitemOpen
  \bibfield  {author} {\bibinfo {author} {\bibfnamefont {M.}~\bibnamefont
  {Minucci}}\ and\ \bibinfo {author} {\bibfnamefont {R.}~\bibnamefont
  {Panosso~Macedo}},\ }\href {https://doi.org/10.1007/s10714-025-03364-7}
  {\bibfield  {journal} {\bibinfo  {journal} {Gen. Rel. Grav.}\ }\textbf
  {\bibinfo {volume} {57}},\ \bibinfo {pages} {33} (\bibinfo {year} {2025})},\
  \Eprint {https://arxiv.org/abs/2411.19740} {arXiv:2411.19740 [gr-qc]}
  \BibitemShut {NoStop}%
\bibitem [{\citenamefont {Stein}(2019)}]{Stein:2019mop}%
  \BibitemOpen
  \bibfield  {author} {\bibinfo {author} {\bibfnamefont {L.~C.}\ \bibnamefont
  {Stein}},\ }\href {https://doi.org/10.21105/joss.01683} {\bibfield  {journal}
  {\bibinfo  {journal} {J. Open Source Softw.}\ }\textbf {\bibinfo {volume}
  {4}},\ \bibinfo {pages} {1683} (\bibinfo {year} {2019})},\ \Eprint
  {https://arxiv.org/abs/1908.10377} {arXiv:1908.10377 [gr-qc]} \BibitemShut
  {NoStop}%
\bibitem [{\citenamefont {Panosso~Macedo}\ \emph {et~al.}(2022)\citenamefont
  {Panosso~Macedo}, \citenamefont {Leather}, \citenamefont {Warburton},
  \citenamefont {Wardell},\ and\ \citenamefont
  {Zengino{\u{g}}lu}}]{PanossoMacedo:2022fdi}%
  \BibitemOpen
  \bibfield  {author} {\bibinfo {author} {\bibfnamefont {R.}~\bibnamefont
  {Panosso~Macedo}}, \bibinfo {author} {\bibfnamefont {B.}~\bibnamefont
  {Leather}}, \bibinfo {author} {\bibfnamefont {N.}~\bibnamefont {Warburton}},
  \bibinfo {author} {\bibfnamefont {B.}~\bibnamefont {Wardell}},\ and\ \bibinfo
  {author} {\bibfnamefont {A.}~\bibnamefont {Zengino{\u{g}}lu}},\ }\href
  {https://doi.org/10.1103/PhysRevD.105.104033} {\bibfield  {journal} {\bibinfo
   {journal} {Phys. Rev. D}\ }\textbf {\bibinfo {volume} {105}},\ \bibinfo
  {pages} {104033} (\bibinfo {year} {2022})},\ \Eprint
  {https://arxiv.org/abs/2202.01794} {arXiv:2202.01794 [gr-qc]} \BibitemShut
  {NoStop}%
\bibitem [{\citenamefont {Zhou}\ and\ \citenamefont
  {Panosso~Macedo}(2025)}]{Zhou:2025xta}%
  \BibitemOpen
  \bibfield  {author} {\bibinfo {author} {\bibfnamefont {Y.}~\bibnamefont
  {Zhou}}\ and\ \bibinfo {author} {\bibfnamefont {R.}~\bibnamefont
  {Panosso~Macedo}},\ }\href {https://doi.org/10.1103/n948-jnfh} {\bibfield
  {journal} {\bibinfo  {journal} {Phys. Rev. D}\ }\textbf {\bibinfo {volume}
  {112}},\ \bibinfo {pages} {084063} (\bibinfo {year} {2025})},\ \Eprint
  {https://arxiv.org/abs/2507.05370} {arXiv:2507.05370 [gr-qc]} \BibitemShut
  {NoStop}%
\end{thebibliography}%

\appendix
\addcontentsline{toc}{section}{Appendices}

\section{Conformal transformation of the current $J^a$}\label{app:conformal}

We begin by identifying two manifolds: the physical spacetime $\{{\cal M}, g_{ab}\}$ and the conformal (unphysical) spacetime $\{\tilde{\cal M}, \tilde g_{ab}\}$, related to each other by a conformal factor $\Omega$ via
\begin{equation}
\label{confresc}
g_{ab} = \Omega^{-2} \tilde g_{ab} \Longrightarrow \boldsymbol{g} = \Omega^{-8} \boldsymbol{\tilde g}. 
\end{equation}
The asymptotic boundary of the spacetime is characterised by $\Omega=0$.

One must also identify a conformal connection $\tilde \nabla_a$ compatible with the conformal spacetime $\tilde g_{ab}$. Hence, given a vector field $V^a$ in the physical spacetime, one finds the relation 
\bea
&&\nabla_a V^b = \tilde \nabla_a\left( V^b \right) - S_{a c}{}^{b d} \left( \nabla_d \ln\Omega \right) \left( V^c\right), \label{eq:app_connection}\\
&&S_{a c}{}^{b d} = \delta_a^b \delta_c^d + \delta_a^d \delta_c^b - \tilde{g}_{a c}\tilde{g}^{b d}. \label{eq:app_connection2}
\eea
In this framework, a massless scalar field $\Psi$ rescales as
\begin{equation}
\Psi = \Omega \,\psi, 
\end{equation}
capturing the $1/r$ decay of the scalar field towards infinity, cf.~\eqref{ansatz}, thus ensuring $\psi \sim {\cal O}(1)$ as $r \to \infty$.

The massless Klein-Gordon equation \eqref{eq:KG} can also be written in terms of the conformally rescaled spacetime. Assuming that the physical spacetime $g_{ab}$ is Ricci flat $R = 0$, one obtains
\begin{align}
\square \Psi = \Omega^3 {L}\, \psi, \quad
{L} = \tilde\square - \dfrac{\tilde R}{6},\label{ConfQuantity3} 
\end{align}
with $\bar R$ the Ricci scalar associated with the conformal spacetime. In practical terms, it corresponds to the term $\sim \rh/r$ within the BH effective potential \eqref{eq:BH_Potential}.
Hence, if the physical massless scalar field $\Psi$ satisfy the Klein-Gordon equation \eqref{eq:KG}, then the conformal scalar field $\tilde \Psi$ satisfies the hyperboloidal conformal wave equation
\be
\label{eq:ConfWaveEq}
L[ \psi] =0.
\ee
which we interpret as an abstract tensorial equation.

This conformal mapping explicitly takes place in terms of the hyperboloidal coordinates discussed in the main text. Specifically, we introduced hyperboloidal coordinates $\bar x^\mu$ and $\check x^\mu$. Thus, instead of the '$\sim$' symbol, we employ in the conformal quantities the ornament associated with each coordinate system, i.e., $\bar g_{\mu \nu}$ and $\check g_{\mu \nu}$ are, respectively, the conformal metric associated with each of these coordinates (with a similar notation for further quantities). It is important to notice that given the previous relations one can construct the conformal unphysical current from Eqn. (\ref{ConfQuantity3}) along with
\begin{subequations}
\begin{align}
\tilde \nabla_{\mu} \tilde J^{\mu} &= \dfrac{1}{\sqrt{-\boldsymbol{\tilde{g}}}} \p_\mu \left( \sqrt{-\boldsymbol{\tilde{g}}} \tilde{J}^\mu\right)
\end{align}
\end{subequations}
so that by performing partial integrations to obtain
\bea
\label{eq:conf_scalarfield_integral_part_int}
\int dx^4 \sqrt{- {\boldsymbol{\tilde{g}}}}\, \psi_2 
L \psi_1 &=& \int dx^4 \sqrt{- {\boldsymbol{\tilde{g}}}}\,  \psi_1 
L \psi_2 \nn \\
&&+ \int dx^4 \sqrt{- {\boldsymbol{\tilde{g}}}}\, \tilde{\nabla}_\mu \tilde{J}^\mu,
\eea
and 
\be
\label{eq:J_conformal}
\tilde J^\mu = \psi_2 \tilde \nabla^\mu \psi_1 - \psi_1 \tilde \nabla^\mu \psi_2.
\ee

This result resembles the one obtained directly in the physical spacetime, for which
\begin{subequations}
\begin{align}
\label{eq:ScalarOperator_physical}
\Psi_2 \square \Psi_1 - \Psi_1 \square \Psi_2 = \nabla_{\mu}J^\mu, \\
\label{eq:J_physical}
J^\mu =  \Psi_2  \nabla^\mu  \Psi_1 -  \Psi_1  \nabla^\mu  \Psi_2.
\end{align}
\end{subequations}
It is interesting to notice that definitions \eqref{eq:J_conformal} and \eqref{eq:J_physical} imply
\be
\label{ConfCurrent}
J^\mu = \Omega^4 \tilde J^\mu,
\ee
which makes it consistent with Eqn.~\eqref{eq:conf_scalarfield_integral_part_int}. Indeed, the left-hand side of Eqn.~\eqref{eq:ScalarOperator_physical} transforms under the conformal mapping to
\be
\Psi_2 \square \Psi_1 - \Psi_1 \square \Psi_2 = \Omega^4 \left(  \psi_2 
L \psi_1 -  \psi_1 
L \psi_2\right).
\ee
On the other hand, the right-hand side of Eqn.~\eqref{eq:ScalarOperator_physical} requires the expression for the conformal connection $\tilde \nabla$ via Eqs.~\eqref{eq:app_connection} and \eqref{eq:app_connection2}. By working out these expressions, terms with $\nabla_\mu \Omega$ cancel out exactly leaving
\be
\nabla_\mu J^\mu = \Omega^4 \tilde \nabla_\mu \tilde J^\mu.
\ee
Thus, the left and right hand side of Eqn.~\eqref{eq:ScalarOperator_physical} pick a factor $\Omega^4$, making the conformal transformation consistent with \eqref{eq:conf_scalarfield_integral_part_int}. For further details regarding Eqn.~\eqref{ConfCurrent} see \cite{ValienteKroon2016}.

\section{Time reversal in hyperboloidal foliations}\label{app:TR_hyperboloidal} 

\begin{figure*}[t!]
 \includegraphics[width=0.68\columnwidth]{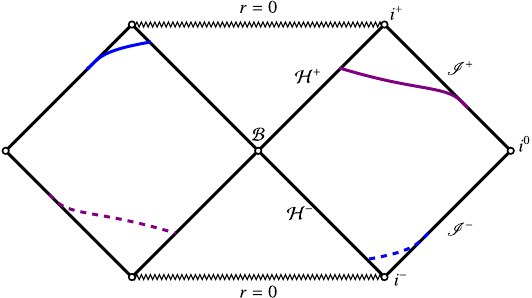}
\includegraphics[width=0.68\columnwidth]{Penrose_Schwarzschild_Future_Hyperboloidal.pdf}
 \includegraphics[width=0.68\columnwidth]{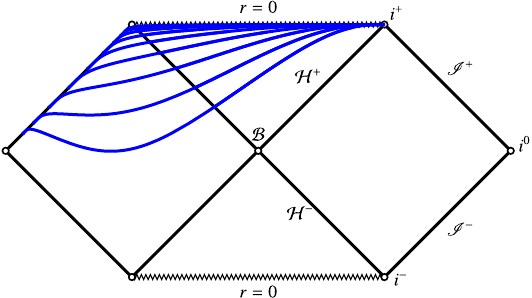}

     \caption{Carter-Penrose diagrams representing the hyperboloidal foliations associated to the coordinates $\bar x^\mu_\pm$, which are related to the Schwarzschild-Droste time via $t\to \tau_{\pm} \mp H(\sigma)$. The sign change with respect to the height function yields the future hyperboloidal (purple) and past (blue) hypersurfaces, identified in both asymptotic flat regions (solid and dashed lines), see left panel. However, surfaces $\tau_{\pm}=$constant are not symmetric with respect to each other (middle and right panels).}
    \label{fig:penrose_diagrams_NoSym}
\end{figure*}

In this section, we discuss a series of manipulations on the time coordinate via the height function technique~\cite{Zenginoglu:2007jw} leading to alternative notions of ``future'' or ``past'' foliations. 
As we shall discuss, there are in total $4$ families of hyperboloidal foliations arising from combining each transformation $t \to \tau \pm H(\sigma)$, with a direct coordinate time reversal $\tau \to -\tau$. 

Among those, we will denote the class of \emph{future hyperboloidal foliations} as those naturally adapted to the QNM boundary conditions at ${\cal H^+}$ and ${\scri^+}$. Similarly \emph{past hyperboloidal foliations} are naturally adapted to the anti-QNM boundary conditions at ${\cal H^-}$ and ${\scri^-}$.
Our goal is to emphasise the unique property of the time reversal transformation introduced in Sec.~\eqref{sec:Hyp_past_foliation}---and captured by the operator ${\cal J}$ in Sec.~\eqref{sec:J_op_Hyp}---as the operation leaving the line element invariant.

For simplicity, we will omit the lenght scale $\lambda$ in the transformations introduced in the next sections. All radial compactifications will follow $\sigma = \rh/r$, and we omit the angular dependence. A similar discussion including the azimuthal angular coordinate $\varphi$ would follow in the Kerr spacetime.

\subsection{The ``+'' and ``$-$'' time foliations}

We consider a generic transformation into coordinates $\bar{x}^\mu_\pm=(\tau_\pm, \sigma_\pm)$ defined by the height function technique~\cite{Zenginoglu:2007jw}
\be
\label{eq:x_future}
t = \tau_{\pm} \mp H(\sigma_\pm).
\ee
It is straightforward to derive the direct transformation  $\bar{x}_+^\mu \leftrightarrow \bar{x}_-^\mu $ as
\be
\label{eq:future_past_trasfo}
\tau_+ = \tau_- + 2\, H(\sigma_-), \quad \sigma_+=\sigma_-.
\ee
Note that if we were to identify the height function with the tortoise coordinate $H(\sigma) = r_*(r(\sigma))$, we would recover the well-known outgoing and ingoing null coordinates $\tau_+ \leftrightarrow u$ and $\tau_- \leftrightarrow v$. Here, we consider $H(\sigma)$  as representing any hyperboloidal foliation, with the plots in Figs.~\ref{fig:penrose_diagrams_NoSym}-\ref{fig:penrose_diagrams_NoSym_timeflip} obtained in the minimal gauge, Eq.~\eqref{eq:MinimalGaugeHeightFunction}.

Changing between $\bar{x}^\mu_+$ and $\bar{x}^\mu_-$ leaves the line element invariant, except for a few metric components related by
\bea
\label{eq:future_past_metric_trasfo}
g_{\tau_+ \sigma_+} = - g_{\tau_- \sigma_-} =  \gamma(\sigma_\pm).
\eea
Besides, the coordinate basis vectors relate via, cf.~\eqref{eq:hyp_derivative_trasfo}
\be
\label{eq:hyp_derivative_trasfo_+-}
 \partial_t = 
\partial_{\tau_\pm},
\quad
\partial_{r_*} = 
\mp\gamma(\sigma_\pm) \partial_{\tau_\pm}  - p(\sigma_\pm) \p_{\sigma_\pm}.
\ee
Thus, derivatives along the null coordinates  read
\bea
&&\nabla_{v} \psi =\dfrac{1}{2} 
\bigg(1\mp \gamma(\sigma_\pm)\bigg)\p_{\tau_\pm} \psi    - p(\sigma_\pm) \p_{\sigma_\pm} \psi,  \\
&&\nabla_{u} \psi =\dfrac{1}{2} 
\bigg(1  \pm \gamma(\sigma_\pm)\bigg)\p_{\tau_\pm} \psi  + p(\sigma_\pm) \p_{\sigma_\pm} \psi.
\eea
Considering the expressions of $p(\sigma_\pm)$ and $\gamma(\sigma_\pm)$, we can see that, at $\sigma_\pm =0$ and $\sigma_\pm=1$,  the coordinates $\bar x^\mu_+$ automatically impose the QNM boundary conditions \eqref{eq:BH_QNM_BC_v3}, whereas $\bar x^\mu_-$ automatically impose the anti-QNM boundary conditions \eqref{eq:WH_QNM_BC_v3}.

Therefore, $\bar x^\mu_+$ belongs to the class of future hyperboloidal foliations, while $\bar x^\mu_-$ to the class of past hyperboloidal foliation. Indeed, $\bar x^\mu_+$ is exactly the coordinates employed in Sec.~\ref{sec:hyp_future}. However, $\bar x^\mu_-$ is not the correct choice for a past foliation to be associated with the operator ${\cal J}$ because it does not lead to an invariant line element, cf.~\eqref{eq:future_past_metric_trasfo}.

Fig.~\eqref{fig:penrose_diagrams_NoSym} displays these hyperboloidal hypersurfaces at a given $\tau_\pm =$constant, with the $\tau_+$ surfaces in purple and the $\tau_-$ surfaces in blue. Even though they are adapted to the geometry of QNM (future foliation) and anti-QNM (past foliation), one observes a loss of symmetry across future and past, or equivalently, across the right and left asymptotically flat regions.

\subsection{The coordinate time-reversal transformations}
\begin{figure*}[t]
 \includegraphics[width=0.68\columnwidth]{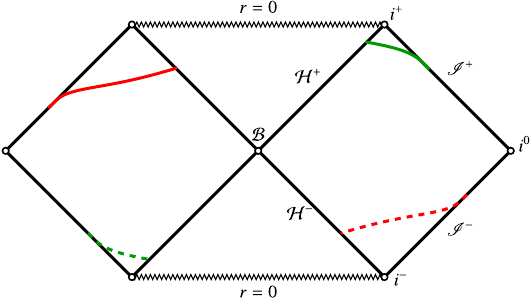}
\includegraphics[width=0.68\columnwidth]{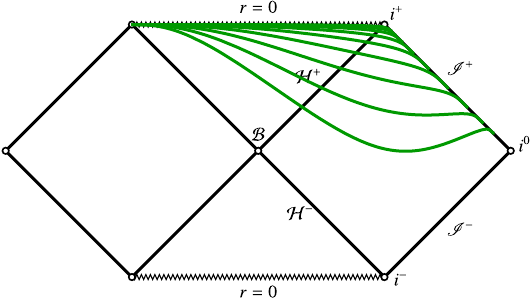}
 \includegraphics[width=0.68\columnwidth]{Penrose_Schwarzschild_Past_Prime_Hyperboloidal.pdf}

     \caption{Carter-Penrose diagrams representing the hyperboloidal foliations associated to the coordinates $\check x^\mu_\pm$, which are obtained by the coordinate time reversal $\tau_\pm = -\check \tau_\pm$. Despite the sign change, $\check \tau_+$ and $\check \tau_-$ still correspond, respectively, to a future (green)  and a past (red) hyperboloidal surfaces (left panel). As in Fig.~\ref{fig:penrose_diagrams_NoSym_timeflip}, surfaces $\check \tau_{\pm}=$constant are not symmetric with respect to each other (middle and right panels). The symmetry accomplished by the time reversal ${\cal J}$ operation relates $\bar x^{\mu}_\pm \leftrightarrow \check x^{\mu}_{\mp}$.
     }
    \label{fig:penrose_diagrams_NoSym_timeflip}
\end{figure*}
Another transformation that achieves a change in the metric similar to Eq.~\eqref{eq:future_past_metric_trasfo} is a direct coordinate time-reversal $\tau \to -\tau$. Hence, we define a new set of coordinates $\check x^\mu_\pm=(\check \tau_\pm, \check \sigma_\pm)$ via
\be
\check \tau_\pm = - \tau_\pm.
\ee
Starting from the original Schwarzchild-Droste coordinates, these transformations would correspond to
\be
\label{eq:x_flip}
t = - \bigg( \check \tau_{\pm} \pm H(\check \sigma_\pm) \bigg).
\ee
Similarly to Eqn.~\eqref{eq:future_past_trasfo}, from the above transformation one obtains the relations
\be
\check{\tau}_+= \check{\tau}_- - 2 H(\check{\sigma}_-), \quad \check{\sigma}_+=\check{\sigma}_- .
\ee
Such mappings will affect the metric components
\begin{subequations}
\label{eq:time_reversal_metric_trasfo}
\begin{align}
g_{\check{\tau}_- \check{\sigma}_-} = - g_{\check{\tau}_+ \check{\sigma}_+} = \gamma(\check \sigma_\pm),
\end{align}
\end{subequations}
where we observe that the functional form of $g_{\check{\tau}_\mp \check{\sigma}_\mp}$ coincide with $g_{{\tau}_\pm {\sigma}_\pm}$, which is a direct manifestation of the ${\cal J}$ operation acting between $\bar x^\mu_\pm$ and $\check x^\mu_\mp$.

Note, however, that this coordinate time-reversal does not alter the causal structure of the resulting hyperboloidal foliations. To appreciate it, we observe that the derivatives along the null coordinates just take an overall minus sign in their temporal component 
\bea
&&\nabla_{v} \psi = - \dfrac{1}{2} 
\bigg(1\mp \gamma(\sigma_\pm)\bigg)\p_{\check\tau_\pm} \psi    - p(\sigma_\pm) \p_{\sigma_\pm} \psi,  \\
&&\nabla_{u} \psi = - \dfrac{1}{2} 
\bigg(1  \pm \gamma(\sigma_\pm)\bigg)\p_{\check\tau_\pm} \psi  + p(\sigma_\pm) \p_{\sigma_\pm} \psi.
\eea
Hence, at $\check \sigma_\pm =0$ and $\sigma_\pm=1$,  the coordinates $\check x^\mu_+$ still imposes the QNM boundary conditions \eqref{eq:BH_QNM_BC_v3}, and $\check x^\mu_-$ the anti-QNM boundary conditions \eqref{eq:WH_QNM_BC_v3}, despite the sign-reversal with respect to $\bar x^\mu_\pm$.
In this way, $\check x^\mu_+$ also belongs to the class of future hyperboloidal foliations, whereas $\check x^\mu_-$ to the class of past hyperboloidal foliations. 

Fig.~\ref{fig:penrose_diagrams_NoSym_timeflip} depicts the time-reversed hyperboloidal hypersurface with the $\check \tau_+$ surface in green and the $\tau_-$ in red. Even though they are adapted to the geometry of QNM (future foliation) and anti-QNM (past foliation), one observe a loss of symmetry across future and past, or equivalently, across the right and left asymptotically flat regions. We observe the same causal features as their counter-part $\bar x^\mu_\pm$ and, in particular, the symmetry between $\bar x^\mu_\pm \leftrightarrow \check x^\mu_\mp$ becomes manifest.  

Thus, the ${\cal J}$ symmetry results from the mapping
\be
\tau_\pm = -\check \tau_\mp + 2 H(\check \sigma_\mp),
\ee
with $\check \tau_-$ being the past hyperboloidal coordinate employed in Sec.~\ref{sec:Hyp_past_foliation}.

\end{document}